\documentclass[twocolumn,hyperref,footinbib]{revtex4-1}

\usepackage{bbm,amssymb}
\usepackage{amsmath}
\usepackage{times}
\usepackage{graphicx}
\usepackage{dcolumn}
\usepackage{bm}
\usepackage{float}
\usepackage{mathrsfs}
\usepackage{color}
\usepackage{hyperref}
\usepackage{booktabs}
\AtBeginDocument{
\heavyrulewidth=.08em
\lightrulewidth=.05em
\cmidrulewidth=.03em
\belowrulesep=.65ex
\belowbottomsep=0pt
\aboverulesep=.4ex
\abovetopsep=0pt
\cmidrulesep=\doublerulesep
\cmidrulekern=.5em
\defaultaddspace=.5em
}

\def\figref#1{Fig.~\ref{#1}}
\def\tableref#1{Table~\ref{#1}}
\def\ket#1{\left|#1\right\rangle}
\def\bra#1{\left\langle#1\right|}

\def\units#1{\,\mathrm{#1}}

\def\CZ{{\textsc{c}\hspace{-0.15pt}\textsc{z}}}

\newenvironment{cmatrix}{\left(\begin{array}{*{4}{r}}}{\end{array}\right)}

\begin{document}

\title{Photonic graph state generation from quantum dots and color centers for quantum communications}
\author{Antonio Russo}
\email{aerusso@ucla.edu}
\author{Edwin Barnes}
\author{Sophia E. Economou}
\email{economou@vt.edu}
\affiliation{Department of Physics, Virginia Polytechnic Institute and State University, Blacksburg, Virginia 24061, USA}
\date{\today}

\begin{abstract}
Highly entangled ``graph'' states of photons have applications in universal quantum computing and in quantum communications. In the latter context, they have been proposed as the key ingredient in the establishment of long-distance entanglement across quantum repeater networks. Recently, a general deterministic approach to generate repeater graph states from quantum emitters was given.  However, a detailed protocol for the generation of such states from realistic systems is still needed in order to guide experiments. Here, we provide such explicit protocols for the generation of repeater graph states from two types of quantum emitters: NV centers in diamond and self-assembled quantum dots. A crucial element of our designs is an efficient controlled-$Z$ gate between the emitter and a nuclear spin, used as an ancilla qubit.  Additionally, a fast protocol for using pairs of exchange-coupled quantum dots to produce repeater graph states is described. Our focus is on near-term experiments feasible with existing experimental capabilities.
\end{abstract}
\maketitle

\section{Introduction}

One of the main intermediate-term goals of quantum information science and technology is the creation of large-scale quantum networks. These would enable both cryptographic tasks, such as quantum key distribution \cite{Gisin_NatPhoton07,Scarani_RMP09,Ursin_NatPhys07,Liao_Nature17} and quantum conference key agreement \cite{Chen_ISIT05,Epping_NJP17}, as well as distributed quantum computation \cite{Beals_PRSA13,VanMeter_Computer16}, including the covert use of a remote quantum computer \cite{Sheikholeslami_IEEE16,Fitzsimons_npjQI17}, and the more ambitious endeavor of a quantum internet \cite{Kimble_Nature08,Pant_arxiv17,1601.00966v4,Azuma_2016,PhysRevA.96.032332}.  The crucial element of these protocols is the ability to share entanglement between remote nodes in the network. The main obstacle to achieving this comes from the limitations of optical components, especially optical fibers, which degrade the quantum state as it propagates, causing errors, including both decoherence and loss. To overcome this issue, concatenated protocols for entanglement distribution, known as quantum repeaters, were devised \cite{PhysRevA.85.062326}. This original protocol for quantum repeaters is based on quantum memories located at some of the nodes, which emit photons that are entangled with the memories.  Entanglement is spread across the repeater network through multiple entanglement swaps between pairs of photons emitted from neighboring memory nodes. While this scheme in principle addressed the problem of entanglement distribution across long distances, the realization of quantum repeaters, and thus quantum networks, is experimentally very challenging. The primary challenges are (i) the requirement of quantum memories with long coherence times and (ii) the probabilistic nature of Bell measurements between photons.

\begin{figure}
\centering
\includegraphics[width=0.92\columnwidth]{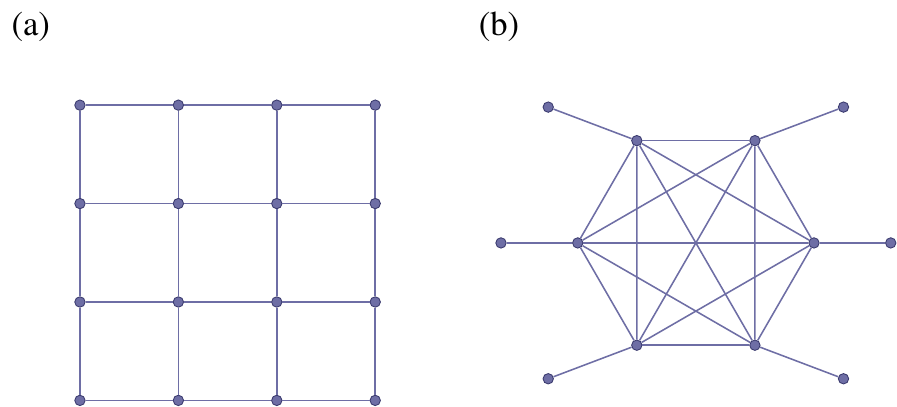}
\caption{Graph states. Each vertex represents a qubit, and edges represent entanglement between qubits. (a) Cluster state used for measurement-based quantum computation \cite{PhysRevLett.86.5188}, and (b) all-photonic repeater state introduced in Ref.~\cite{Azuma_2015}.\label{fig:rgsandcluster}}
\end{figure}

To overcome these limitations, new types of quantum repeaters have been suggested \cite{PhysRevA.85.062326,Azuma_2015,Muralidharan_SciRep16}. One of these \cite{Azuma_2015} completely eliminates the quantum memory, and instead uses highly entangled states of photons known as graph states. Graph states are entangled states that can be represented graphically. Perhaps the most well-known example of a graph state is the two-dimensional (2D) cluster state  [\figref{fig:rgsandcluster}(a)], a state which, when sufficiently large, is universal for quantum computing \cite{PhysRevLett.86.5188}. In this model, all the entanglement is created up front, and the computation proceeds with single-qubit measurements and feed forward, consuming the resource state in the process \cite{PhysRevLett.86.5188,PhysRevLett.86.910}. The graph states introduced in Ref. \onlinecite{Azuma_2015} as the main elements of a quantum repeater network are shown \figref{fig:rgsandcluster}(b). The special properties of graph states, reviewed in the next section, allow the establishment of entanglement between nodes in a repeater network, even when the probabilistic nature of photonic Bell measurements \cite{PhysRevLett.95.010501} is taken into consideration, essentially by building redundancy into the states (the multiple arms sticking out).  Moreover, in this scheme there is no need for long-lived quantum memories. In both quantum computing and communications, graph states can also be protected against photon loss by appropriately modifying the graph \cite{PhysRevLett.97.120501}.

The idea in \onlinecite{Azuma_2015} is to generate such a repeater graph state (RGS) at `primary' nodes of the network and to send half the photons in the state toward the neighboring node on the left and the other half toward the neighboring node on the right. The photons from adjacent (primary) nodes are then collected at intermediate, secondary nodes and subjected to Bell measurements, where each measured photon pair involves photons that originated from different primary nodes. By performing Bell measurements at these intermediate nodes all the way down the repeater network, long-distance entanglement spanning the entire network is created.

The key challenge with graph-state quantum computing and communications is the creation of the graph state. Because photons do not interact with one another, this process has to either be assisted by a nonlinear interaction, i.e., via the mutual interaction of two qubits with matter \cite{Munro_NJP2005,PhysRevA.73.062305,Hacker_Nature2016}, or implemented with a combination of linear optical elements and measurement, using so-called fusion gates \cite{PhysRevLett.95.010501}. The first approach has mostly yielded weak effective interactions. The latter is the approach researchers have mostly taken to create modest-sized photonic graph states, starting with Bell pairs from parametric down-conversion \cite{Zhao_2004,PhysRevLett.104.020501,Gao_2010}. Because this approach is inherently probabilistic, only about 10 photons have been entangled into a graph state to date. \cite{Gao_2010}

\begin{figure}
\centering
\includegraphics[width=0.45\columnwidth]{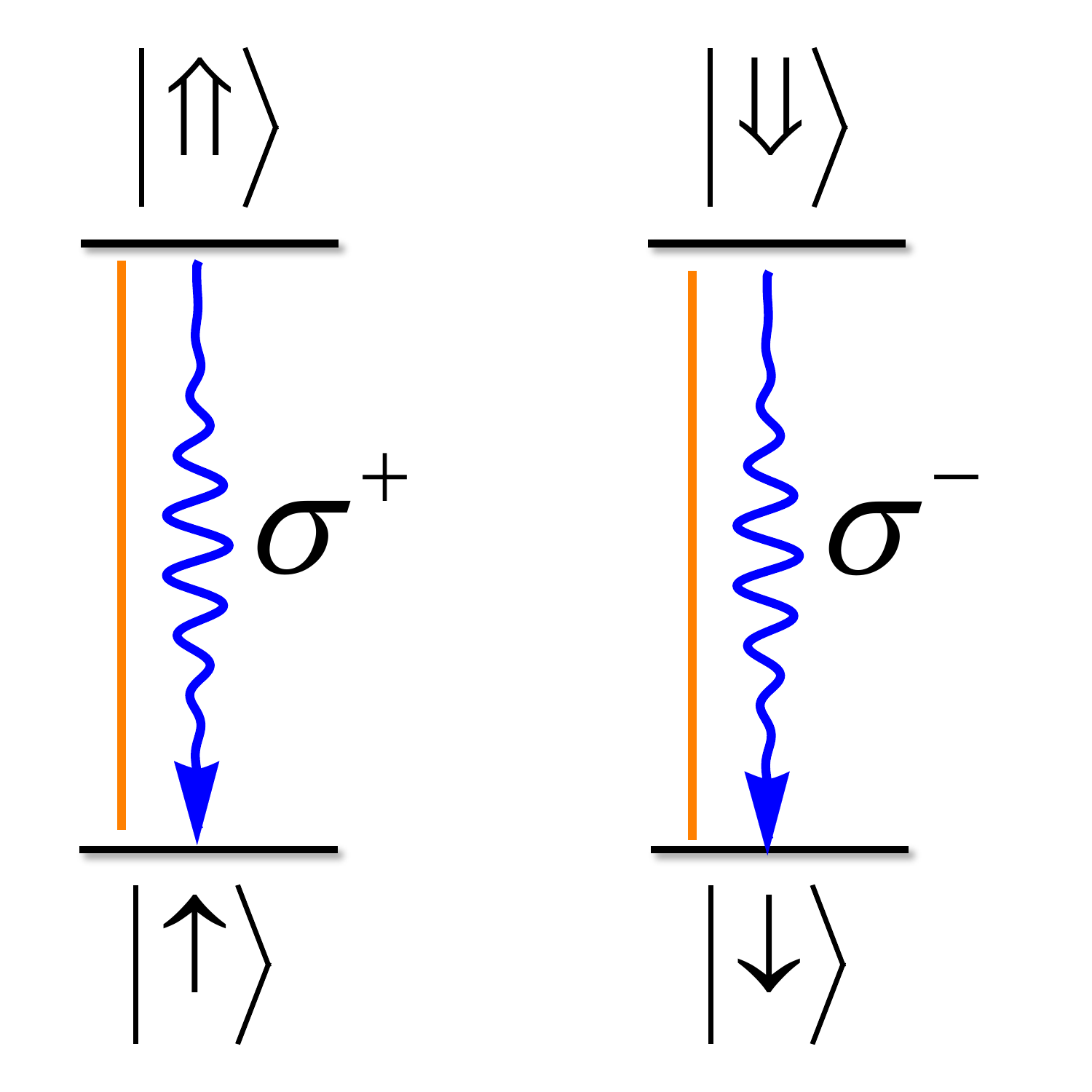}
\caption{Quantum emitter level structure needed to produce graph states. Two ground states each separately couple to one excited state via circularly polarized light of opposite polarization.\label{fig:levelstructure}}
\end{figure}

To address the challenge of constructing graph states in a more efficient way, a scheme was proposed in Ref.~\onlinecite{PhysRevLett.103.113602} that uses quantum emitters with a particular level structure and selection rules, as shown in \figref{fig:levelstructure}. By periodically pumping such an emitter and collecting the emitted photons, certain types of graph states can be obtained. For example, Greenberger-Horne-Zeilinger (GHZ) states \cite{Greenberger_1989} can be generated by periodic pumping alone, while applying unitary operations on the emitter in-between pumping cycles can create one-dimensional (1D) cluster states. A crucial possible advantage of this approach is that in the limit of a very efficient photon emission process, the protocol is essentially deterministic, assuming a few more requirements are satisfied, such as long coherence times in the ground state and the ability to perform unitary operations between the ground states of the emitter. In a recent breakthrough experiment \cite{Schwartz_2016}, this protocol was used to create a 1D linear photonic cluster state from a quantum dot (QD).

Because 1D cluster states are not universal for quantum computing, it is essential to grow the cluster along a second dimension. To generate more complex graphs such as a 2D cluster state, additional capabilities are needed compared to the 1D case. In Ref.~\onlinecite{PhysRevLett.105.093601}, it was shown that using two emitters, which can be controllably entangled through the use of a controlled-Z ($\CZ$) gate, a 2$\times N$ cluster state can be generated. To scale it up to an arbitrary sized $N\times N$ cluster state, $N$ emitters would be required, largely increasing the required overhead and capabilities.

Recently, we have discovered that the scaling is dramatically more favorable in the case of RGSs \cite{PhysRevX.7.041023}. In particular, we showed that an arbitrary-sized RGS can be generated using only one emitter of the structure of \figref{fig:levelstructure} coupled to one additional (ancilla) qubit, which in fact does not need to be an emitter. These modest requirements bring the generation of such states into an experimentally feasible regime with existing quantum emitters and photonic circuit capabilities. What is still required for an experimental generation of RGS states is a detailed protocol taking into account the particular quantum emitter's constraints and capabilities.

In this paper, we address this problem by providing explicit schemes for the generation of RGSs from NV centers in diamond and from self-assembled QDs.  Both these systems are natural for the generation of graph states, as they have the correct level diagram (\figref{fig:levelstructure}), they can be integrated with photonic elements, such as cavities \cite{Toishi_OE09,Englund_NanoLett10,Calusine_APL14,Grange_PRL15,Vora_NC15,Somaschi_NatPhoton16,Schroder_OME17,Bracher_PNAS17} and waveguides \cite{Luxmoore_PRL13,PhysRevLett.113.093603,Lodahl_RMP15,Lodahl_Nature17,Lang_JOpt17}, and they have been used to demonstrate spin-photon entanglement \cite{Togan_2010,DeGreve_Nature12,Gao_Nature12,Hensen_2015,Stockill_PRL17}.  Our focus here is on exploiting the capabilities of state-of-the-art systems and developing recipes that can be readily adapted in the laboratory. In this way, we hope to motivate experiments that demonstrate the creation of small or modest-sized photonic graph states. Such an experimental endeavor will help uncover challenges and opportunities pertaining to entangled graph state generation in these systems, which in turn will guide future theoretical efforts.

This paper is organized as follows. In Sec. \ref{graphstates}, we define graph states and briefly review their properties. In Sec. \ref{deterministicreview} we review previous protocols for the generation of particular types of graph states from quantum emitters. In Sec. \ref{qds} we develop new protocols for the generation of RGSs from quantum dots. We consider two tunnel-coupled QDs (``QD molecules'') and present a protocol for the generation of a RGS with six external arms. In Sec. \ref{nvs}, we develop in detail a protocol for the generation of a RGS from a NV center in diamond. The electronic transitions of the NV are exploited for the generation of the photonic graph states, while a nearby ${}^{13}$C nuclear spin is utilized as the necessary ancilla qubit. We develop a $\CZ$ gate between the electron and nuclear spins, as required for our protocol, and show that it is both fast and of high fidelity, even when the $^{13}$C happens to be several sites away from the NV. Finally, Sec.  \ref{final} provides a discussion and outlook.

\section{Graph states} \label{graphstates}

\begin{figure}
\centering
\includegraphics[width=\columnwidth]{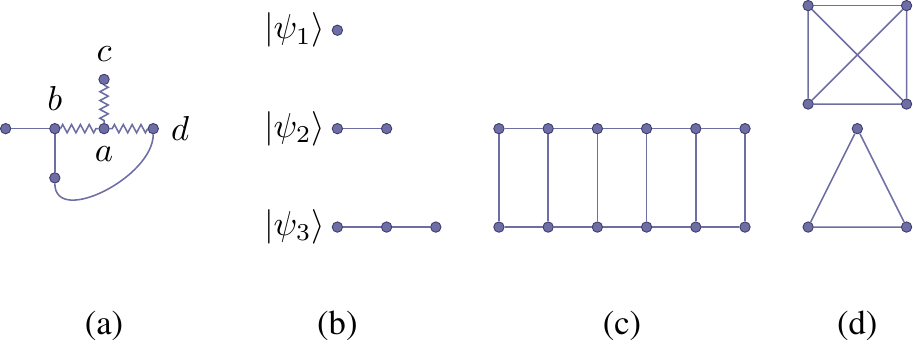}
\caption{Graph states and their construction. We use the notation $\ket{\pm}=(\ket{0}\pm\ket{1})/\sqrt{2}$, and remark that $\CZ\ket{\pm +}=(\ket{0+}\pm\ket{1-})/\sqrt{2} =(\ket{\pm0}+\ket{\mp1})/\sqrt{2}$.
(a) A graph state and stabilizer $K_a=X_a Z_b Z_c Z_d$, with edges shown in zig-zag pattern.
(b) Generation of linear cluster states: the emitter is initialized in $\ket{\psi_1}=\ket{+}$; after one pumping cycle, the emitter is entangled with the emitted photon such that the total state is $\ket{\psi_2}=\left(\ket{0+}+\ket{1-}\right)/\sqrt{2}$; after the second pumping cycle (including Hadamards), the second photon is entangled with both the first photon and the emitter so that the total state is $\ket{\psi_3}=\left(\ket{00+}+\ket{01-}+\ket{10+}-\ket{11-}\right)/2$, and so on.
(c) A ladder graph state, a simple two-dimensional cluster state.
(d) Fully connected, or ``complete,'' graphs.
\label{fig:graph-state}}
\end{figure}

A graph state \cite{PhysRevLett.86.5188,PhysRevLett.86.910,Hein_2004,PhysRevA.68.022312,PhysRevA.71.042323,PhysRevLett.91.107903,PhysRevA.65.012308} is defined as the simultaneous eigenstate (with eigenvalues equal to 1) of the stabilizer generators
\begin{equation}
 K_{G,a} = X_a \prod_{b\in V} Z_b^{\Gamma_{ab}},\label{graphstabilizers}
\end{equation}
where $G$ is a graph consisting of a set of vertices $V$ connected by edges according to the adjacency matrix $\Gamma_{ab}$, $a\in V$ is one particular vertex, and $X_a$ and $Z_b$ are single-qubit Pauli operators. Each qubit is represented as a vertex in the graph, and edges in the graph represent entanglement. As shown visually in \figref{fig:graph-state}(a), the stabilizer $K_{G,a}$ implements a Pauli $X$ on qubit $a$ and (simultaneously) $Z$ operations on all adjacent vertices $b$. Because there is one stabilizer for each qubit (or vertex) in the graph, the full set of stabilizers defined in (\ref{graphstabilizers}) specify a single multiqubit state. Both RGSs [\figref{fig:rgsandcluster}(b)] and cluster states [\figref{fig:rgsandcluster}(a)] are special types of graph states corresponding to particular choices of $\Gamma_{ab}$.

A graph state can also be defined constructively by first setting each qubit in the state $(\ket{0}+\ket{1})/\sqrt{2}$ and then applying a $\CZ$ gate between each pair of qubits connected by an edge of the graph \cite{Hein_2004,PhysRevLett.91.107903}. A graph consisting of two qubits connected by an edge corresponds to an entangled Bell pair (up to local operations on one of the qubits). Because each edge of the graph corresponds to entanglement, there is a clear sense in which entanglement extends throughout the entire multiqubit state. However, the entanglement depends on the specific layout of edges. For example, a fully connected graph consisting of edges between every pair of qubits corresponds to a GHZ state. The entanglement can be quantified by determining how many local measurements are needed to completely disentangle the state \cite{PhysRevLett.86.910}. A single local measurement can disentangle a GHZ state, while local measurements in a cluster state [\figref{fig:rgsandcluster}(a)] only affect the entanglement in a small region of the graph \cite{PhysRevLett.86.5188,PhysRevLett.86.910}, a feature which is crucial for measurement-based quantum computing and for quantum repeaters.

\section{Deterministic graph state generation} \label{deterministicreview}

\begin{figure}
\centering
\includegraphics[width=\columnwidth]{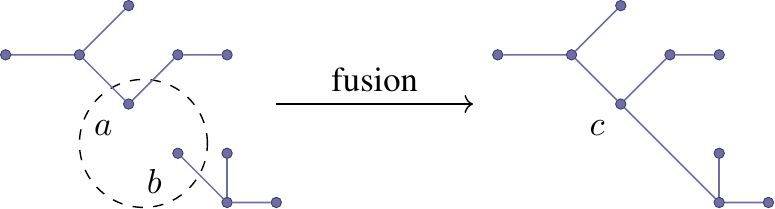}
\caption{A joint measurement is performed on photons $a$ and $b$ by passing them through a polarizing beam splitter, which reflects only one mode (say horizontal) of the photons. When only a single photon $c$ emerges, a new graph is formed, with $c$ inheriting all the edges of $a$ and $b$.  \label{fig:qubit-fusion}}
\end{figure}

The constructive definition of a graph state reviewed in the previous section suggests a conceptually simple way to create these states in the laboratory; however, this approach in practice is quite challenging to implement experimentally because a $\CZ$ gate between photons is not readily available.  A more practical method to create a photonic graph state is to combine existing photon Bell pairs via a process known as ``fusion'' \cite{PhysRevLett.95.010501}, which is summarized in \figref{fig:qubit-fusion}.  Under fusion, two existing graph states of photons are, with some probability, combined into a single graph state obtained by ``fusing'' a pair of vertices taken from each parent graph. Although this technique constituted a significant advancement over prior experimental protocols \cite{Knill_2001}, it has a high overhead in terms of resource requirements because the fusion process only succeeds probabilistically. For both quantum communication and computing, large graph states are required, and these would be prohibitively difficult to produce using just this method; experiments to date \cite{Zhao_2004,PhysRevLett.104.020501,Gao_2010,Pan_RMP12,PhysRevLett.117.210502} have managed to achieve 10 entangled photons using this fusion technique.

\begin{figure*}
\centering
\includegraphics[width=1.7\columnwidth]{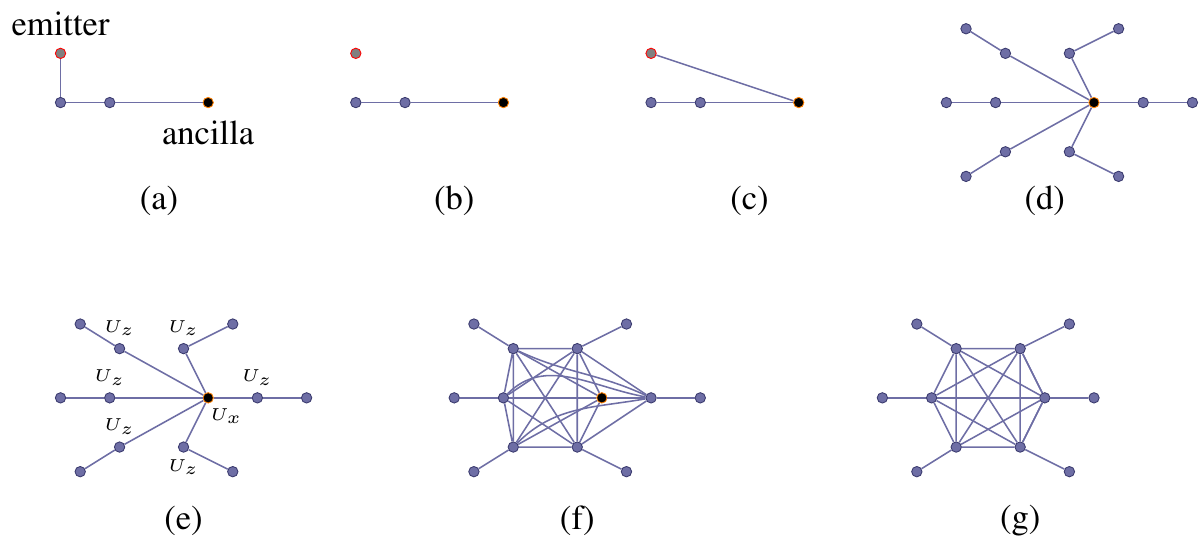}
\caption{ (a) An emitter entangled with an ancilla is pumped twice, interspersed with Hadamard gates. (b) The emitter is detached by measuring in the $z$ direction [see \figref{fig:graph-state} (b) for simplified wavefunction]. (c) The ancilla is entangled with the emitter by a $\CZ$ gate.  (d) Process is repeated to create six arms. (e) ``Local complementation'' is performed on the ancilla and core photons creating (f) a completely connected subgraph. (g) The ancilla is measured, removing it from the graph, creating the final repeater graph state. \label{fig:ancilla-anchor}}
\end{figure*}

This unfavorable scaling can be overcome by a deterministic generation protocol \cite{PhysRevLett.103.113602}. This protocol employs a four-state emitter with the level structure shown in \figref{fig:levelstructure}. The emitter consists of two ground states, $\ket{\uparrow}$ and $\ket{\downarrow}$, each of which couples optically to one excited state, $\ket{\Uparrow}$ and $\ket{\Downarrow}$, respectively. We assume that the selection rules are such that $\ket{\uparrow}$ couples to $\ket{\Uparrow}$ via $\sigma^+$ polarized light, while $\ket{\downarrow}$ couples to $\ket{\Downarrow}$ via $\sigma^-$ light (more generally, any set of orthogonal polarizations would do). In the following sections, we describe how this level structure and set of selection rules are realized in both semiconductor quantum dots and NV centers in diamond. If the emitter is first prepared in the superposition state $\left(\ket{\uparrow}+\ket{\downarrow}\right)/\sqrt{2}$ and then pumped with linearly polarized light (so that both transitions are excited), then the system ends up in the state
$
\left(\ket{\Uparrow}+\ket{\Downarrow}\right)/\sqrt{2}.
$
This excited state then decays back to the ground state, and since the selection rules are obeyed during this process, the final state describing both the emitter and the emitted photon is
\begin{equation}
\left(\ket{\uparrow}\ket{\sigma^+}+\ket{\downarrow}\ket{\sigma^-}\right)/\sqrt{2}.
\end{equation}
Repeating the pumping process $n$ times results in the maximally entangled GHZ state
\begin{equation}
\left(\ket{\uparrow}\ket{\sigma^+}^{\otimes n}+\ket{\downarrow}\ket{\sigma^-}^{\otimes n}\right)/\sqrt{2}.
\end{equation}
This corresponds to star graphs (or complete graphs with local complementation, which are illustrated in \figref{fig:graph-state}(d)).

If we also include an additional Hadamard gate on the emitter between every pumping step, then a linear cluster state is produced instead of a GHZ state \cite{PhysRevLett.103.113602}. Each pumping process adds one more photon to the linear chain, and the size of the final cluster state is ultimately determined by the number of photons that can be reliably produced before the emitter decoheres. By including a second emitter that couples to the first, this procedure can be generalized further to produce a ``ladder'' graph state (\figref{fig:graph-state}(c)) by including an entangling $\CZ$ gate on the emitters between each pumping cycle \cite{PhysRevLett.105.093601}. Additional classes of graphs can be obtained by varying when the $\CZ$ gates are applied and by performing additional, appropriately timed single-qubit operations \cite{PhysRevX.7.041023} (see \figref{fig:ancilla-anchor}). In the following sections, we describe how these operations can be performed in quantum dot and NV center systems to produce the RGSs needed for long-distance quantum communication, and we discuss some of the experimental considerations that arise.

\section{Quantum Dots} \label{qds}

QDs have a particularly high spontaneous emission rate compared to other quantum emitters, and their selection rules are robust \cite{Senellart_NatNano17}. Single QDs cannot generate RGSs or 2D cluster states directly---coupling to additional emitters and/or ancillas is required.  For instance, for the 2D ladder cluster state shown in \figref{fig:graph-state}(c), the scheme of \cite{PhysRevLett.105.093601} employs two QDs. This protocol is similar to that for producing a 1D cluster state from a single QD in terms of the pumping and Hadamard sequence; the additional element here is an entangling $\CZ$ gate between the QDs at every step of the 1D protocol. If such a $\CZ$ gate is available between the two QDs, then an arbitrary-sized RGS can also be generated by following the protocol of Ref.~\cite{PhysRevX.7.041023} and summarized in \figref{fig:ancilla-anchor}.  Here, only one QD would be pumped to produce photons, and the other one would be used as an ``anchor'' for the arms of the RGS while the remaining arms are being produced.

\begin{figure*}[htp]
\centering
\includegraphics[width=1.6\columnwidth]{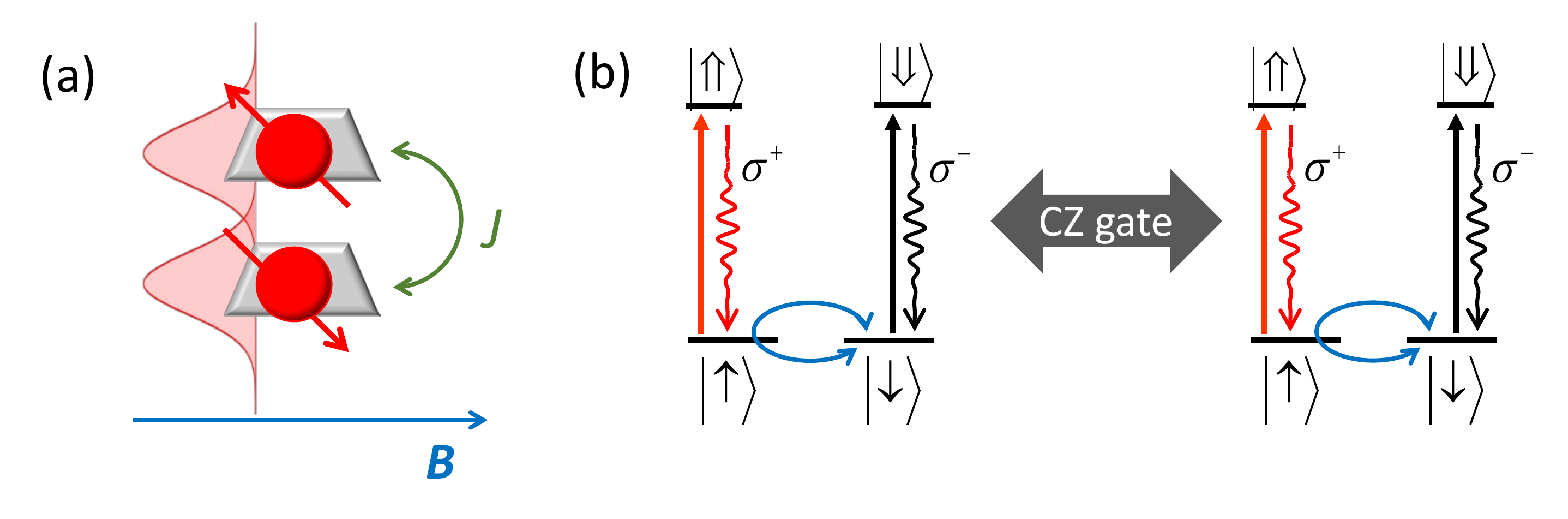}
\caption{(a) A stacked double quantum dot with one electron in each dot. The two electron spins are coupled via exchange interaction $J$ and subject to a magnetic field applied orthogonally to the growth direction. (b) Free evolution under the exchange interaction can generate the $\CZ$ gate necessary to produce RGSs. The magnetic field implements the single-spin Hadamard gates.\label{fig:qdcartoon}}
\end{figure*}

Here, we show that the necessary coupling for the $\CZ$ gate can be realized by taking advantage of progress in growing two or more QDs on top of each other such that they are tunnel-coupled \cite{PhysRevLett.94.057402,Stinaff636,Bayer451}. We show that an always-on exchange coupling between two spins in vertically stacked quantum dots combined with single-qubit gates can generate the required $\CZ$ gate (see \figref{fig:qdcartoon}), and we provide the full protocol for the generation of a modified RGS of eight photons.

\begin{figure}
\centering
\includegraphics[width=0.4\columnwidth]{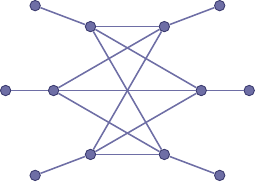}
\caption{A modified RGS that is equivalent to the original RGS of \cite{Azuma_2015}: the missing connections are local to each secondary node and are thus not required for the repeater protocol. \label{fig:modifiedRGS}}
\end{figure}

\begin{figure}
\centering
\includegraphics[width=1\columnwidth]{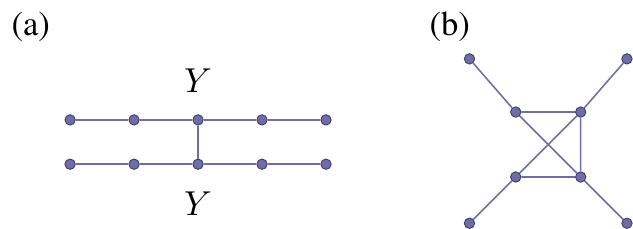}
\caption{(a) Single-rung ladder state produced from two coupled QDs.  Performing two $Y$ measurements on the central photons produces (b) an ``almost'' RGS. The missing edge does not affect the functionality of the state for quantum repeaters. \label{fig:almostrgs}}
\end{figure}

\begin{figure*}
\centering
\includegraphics{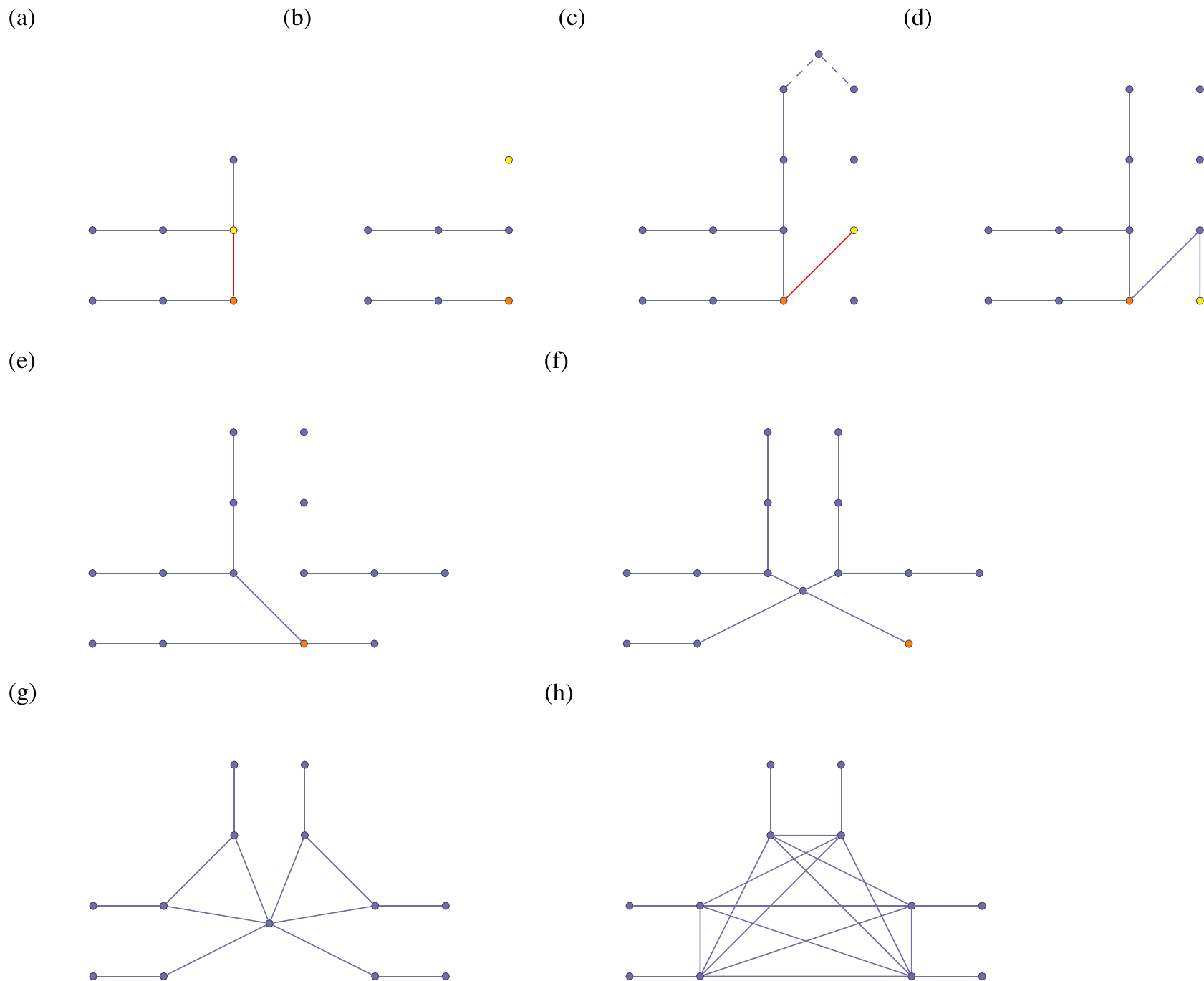}
\caption{(a) Two quantum dots (yellow and red) are pumped to produce a pair of entangled photons each. The yellow dot is pumped a third time, and a $\CZ$ gate (red line) is applied between the dots. (b) Hadamards are performed on the last photon and yellow dot, realizing two local complementations: one on the quantum dot, and another on the last photon. (c) The yellow dot is alternatingly pumped and Hadamard-rotated five times to produce a linear cluster state, and the third (top) photon is measured in the $z$ basis to separate the ``loop'' into two arms; an additional sixth pumping, without Hadamard, is performed\footnote{The Hadamard performed in step (d) can be performed at this point instead, because the $\CZ$ gate does not affect the photon. The visualization, however, is simpler. Notice that the Hadamard on the \emph{quantum dot} must be performed in the stated order.}. The $\CZ$ gate between dots is then applied a second (and final) time. (d) Hadamards are performed on the last photon and yellow dot, again creating two local complementations as in (b). (e) The yellow dot is pumped to produce a linear cluster of length two, and then disconnected (by either direct measurement or measurement of another emitted photon). The red dot is pumped once. (f) Hadamards are performed on the last photon and the red dot, again realizing two local complementations as in (b). (g) Local complementations (realized by wave plate operations) are performed on the two ``corner'' photons, and then measured in the $z$ basis. Simultaneously, the red dot is pumped to produce a linear cluster state of two photons, and then disconnected. (h) A local complementation is performed on the ``core'' photon (realized by local operations on the core photon and six attached interior photons) and then measured in the $z$ basis. The resulting RGS is only missing two edges, and can be used in the repeater protocol without any limitation.  \label{fig:imperfect-sixleg-rgs}}
\end{figure*}

Since it is generally challenging to create connections (entanglement) between the emitted photons, it is natural to ask whether the inner photons in e.g., \figref{fig:ancilla-anchor}(g) need to be fully connected. This is in fact not the case. Consider a modified RGS, where each photon sent to the left (right) node is fully connected to each photon sent to the right (left) node. Such a state, depicted for 12 photons in \figref{fig:modifiedRGS}, has all the desirable properties of the original RGS introduced in Ref. \cite{Azuma_2015}.  This is because no inter-node entanglement needs to be established, and the local connections are thus redundant. Two such RGS generation protocols are described, one for four photons, \figref{fig:almostrgs}, and one for six photons, \figref{fig:imperfect-sixleg-rgs}, representing short- and intermediate-term objectives, with a commensurate increase in difficulty and utility.

The four-armed RGS requires only one use of a $\CZ$ entangling gate between the emitters, along with the standard requirements of pumping and a Hadamard or $\pi/2$ rotation about an axis in the $xy$ plane. The protocol proceeds as follows: both emitters are pumped as in Ref.~\cite{PhysRevLett.103.113602} for two periods, so that each QD generates a linear cluster state with two photons.  Next, each QD undergoes a Hadamard ($\pi/2$ rotation), followed by a $\CZ$ gate between the QD spins and a pumping operation, so that one more photon per QD is generated, and the photons are entangled with each other.  After this step, the first part of the protocol is repeated, producing two more photons in the cluster state of each QD. Note that to decouple the QDs from the cluster, one extra photon per dot can be produced and measured (since single-photon measurements are simpler than single-shot spin measurements).  Next, each of the central photons (the only ones that are connected to each other from a different chain) are measured in the $y$ basis (see \figref{fig:almostrgs}(a)).  This in turn produces enough connectivity that a modified photonic RGS with four (not fully connected) core photons is produced (\figref{fig:almostrgs}(b)). The photons with a missing link are sent to the same secondary node for entanglement swapping with photons from another RGS.

To implement this protocol, the recent proposal of \cite{Gimeno_Segovia_2017} may be adapted, which makes use of the electron exchange coupling between the two QDs (along with a Zeeman term) to create the entangling gate (see \figref{fig:qdcartoon}). Moreover, the simplest single-qubit gates to implement in QDs are $Z$ gates due to the reduced symmetry along the growth ($z$) axis \cite{PhysRevB.74.205415,Press_Nature08,Greilich_NP09}. An in-plane field (which defines the $x$ axis) is also required. The system in the ground state is governed by the Hamiltonian
\begin{equation}
 H_{qd} = J s_1\cdot s_2 + \omega s_{1x} + \omega s_{2x},
\end{equation}
where $J$ is the exchange coupling, and $\omega$ denotes the Zeeman splitting due to a magnetic field along the $x$ direction. Free evolution for time $\Delta t=2\pi/J$ yields an $x$ rotation $R_x(-2\pi\omega/J)$, which combined with $z$ rotations can give arbitrary $X$ gates \cite{Gimeno_Segovia_2017}. As a result, we can implement individual $R_x(\pi/2)$ gates on both spins, even though the system is coupled. Free evolution for a quarter of the time, $\Delta t=\pi/(2J)$, gives an entangling operation:
\begin{equation}
 U(\Delta t) = \left( e^{-i\pi \omega/(2J) s_x}\otimes e^{-i\pi \omega/(2J) s_x}\right) \sqrt{\mathrm{SWAP}},
\end{equation}
enabling the implementation of a $\CZ$ $H\otimes H$ gate via a combination of free evolution and single-qubit gates \cite{Gimeno_Segovia_2017}:
\begin{eqnarray}
 \CZ(H'\otimes H') =\hspace{-7em}&\hspace{7.3em} (e^{-i\frac{\pi}{2} s_{z}}\otimes e^{i \frac{\pi}{2} s_{z}})(e^{-i \frac{\pi \omega}{2J} s_{x}}\otimes e^{-i \frac{\pi \omega}{2J} s_{x}}) U(\Delta t)\nonumber \\
&\times (\mathbbm{1} \otimes e^{-i\pi s_{z}})\left( e^{i\frac{\pi}{2}(\frac{\omega}{J} + 1) s_{x}}\otimes e^{-i\frac{\pi}{2}(\frac{\omega}{J} + 1) s_{x}} \right) U(\Delta t),
\end{eqnarray}
where $H'$ is a rotation around the $x$ axis by $\pi/2$ and is a sufficient substitute for a Hadamard for the purposes of the protocol. Accordingly, single-qubit rotations interspersed with timed free evolution provide a $\CZ$ gate. The realistic case with unequal Zeeman coupling is addressed in \cite{Gimeno_Segovia_2017} and proceeds analogously. Additionally, a careful choice of parameters can remove the need for some of the single-qubit gates.

The six-armed variant requires two uses of entangling gates (in contrast to the original RGS shown in  \figref{fig:ancilla-anchor}, which requires \emph{six} entangling operations) along with standard single-photon unitaries and Hadamard operations on the QDs. Roughly, where the four-armed version attaches two 1D cluster states at one point, the six-armed protocol attaches the 1D cluster states at two points, adding a ``loop'' which is cut to produce the two additional arms, though care must be taken to ensure the proper connectivity of the core photons of the RGS. See \figref{fig:imperfect-sixleg-rgs} for a detailed motivation of the protocol:
\begin{enumerate}
\item Two QDs are initialized into $\ket{+}\otimes\ket{+}$.
\item Each QD is pumped twice.
\item The first QD is optically excited and allowed to emit.
\item The entanging $\CZ$ gate is performed on the QDs.
\item The first QD is pumped five times, and the third photon is measured in the $z$ basis.
\item The first QD is optically excited and allowed to emit.
\item The entanging $\CZ$ gate is performed on the QDs.
\item The first QD is pumped three times, and the last photon is measured in the $z$ basis.
\item The second QD is pumped four times. The first photon is measured in the $y$ basis, and the last photon is measured in the $z$ basis.
\end{enumerate}

Both of the above protocols require that the QDs emit much faster (by a factor of 5 to 10) than both the Larmor precession frequency \cite{PhysRevLett.103.113602} and the exchange timescale $1/J$. Each iteration of the protocol itself must be short enough that many cycles can be completed well before the spin coherence time is reached. The QD emission time, $T_1$, can be reduced from $\sim\!1\units{ns}$ \cite{PhysRevB.90.155303} to $\sim\!100$ ps by coupling the QD to a cavity via the Purcell effect \cite{Birowosuto_2012,Kelaita_2016}.  Off-resonant pulses can be used to optically control single qubits \cite{PhysRevB.74.205415,Press_Nature08,Greilich_NP09,Carter_2013}. The sequence takes time on the order of $32\pi/J$, corresponding to $\sim\!20\units{ns}$ for couplings on the order of $J\sim\omega\sim 5\units{GHz}$. The electron coherence time, which is typically $T_2\sim 1\units{\mu s}$ and can be improved further by decoherence pulses \cite{PhysRevLett.116.033603}.  Moreover, the fidelity of the $\CZ$ gate can be shown to be $\gtrsim0.99$ with frequency errors on the order of $10\%$ for the equal-Zeeman coupling case, and order $0.1\%$ for the unequal case \cite{Gimeno_Segovia_2017}.  \tableref{table:qd-parameters} summarizes the physical requirements for our protocol.
\begin{table}
\caption{Requirements for RGS generation in QDs\label{table:qd-parameters}}
\begin{tabular}{ll}
\toprule
\multicolumn{1}{c}{Quantity}& \multicolumn{1}{c}{Descriptions}\\
\midrule
$T_1$& Spontaneous emission time\\
$T_2$ & Electron coherence time\\
$J$ & Exchange Coupling\\
$\omega$ & Zeeman Splitting\\
\midrule
\multicolumn{2}{c}{$T_1\ll \frac{2\pi}{J}\sim\frac{2\pi}{\omega}\ll T_2$}\\
\bottomrule
\end{tabular}
\end{table}

An important consideration that must be addressed for practical implementations is how errors propagate throughout the RGS generation protocol. If a single error in either a gate or a pumping process leads to catastrophic failure of the RGS, the success of the protocol will depend on achieving exponentially small error rates. We find that the presence of a single error usually leads to the corruption of only a limited number of photons. As in \cite{PhysRevLett.103.113602}, any single Pauli error on either QD leads to at most two corrupted emitted photons during the pumping protocol: each successive pumping-Hadamard operation converts a $X$ (or $Y$) error on the emitter to a $Z$ error, and finally the \mbox{(non-)error} $I$, while causing the emitted photons to suffer $X$ (or $Y$), and then $Z$ errors, respectively. Our protocol differs from the 1D cluster protocol in one critical respect: the entangling operation. Unlike the pumping protocol, in which errors naturally ``localize,'' there is no natural limit to the degree of contamination from corrupted QDs entangling with the other QD: $X$ and $Y$ errors on one QD lead to $Z$ errors on the other ($Z$ errors commute with $\CZ$). The precise evolution of the wave function is detailed in Appendix~\ref{appendix:error}. This highlights the importance of reducing the number of requisite $\CZ$ gates by using the ``almost'' RGSs shown in Figs.~\ref{fig:almostrgs} and \ref{fig:imperfect-sixleg-rgs} instead of the full RGSs.

The consequence of a $Z$ error on either emitter will be a single photon with a $Z$ error. If that photon is along the line of photons connecting successful Bell measurements on opposite sides of the RGS, the entire line of photons will be corrupted, and the repeater protocol fails. If, on the other hand, it does not lie along this line, then the $Z$ error commutes with the $Z$ measurement on the affected photon that disconnects the photon from the graph, and the error contributes only an irrelevant and undetectable overall phase.  On the other hand, measuring any photon with $X$ or $Y$ errors corrupts the protocol.  Alternatively, an error on the QD during the entangling process ($\CZ$ gate) contaminates the other QD: $X$ or $Y$ errors lead to $Z$ errors on the other (see Appendix~\ref{appendix:error}). Any errors on such central nodes, e.g., the central photon of \figref{fig:imperfect-sixleg-rgs}(f) or either $Y$-measured photons of \figref{fig:almostrgs}(a) will result in protocol failure. The dominant source of error in these QD systems is expected to be from spin dephasing, corresponding to $Z$ errors on the emitters; fortunately, these errors affect at most one emitted photon.

Finally, to use these dots as photonic sources, there must be some distinction between the input photons used to manipulate the system, and the output photons, and there must be some way to collect the photons. To distinguish between input and output photons, simply turning off the driving laser quickly can temporally separate the photons provided the laser is fast compared to the spontaneous emission time of the QDs ($T_1\sim 1\units{ns}$). To collect the photons, distributed Bragg reflectors \cite{2016arXiv160902851A} or micropillar resonators \cite{Gazzano_NC13,Arnold_NC15} can be used.  Alternatively, QDs can be coupled to photonic crystal structures and thus integrated on-chip \cite{PhysRevLett.95.013904,Lodahl_Nature2004,Faraon_APL2007}. Recently, chiral photonic-crystal waveguides have been used to demonstrate collection efficiencies from QDs exceeding $98\%$ \cite{PhysRevLett.113.093603}, making them an interesting choice for photon collection. Moreover, the propagation direction in these waveguides is determined by the chirality, opening the possibility for quantum networks based on these light-matter interactions \cite{PhysRevLett.117.240501,Lodahl_Nature17}, allowing, e.g., for on-chip polarization measurements.

\section{Nitrogen-Vacancy Color Centers} \label{nvs}

Negatively charged NV centers in diamond also have the correct level structure at zero B field, along with extremely long spin coherence times. Moreover, spin-photon entanglement has been demonstrated \cite{Togan_2010,Bernien_2013} and used for heralded spin-spin entanglement \cite{Hensen_2015}. An additional attractive property of this system is that the diamond crystal features highly stable nuclear spins ($^{13}$C isotopes), in close enough proximity to allow for hyperfine interaction (and concomitant entanglement) with the NV spin \cite{Taminiau_NatNano2014}. Here, we consider such a nuclear spin to play the role of the ancilla qubit in our RGS generation protocol. While considerable progress in controlling nuclear spins coupled to NV centers has been made over the past decade \cite{Dutt1312,Neumann1326,Fuchs_NatPhys2011,Taminiau_NatNano2014,Cramer_NatComm2016},
the required $\CZ$ gate for the RGS generation protocol is lacking. Below, we provide an experimentally friendly microwave-based $\CZ$ gate between the two spins and show how to speed it up by an order of magnitude compared to a naive design by using pulse shaping.

\begin{figure}
\centering
\includegraphics{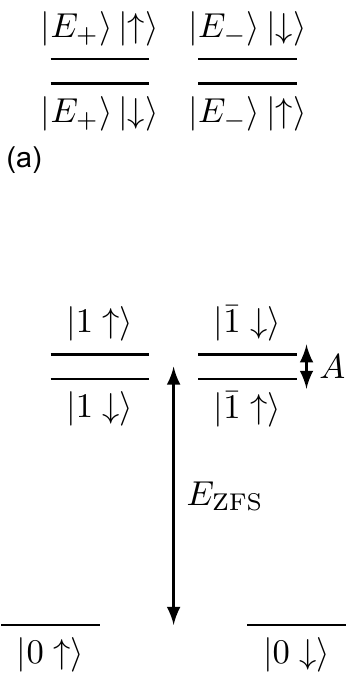}\qquad
\includegraphics[width=0.8\columnwidth]{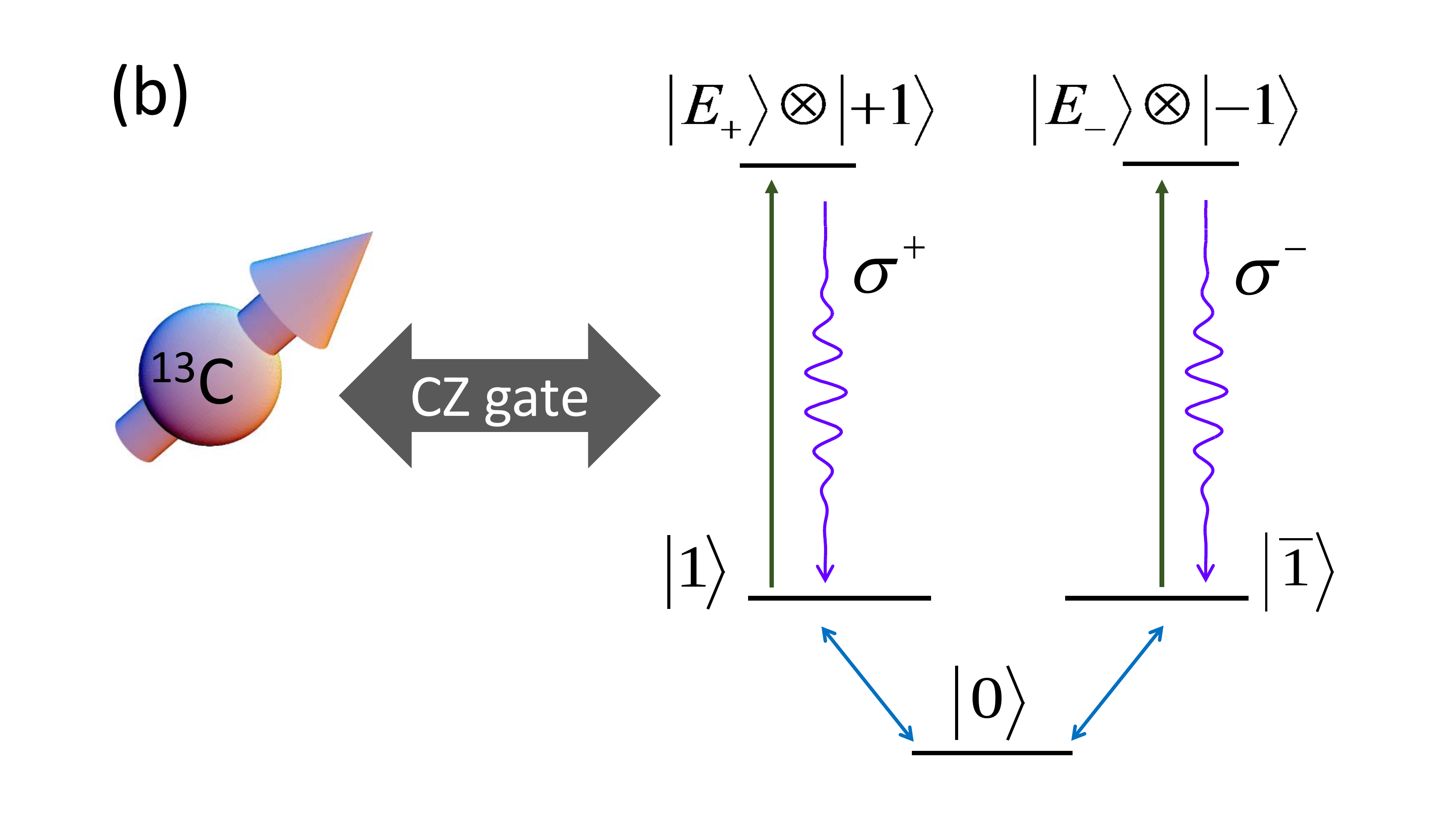}
\caption{(a) Energy level diagram for NV centers in diamond. Hyperfine coupling $A$ is much smaller than the zero-field splitting $E_\mathrm{ZFS}$.  Selection rules guarantee that decays to the $\ket{1,\uparrow}$ or $\ket{1,\downarrow}$ states have different polarizations. (b) A nearby $^{13}$C nuclear spin can be used as the ancilla qubit in the RGS generation protocol. Microwave driving (blue arrows) between the $S_z=0$ and $|S_z|=1$ states can be used to implement Hadamard gates and entangling $\CZ$ gates.  \label{fig:color-center-level-diagram}}
\end{figure}

Negatively charged NV centers in diamond have two degenerate states with spin projection $|S_z|=1$ at zero B field and several optically excited states. Two of these excited states have the correct quantum numbers to provide the desired level diagram \cite{Togan_2010}. A remarkable property of the NV is its long spin coherence time. While the optical decay rate $T_1$ is about an order of magnitude less than that of QDs ($\sim\!0.1$ GHz), the spin coherence times $T_2^\mathrm{spin}$ can more than make up for that, exceeding a millisecond at room temperature \cite{Balasubramanian_2009}. In addition, the NV ground-state manifold includes a $S_z=0$ state, separated in energy from the $|S_z|=1$ states. This state provides a way to couple the two active states to each other to implement the necessary single-qubit gates \cite{Fuchs_2009}.

Either the $^{15}$N or a nearby ${}^{13}$C nucleus ($I=1/2$ nuclear spin) can be used as the ancilla of the protocol in \cite{PhysRevX.7.041023}. The energy level structure of the combined system is modified by the hyperfine Hamiltonian,
\begin{equation}
 H_\mathrm{HF} = A \vec S\cdot\vec I = AS_zI_z + A/2(S^+I^-+S^-I^+),
\end{equation}
where $A\sim 50 \mathrm{kHz}\ll E_\mathrm{ZFS}\sim \mathrm{GHz}$, i.e., the hyperfine coupling is much smaller than the zero-field splitting. The flip-flop terms can therefore be neglected, and we get the energy levels of \figref{fig:color-center-level-diagram}. Using these NV centers as an entangled photon source requires initialization, pumping, and then (if a linear cluster state is desired) a Hadamard gate, which will be described here. The states $\ket{1},\ket{E_+},\ket{\bar 1},\ket{E_-}$ play the roles of $\ket{\uparrow},\ket{\Uparrow},\ket{\downarrow},\ket{\Downarrow}$ in \figref{fig:levelstructure}, respectively. In the remainder of this section, we will use $\ket{\uparrow}$ and $\ket{\downarrow}$ to denote the two {\it nuclear} spin states.

\begin{figure}
\centering
\includegraphics[width=0.85\columnwidth]{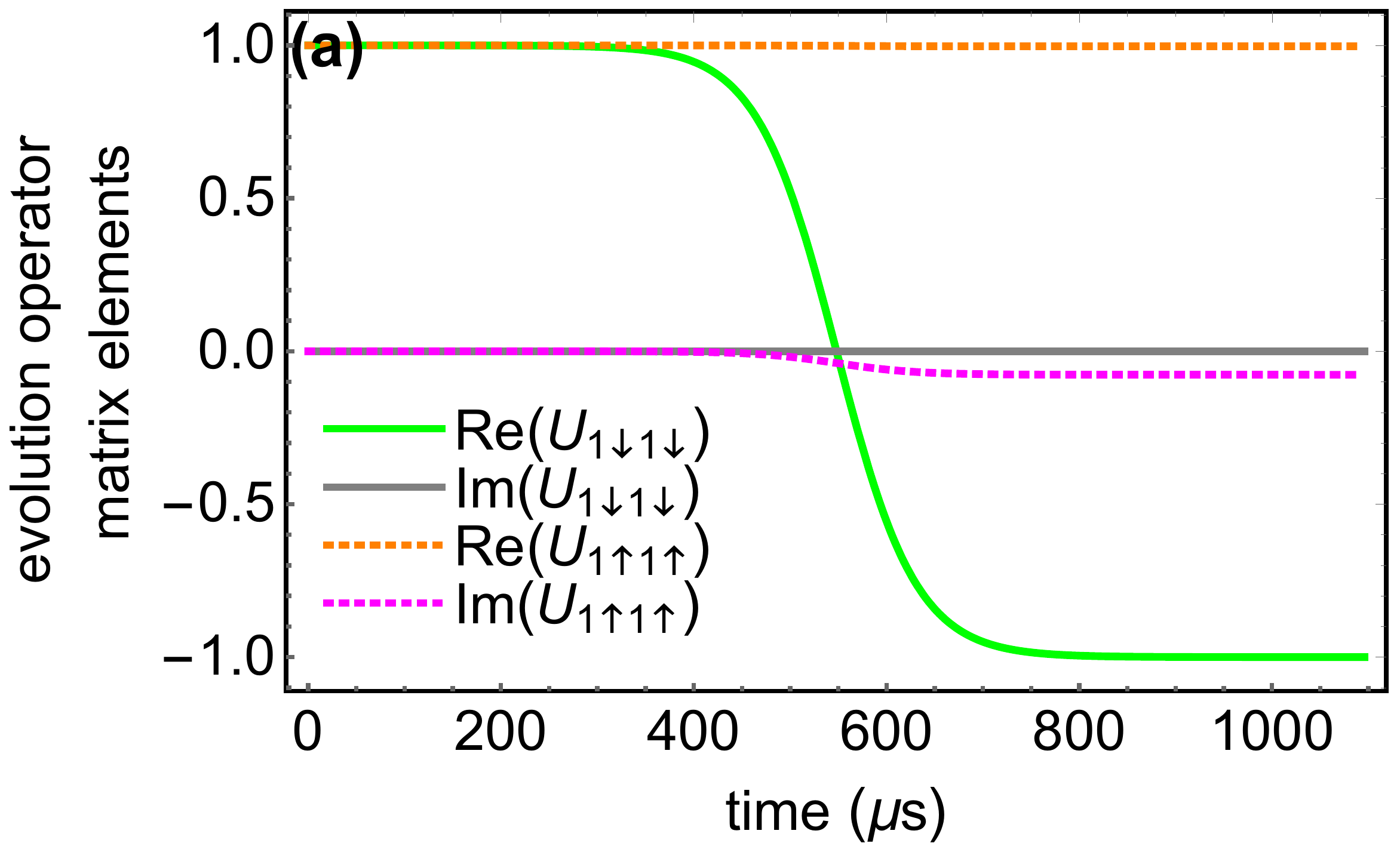}\qquad
\includegraphics[width=0.85\columnwidth]{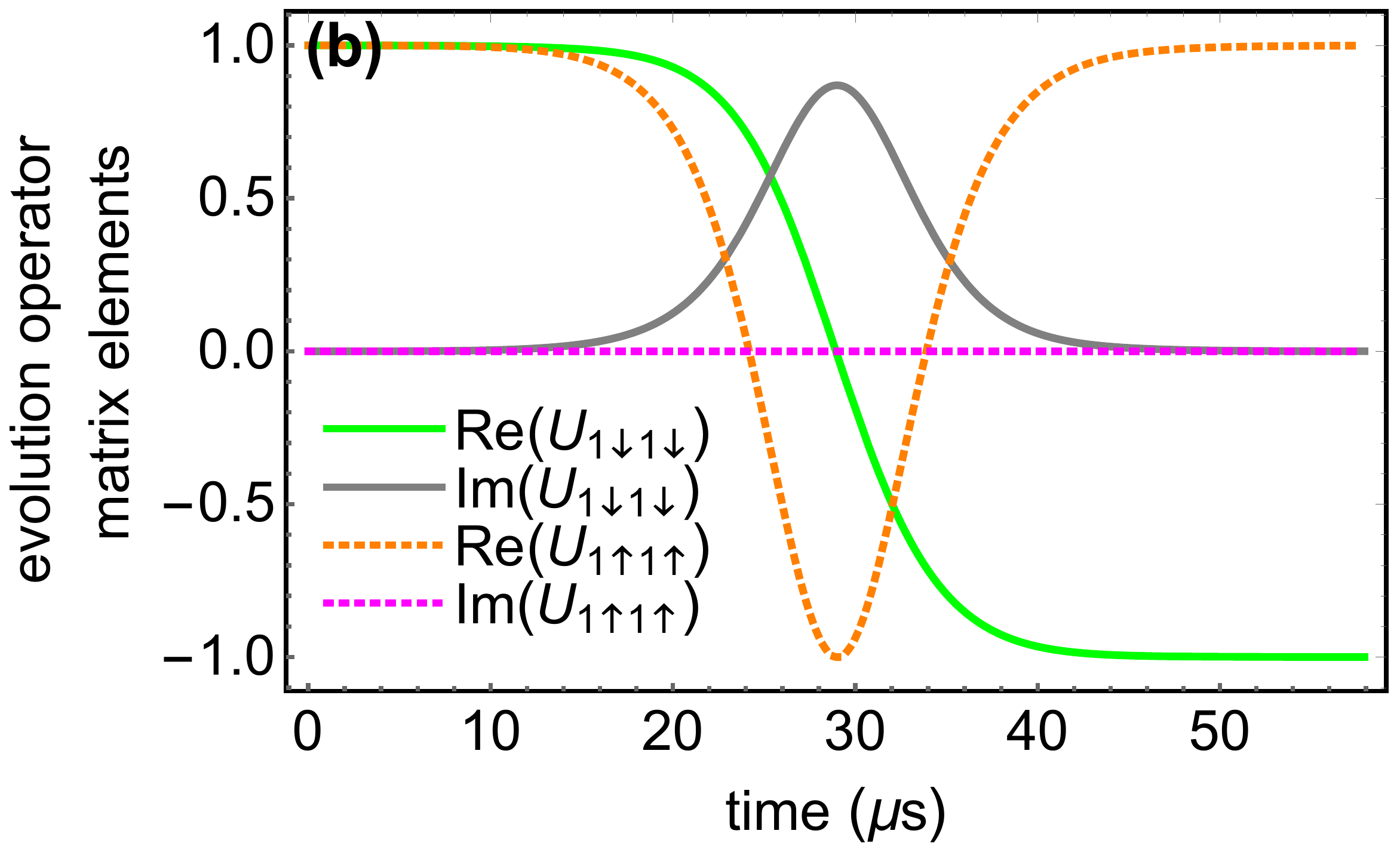}
\caption{NV-nuclear spin entangling $\CZ$ gate implemented with right-circularly polarized sech pulse, $\Omega(t)=\Omega_0\hbox{sech}(\sigma (t-t_0))$. (a) Frequency-selective, resonant 2$\pi$ pulse ($\Omega_0=\sigma$) induces a minus sign on the target transition $\ket{0\downarrow}\leftrightarrow\ket{1\downarrow}$ while approximately avoiding the competing transition $\ket{0\uparrow}\leftrightarrow\ket{1\uparrow}$.  (b) Designed $4\pi$ pulse ($\Omega_0=2\sigma$) that provides a twentyfold speedup compared to (a) by driving both transitions and inducing a $-1$ phase factor to the target transition and a $+1$ phase factor to the competing transition (with which the pulse is resonant). The other logical states, $\ket{\bar 1\downarrow}$ and $\ket{\bar 1\uparrow}$ are not affected by the pulse. Here, we have taken $A=50\units{kHz}$, $E_\mathrm{ZFS}=2.88\units{GHz}$, $\sigma=1.9\units{kHz}$ (a) and $\sigma=28.9\units{kHz}$ (b), and gate durations $2t_0=1.1\units{ms}$ (a) and $2t_0=58\units{\mu s}$ (b), respectively.\label{fig:CZgates}}
\end{figure}

\begin{figure}
\centering
\includegraphics[width=0.85\columnwidth]{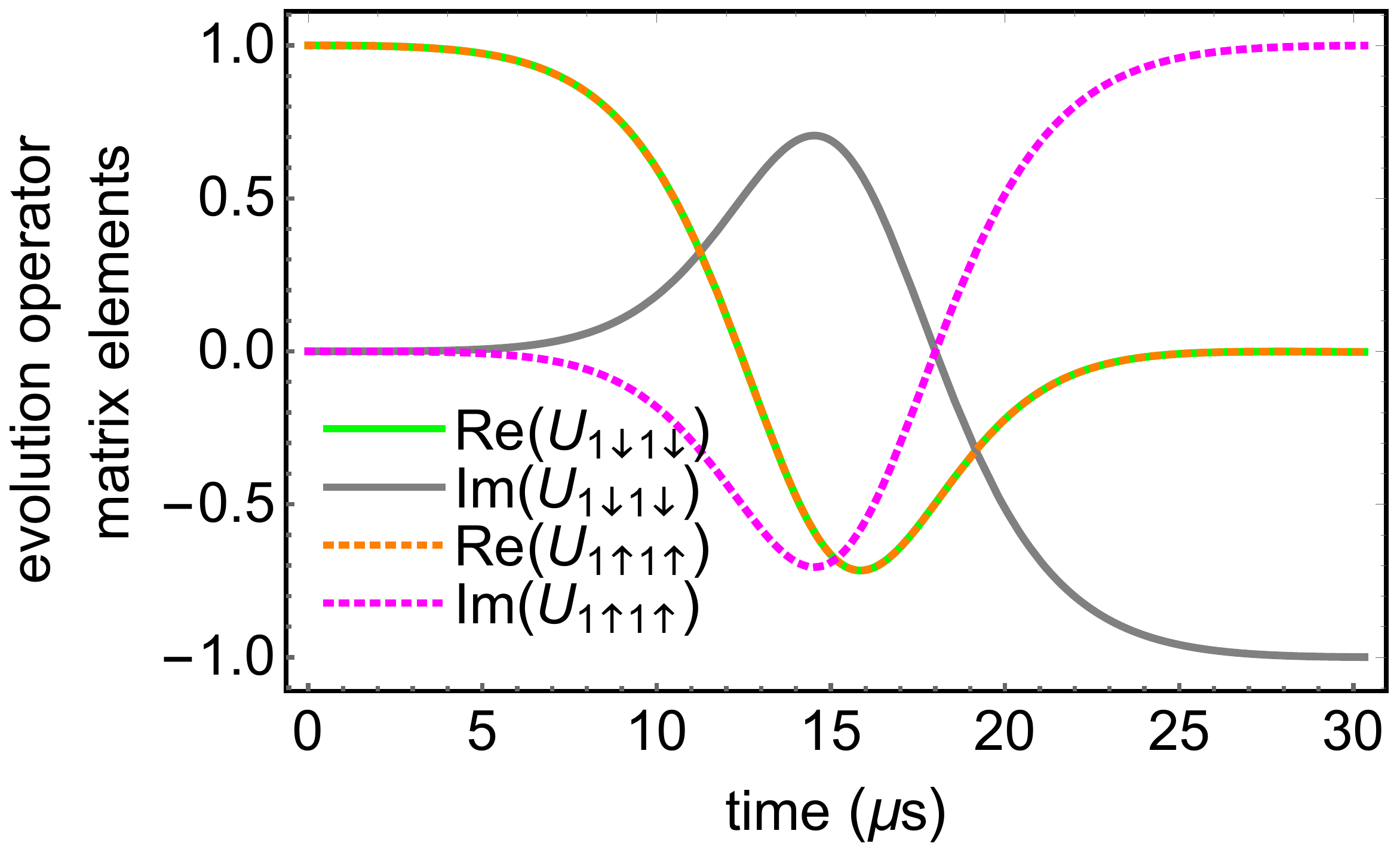}
\caption{Fastest NV-nuclear spin $\CZ$ gate, obtained by setting the pulse frequency halfway between the $\ket{0,\downarrow}\leftrightarrow \ket{1,\downarrow}$ and $\ket{0,\uparrow}\leftrightarrow \ket{ 1,\uparrow}$ transitions and using a right-circularly polarized $4\pi$ sech pulse. The other logical states, $\ket{\bar 1\downarrow}$ and $\ket{\bar 1\uparrow}$ are not affected by the pulse. Here, we have taken $A=50\units{kHz}$, $E_\mathrm{ZFS}=2.88\units{GHz}$, $\sigma=38.7\units{kHz}$, and gate time $2t_0=30\units{\mu s}$. \label{fig:fastestCZgate}}
\end{figure}

The NV spin (already prepared in the ground state $\ket{0}$) can be brought into the long-lived state $(\ket{1}+\ket{\bar 1})/\sqrt{2}$ by a microwave $\pi$ pulse. Once the NV spin is initialized, the pumping procedure of \cite{PhysRevLett.103.113602,Economou_2016} can be performed: a linearly polarized optical pulse converts the population to $(\ket{E_+}+\ket{E_-})/\sqrt{2}$ which then spontaneously decays with $T_1\sim 10\units{ns}$ to the state $(\ket{1}\ket{L}+\ket{\bar 1}\ket{R})/\sqrt{2}$.  The appropriate Hadamard-type gate, $H'$, is
\begin{equation}
H' = \frac{1}{\sqrt{2}}\begin{cmatrix} 1&-1\\1&1\end{cmatrix}
\end{equation}
in the $\ket{1},\ket{\bar 1}$ basis. It is diagonal in the basis $\ket{1}\pm i\ket{\bar 1}$ with eigenvalues $e^{\pm i\pi/4}$; up to a phase, it is a rotation about the $y$-axis by $-\pi/2$. Discarding an overall phase, this gate can be realized by applying a $\pi/2$ pulse driving the transition $\ket{0}\leftrightarrow(\ket{1}+\ket{\bar1})/\sqrt{2}$. This can be done with a microwave hyperbolic secant pulse with appropriately chosen detuning $\Delta=\sigma$ \cite{Economou_2016,PhysRevB.74.205415,Ku_PRA17}.

The $\CZ$ gate between the NV and nuclear spins can be performed as follows.  The initialization step should first bring the nuclear spin into the state $(\ket{\uparrow}+\ket{\downarrow})/\sqrt{2}$, for example, by using techniques developed by the Delft group \cite{Taminiau_NatNano2014,Cramer_NatComm2016}.  To entangle the nuclear and electronic states, ideally, a right-circularly polarized microwave \cite{PhysRevB.76.165205} pulse (we also consider the more commonly used linearly polarized pulses below) would drive the transition $\ket{0,\downarrow}\leftrightarrow \ket{1,\downarrow}$. If no other transitions were driven by this pulse, the pulse could implement a $\CZ$ gate.  However, the competing transition $\ket{0,\uparrow}\leftrightarrow \ket{ 1,\uparrow}$ can have a frequency shift of only $A\sim$ 50 kHz from the desired transition.  We choose this value conservatively, i.e., to be consistent with the measurements for a rather remote ${}^{13}$C  \cite{Cramer_NatComm2016}. This competing transition will therefore also be strongly driven by such a pulse, necessitating very long pulses.

To quantify this, we will find the average fidelity \cite{Pedersen_2007} of the gate $U$:
\begin{equation}
F=\frac{1}{20}\left[ \mathrm{Tr}\,(\tilde U\tilde U^\dag) + |\mathrm{Tr}\,(\tilde U^\dag\CZ)|^2 \right],
\end{equation}
where $\tilde U$ is the projection of $U$ into the $\ket{\pm1\uparrow/\downarrow}$ subspace.\footnote{We allow for arbitrary local operations, chosen ahead of time, to maximize fidelity.} We find that using a narrow-band hyperbolic secant pulse of area $2\pi$ to induce the desired $-1$ phase \cite{PhysRevB.74.205415,Economou_PRL07,Economou_PRB12} on one of the states (here chosen to be $\ket{1\downarrow}$) takes $1.1\units{ms}$ ($3.7\units{ms}$, not shown) to achieve a fidelity of $99.9\%$ ($99.99\%$) [see \figref{fig:CZgates}(a)]. This timescale begins to compete with the spin coherence time, making it problematic.

The so-called SWIPHT method \cite{PhysRevB.91.161405} provides a way to accelerate these gates without sacrificing fidelity. The protocol is implemented by applying a pulse resonantly to the harmful transition, driving it, but in such a way that only a trivial $2\pi$ phase is induced on it. By carefully shaping the pulse, the phase on the desired transition can be controlled: a hyperbolic secant pulse $\Omega(t)=\Omega_0\,\mathrm{sech}(\sigma (t-t_0))$ with an appropriately chosen $\sigma$ and $\Omega_0=2\sigma$ induces the desired $-1$ phase (up to local unitary operations). The simulation for this case is shown in \figref{fig:CZgates}(b), where the time to achieve a fidelity of $99.9\%$ ($99.99\%$) is $58\units{\mu s}$ ($71\units{\mu s}$, not shown), well below the spin coherence time. (Notice that, in our protocol, the NV spin only needs to remain coherent during the photon production of each arm and the subsequent $\CZ$ gate, which attaches the arm to the ancilla nuclear spin.) An additional improvement in gate time can be achieved by driving off-resonantly halfway between the two transitions: fidelities of $99.9\%$ ($99.99\%$) are obtained in $30\units{\mu s}$ ($53\units{\mu s}$, not shown), as shown in \figref{fig:fastestCZgate}.

\begin{figure}
\centering
\includegraphics[width=0.85\columnwidth]{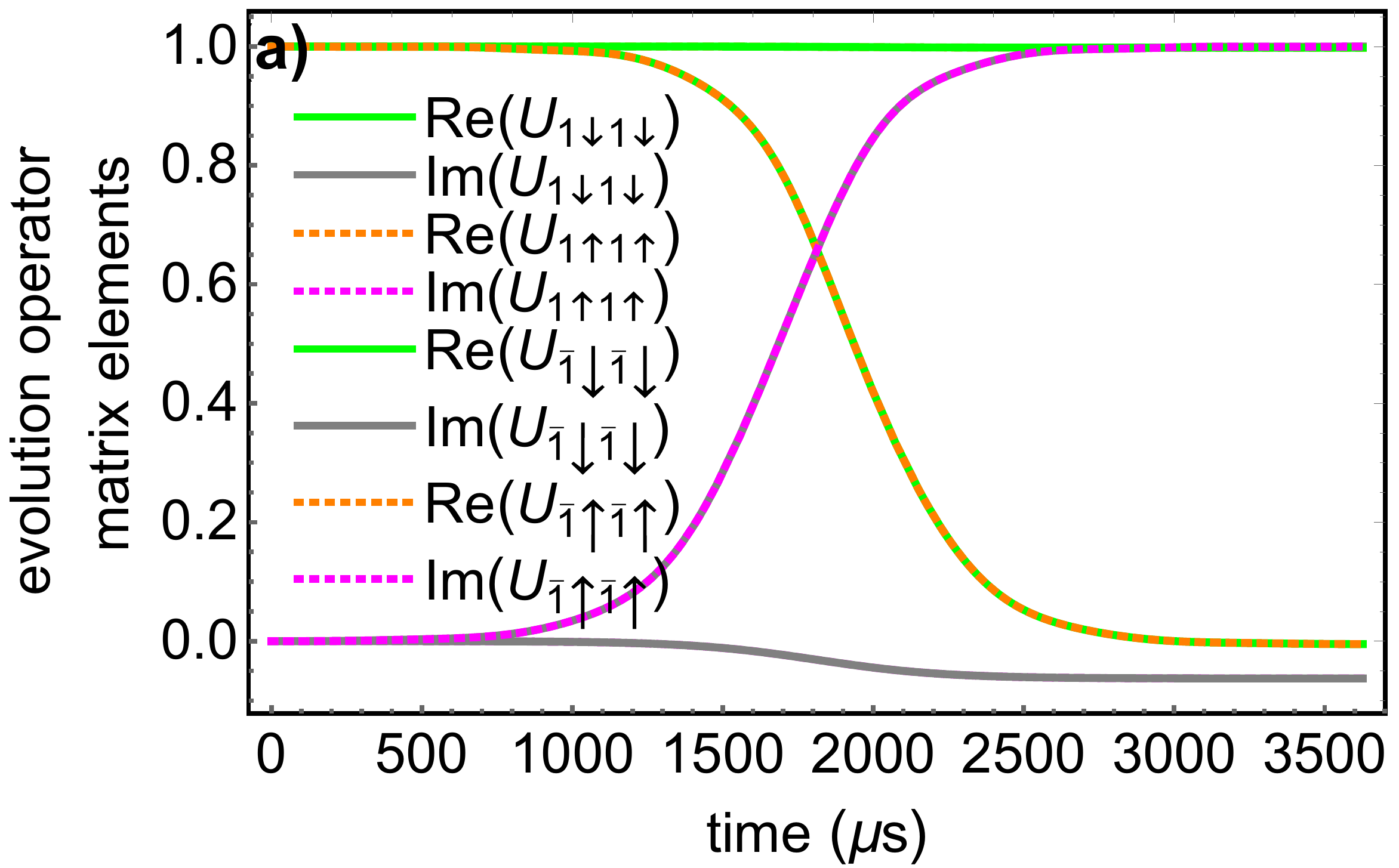}\qquad
\includegraphics[width=0.85\columnwidth]{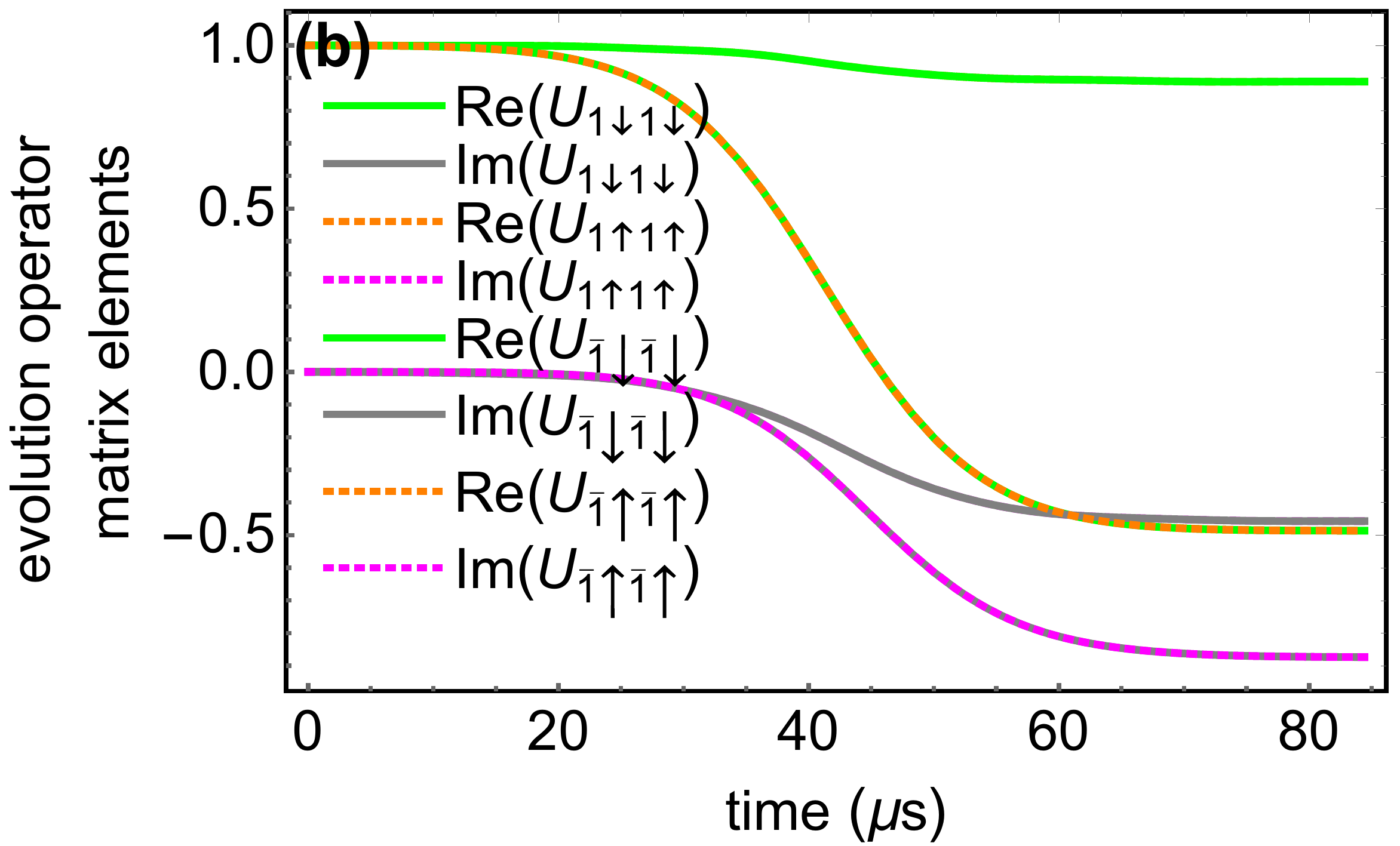}
\caption{NV-nuclear spin $\CZ$ gates implemented with linearly polarized microwave sech pulses, achieving $99.9\%$ fidelity. (a) Gate produced by frequency-selective 4$\pi$ sech pulse that induces a net minus sign between the transitions $\ket{0\downarrow}\leftrightarrow\ket{1\downarrow}$ and $\ket{0\uparrow}\leftrightarrow\ket{\bar 1\uparrow}$ while approximately avoiding the competing transitions $\ket{0\uparrow}\leftrightarrow\ket{1\uparrow}$ and $\ket{0\downarrow}\leftrightarrow\ket{\bar 1\downarrow}$.  (b) Gate produced by numerically identified $4\pi$ sech pulse that provides a 40-fold speedup compared to (a) by driving \emph{all} transitions. Here, we have taken $A=50\units{kHz}$, $E_\mathrm{ZFS}=2.88\units{GHz}$, and the pulse bandwidth $\sigma$ is $1.8\units{kHz}$ and $12.8\units{kHz}$ for (a) and (b), respectively. Pulse (b) drives at the detuned frequency $\Delta=-5.2\units{kHz}$ for a total time of $2t_0=84\units{\mu s}$. All fidelity calculations assume the hyperbolic secant is truncated around its maximum, but, unlike the other gates, this gate is sensitive to increased driving duration; if the drive is applied for longer, the fidelity \emph{drops}.\label{fig:linearCZgates}}
\end{figure}

In the event that circularly polarized microwave pulses are not available, the situation at first appears quite challenging: instead of driving two independent two-level systems, we now have exactly degenerate three-level systems. Indeed, naively driving a $\pi/2$ phase on two degenerate transitions (e.g., $\ket{0,\downarrow}\leftrightarrow \ket{1,\downarrow}$ and $\ket{0,\uparrow}\leftrightarrow \ket{\bar 1,\uparrow}$) with a (slightly detuned, so as to obtain $\pi/2$ phases) narrow energy pulse takes $3.6\units{ms}$ to achieve a fidelity of $99.9\%$. However, careful pulse shaping can shorten this time on a case-by-case basis, making it competitive with circularly polarized microwave pulses; see \figref{fig:linearCZgates} for a detailed description of the pulse.
\tableref{table:nv-parameters} summarizes the physical requirements for our protocol in NV centers.  For completeness, we restate the steps required to produce the RGS in \figref{fig:ancilla-anchor}:
\begin{enumerate}
\item The emitter and ancilla are initialized into $\ket{+}\otimes\ket{+}$
\item The emitter and ancilla are entangled via the $\CZ$ gate.
\item The emitter is pumped three times, and the last photon is measured in the $z$ basis.
\item Steps 2 and 3 are repeated for a total of 6 iterations.
\item The ancilla is measured in the $y$ basis.
\end{enumerate}

\begin{table}
\caption{Requirements for RGS generation in NV centers\label{table:nv-parameters}}
\begin{tabular}{ll}
\toprule
\multicolumn{1}{c}{Quantity}&\multicolumn{1}{c}{Descriptions}\\
\midrule
$T_1$& Spontaneous emission time\\
$T_2^\mathrm{spin}$ & NV spin coherence time\\
$T_2^\mathrm{nuclear}$ & Nuclear coherence time\\
$E_\mathrm{ZFS}$ & Zero-field splitting\\
$A$ & Hyperfine Coupling\\
\midrule
\multicolumn{2}{c}{$T_1\ll \frac{2\pi}{A}\ll T_2^\mathrm{spin} < T_2^\mathrm{nuclear}$}\\[0.5em]
\multicolumn{2}{c}{$50\units{kHz}\lesssim A\ll E_\mathrm{ZFS}$}\\
\bottomrule
\end{tabular}
\end{table}

Errors induced on the wave function during this protocol will propagate similarly to the QD case; in fact, all of the analysis of entanglement and pumping errors is the same (see Appendix~\ref{appendix:error}). An important difference here is that the ancilla qubit is not an emitter, so that the ``almost'' RGSs shown in Figs.~\ref{fig:almostrgs} and \ref{fig:imperfect-sixleg-rgs} cannot be created, and the full RGS must be generated as in \figref{fig:ancilla-anchor}. Although this requires more $\CZ$ gates, the ancilla is a nuclear spin, which has very long coherence times \cite{Taminiau_NatNano2014,Cramer_NatComm2016}. Moreover, the risk of errors propagating to the nuclear spin via $\CZ$ gates is relatively low: only $X$ or $Y$ errors will spread to the ancilla from the emitter, and these must occur immediately after the last photon pumping procedure (otherwise, they will evolve into a $Z$ error, and not spread via $\CZ$). Similarly to the QD case, the protocol withstands errors on unused exterior arms, and $Z$ errors on unused interior photons. Analogous to the QD case, $Z$ errors, corresponding to spin dephasing, are expected to be the dominant source of error. Again, such errors lead to only a single corrupted photon.

As for the QDs, light must be appropriately collected. Isolating the input and output photons can be done via temporal separation of emission from the optical driving from the spontaneous emission, on the timescale $T_1\sim 10\units{ns}$.  A notorious challenge with NV centers is the high probability of emission into the phonon sidebands. Using the Purcell effect, emission into the zero-phonon line (ZPL) can be enhanced from $3\%$ to $46\%$ \cite{PhysRevX.7.031040}.  Out coupling is conventionally performed by coupling to Bragg reflectors \cite{PhysRevX.7.031040}, though recent work has also coupled NV centers to photonic crystal waveguides \cite{Hausmann_2013,Li_NatCommun15}, with similarly impressive improvements to the ZPL. It is not unreasonable, then, to also anticipate progress on chiral waveguides \cite{PhysRevLett.113.093603}, that are both highly efficient, and can lead to high ZPL emission probability.

\section{Conclusions and outlook} \label{final}

In conclusion, we have developed explicit protocols of repeater graph state generation from both self-assembled quantum dots and in NV centers in diamond, providing all the necessary steps required for an experimental demonstration.  For a QD molecule, we provide two protocols, able to yield either a four- or six-external arm RGS. These numbers of photons are already sufficient to test the protocol with a highly nontrivial, modest-sized state. In the case of NV centers, the RGS can be of arbitrary size, only limited by the \emph{nuclear} spin coherence time. The spin-ancilla $\CZ$ gate we design is both fast and of high fidelity, as we demonstrated with simulations, and could be of interest to other quantum information technologies as well, such as quantum memories and quantum error correction \cite{Taminiau_NatNano2014,Cramer_NatComm2016}.  Our detailed schemes are consistent with existing experimental capabilities for modest-sized graph states. We hope to motivate such experiments, which would not only pave the way toward quantum networks by developing experimental capabilities further, but would also provide useful feedback to the theory to further optimize the designs. For example, understanding the effects of the environment on these graph state generation protocols would feed back to the theory efforts to design schemes that combat decoherence effects. Our protocols can also be straightforwardly adapted to other systems, such as vacancy defects in SiC. In particular, silicon carbide NV-centers could be especially interesting, as they have the desirable property of emitting in the telecommunications frequency band \cite{PhysRevB.92.064104}. Emitters in 2D materials are another interesting system due to the versatility in layering the host 2D crystals and integrating into devices. With the rapid progress in the development of quantum emitters \cite{Aharonovich_2016} and that in the field of nanophotonics, large-scale photonic graph states should be within reach in the near future.

\section{Acknowledgements}

This research was supported by the NSF (Grant No. 1741656).

\appendix
\section{Error Propagation\label{appendix:error}}

To propagate the errors through the generation procedures, a few commutation relations are used:
\begin{align*}
(X\otimes Z) \CZ &= \CZ (X\otimes I)\\
(Y\otimes Z) \CZ &= \CZ (Y\otimes I)\\
(Z\otimes I) \CZ &= \CZ (Z\otimes I)
\end{align*}
$X$ and $Y$ errors on one qubit induce a $Z$ error on the other when a $\CZ$ gate is applied. Notice the preferred $Z$ axis of the $\CZ$ aligns with the $Z$ preferred axis of growth in the QD and computational basis in the NV centers.  The pumping of photons can be modeled as an imperfect $\CZ$ entangling on a (guaranteed) $\ket{+}$ state. The appropriate commutation relations are
\begin{align*}
(X\otimes Z) \CZ ( I \otimes P_{+}) &= \CZ (X\otimes P_{+})\\
-(X\otimes Y) \CZ ( I \otimes P_{+}) &= \CZ (Y\otimes P_{+})\\
(I\otimes X) \CZ ( I \otimes P_{+}) &= \CZ (Z\otimes P_{+})
\end{align*}
where $P_+$ is the projector onto the $+x$ state, $\ket{+}\bra{+}$. A degree of freedom exists in the choice of operators compensating for the errors on the emitter due to this restriction. We have chosen, for clarity, that the error on the emitter is either $I$ (no error), or $X$. Our protocols always involve a Hadamard-type gate $H'$ which can be taken to be a rotation by $\pi/2$ around the $Y$ axis. Thus, a $Z$ error on the emitter is converted into an $X$ error on the emitted photon. An $X$ (or $Y$) emitter error, on the other hand, will convert into an $X$ error on the emitter, and a $Z$ (or $Y$) error on the emitted photon. The subsequent local gate on the emitter converts the $X$ emitter error into a $Z$ error.

%


\begin{thebibliography}{100}%
\makeatletter
\providecommand \@ifxundefined [1]{%
 \@ifx{#1\undefined}
}%
\providecommand \@ifnum [1]{%
 \ifnum #1\expandafter \@firstoftwo
 \else \expandafter \@secondoftwo
 \fi
}%
\providecommand \@ifx [1]{%
 \ifx #1\expandafter \@firstoftwo
 \else \expandafter \@secondoftwo
 \fi
}%
\providecommand \natexlab [1]{#1}%
\providecommand \enquote  [1]{``#1''}%
\providecommand \bibnamefont  [1]{#1}%
\providecommand \bibfnamefont [1]{#1}%
\providecommand \citenamefont [1]{#1}%
\providecommand \href@noop [0]{\@secondoftwo}%
\providecommand \href [0]{\begingroup \@sanitize@url \@href}%
\providecommand \@href[1]{\@@startlink{#1}\@@href}%
\providecommand \@@href[1]{\endgroup#1\@@endlink}%
\providecommand \@sanitize@url [0]{\catcode `\\12\catcode `\$12\catcode
  `\&12\catcode `\#12\catcode `\^12\catcode `\_12\catcode `\%12\relax}%
\providecommand \@@startlink[1]{}%
\providecommand \@@endlink[0]{}%
\providecommand \url  [0]{\begingroup\@sanitize@url \@url }%
\providecommand \@url [1]{\endgroup\@href {#1}{\urlprefix }}%
\providecommand \urlprefix  [0]{URL }%
\providecommand \Eprint [0]{\href }%
\providecommand \doibase [0]{http://dx.doi.org/}%
\providecommand \selectlanguage [0]{\@gobble}%
\providecommand \bibinfo  [0]{\@secondoftwo}%
\providecommand \bibfield  [0]{\@secondoftwo}%
\providecommand \translation [1]{[#1]}%
\providecommand \BibitemOpen [0]{}%
\providecommand \bibitemStop [0]{}%
\providecommand \bibitemNoStop [0]{.\EOS\space}%
\providecommand \EOS [0]{\spacefactor3000\relax}%
\providecommand \BibitemShut  [1]{\csname bibitem#1\endcsname}%
\let\auto@bib@innerbib\@empty
\bibitem [{\citenamefont {Gisin}\ and\ \citenamefont
  {Thew}(2007)}]{Gisin_NatPhoton07}%
  \BibitemOpen
  \bibfield  {author} {\bibinfo {author} {\bibfnamefont {N.}~\bibnamefont
  {Gisin}}\ and\ \bibinfo {author} {\bibfnamefont {R.}~\bibnamefont {Thew}},\
  }\href {\doibase 10.1038/nphoton.2007.22} {\bibfield  {journal} {\bibinfo
  {journal} {Nat. Photon.}\ }\textbf {\bibinfo {volume} {1}},\ \bibinfo {pages}
  {165} (\bibinfo {year} {2007})}\BibitemShut {NoStop}%
\bibitem [{\citenamefont {Scarani}\ \emph {et~al.}(2009)\citenamefont
  {Scarani}, \citenamefont {Bechmann-Pasquinucci}, \citenamefont {Cerf},
  \citenamefont {{Du\ifmmode \check{s}\else {\v s}\fi{}ek}}, \citenamefont
  {L{\"u}tkenhaus},\ and\ \citenamefont {Peev}}]{Scarani_RMP09}%
  \BibitemOpen
  \bibfield  {author} {\bibinfo {author} {\bibfnamefont {V.}~\bibnamefont
  {Scarani}}, \bibinfo {author} {\bibfnamefont {H.}~\bibnamefont
  {Bechmann-Pasquinucci}}, \bibinfo {author} {\bibfnamefont {N.~J.}\
  \bibnamefont {Cerf}}, \bibinfo {author} {\bibfnamefont {M.}~\bibnamefont
  {{Du\ifmmode \check{s}\else {\v s}\fi{}ek}}}, \bibinfo {author}
  {\bibfnamefont {N.}~\bibnamefont {L{\"u}tkenhaus}}, \ and\ \bibinfo {author}
  {\bibfnamefont {M.}~\bibnamefont {Peev}},\ }\href@noop {} {\bibfield
  {journal} {\bibinfo  {journal} {Rev. Mod. Phys.}\ }\textbf {\bibinfo {volume}
  {81}},\ \bibinfo {pages} {1301} (\bibinfo {year} {2009})}\BibitemShut
  {NoStop}%
\bibitem [{\citenamefont {Ursin}\ \emph {et~al.}(2007)\citenamefont {Ursin},
  \citenamefont {Tiefenbacher}, \citenamefont {Schmitt-Manderbach},
  \citenamefont {Weier}, \citenamefont {Scheidl}, \citenamefont {Lindenthal},
  \citenamefont {Blauensteiner}, \citenamefont {Jennewein}, \citenamefont
  {Perdigues}, \citenamefont {Trojek}, \citenamefont {Omer}, \citenamefont
  {Furst}, \citenamefont {Meyenburg}, \citenamefont {Rarity}, \citenamefont
  {Sodnik}, \citenamefont {Barbieri}, \citenamefont {Weinfurter},\ and\
  \citenamefont {Zeilinger}}]{Ursin_NatPhys07}%
  \BibitemOpen
  \bibfield  {author} {\bibinfo {author} {\bibfnamefont {R.}~\bibnamefont
  {Ursin}}, \bibinfo {author} {\bibfnamefont {F.}~\bibnamefont {Tiefenbacher}},
  \bibinfo {author} {\bibfnamefont {T.}~\bibnamefont {Schmitt-Manderbach}},
  \bibinfo {author} {\bibfnamefont {H.}~\bibnamefont {Weier}}, \bibinfo
  {author} {\bibfnamefont {T.}~\bibnamefont {Scheidl}}, \bibinfo {author}
  {\bibfnamefont {M.}~\bibnamefont {Lindenthal}}, \bibinfo {author}
  {\bibfnamefont {B.}~\bibnamefont {Blauensteiner}}, \bibinfo {author}
  {\bibfnamefont {T.}~\bibnamefont {Jennewein}}, \bibinfo {author}
  {\bibfnamefont {J.}~\bibnamefont {Perdigues}}, \bibinfo {author}
  {\bibfnamefont {P.}~\bibnamefont {Trojek}}, \bibinfo {author} {\bibfnamefont
  {B.}~\bibnamefont {Omer}}, \bibinfo {author} {\bibfnamefont {M.}~\bibnamefont
  {Furst}}, \bibinfo {author} {\bibfnamefont {M.}~\bibnamefont {Meyenburg}},
  \bibinfo {author} {\bibfnamefont {J.}~\bibnamefont {Rarity}}, \bibinfo
  {author} {\bibfnamefont {Z.}~\bibnamefont {Sodnik}}, \bibinfo {author}
  {\bibfnamefont {C.}~\bibnamefont {Barbieri}}, \bibinfo {author}
  {\bibfnamefont {H.}~\bibnamefont {Weinfurter}}, \ and\ \bibinfo {author}
  {\bibfnamefont {A.}~\bibnamefont {Zeilinger}},\ }\href@noop {} {\bibfield
  {journal} {\bibinfo  {journal} {Nat. Phys.}\ }\textbf {\bibinfo {volume}
  {3}},\ \bibinfo {pages} {481} (\bibinfo {year} {2007})}\BibitemShut {NoStop}%
\bibitem [{\citenamefont {Liao}\ \emph {et~al.}(2017)\citenamefont {Liao},
  \citenamefont {Cai}, \citenamefont {Liu}, \citenamefont {Zhang},
  \citenamefont {Li}, \citenamefont {Ren}, \citenamefont {Yin}, \citenamefont
  {Shen}, \citenamefont {Cao}, \citenamefont {Li}, \citenamefont {Li},
  \citenamefont {Chen}, \citenamefont {Sun}, \citenamefont {Jia}, \citenamefont
  {Wu}, \citenamefont {Jiang}, \citenamefont {Wang}, \citenamefont {Huang},
  \citenamefont {Wang}, \citenamefont {Zhou}, \citenamefont {Deng},
  \citenamefont {Xi}, \citenamefont {Ma}, \citenamefont {Hu}, \citenamefont
  {Zhang}, \citenamefont {Chen}, \citenamefont {Liu}, \citenamefont {Wang},
  \citenamefont {Zhu}, \citenamefont {Lu}, \citenamefont {Shu}, \citenamefont
  {Peng}, \citenamefont {Wang},\ and\ \citenamefont {Pan}}]{Liao_Nature17}%
  \BibitemOpen
  \bibfield  {author} {\bibinfo {author} {\bibfnamefont {S.-K.}\ \bibnamefont
  {Liao}}, \bibinfo {author} {\bibfnamefont {W.-Q.}\ \bibnamefont {Cai}},
  \bibinfo {author} {\bibfnamefont {W.-Y.}\ \bibnamefont {Liu}}, \bibinfo
  {author} {\bibfnamefont {L.}~\bibnamefont {Zhang}}, \bibinfo {author}
  {\bibfnamefont {Y.}~\bibnamefont {Li}}, \bibinfo {author} {\bibfnamefont
  {J.-G.}\ \bibnamefont {Ren}}, \bibinfo {author} {\bibfnamefont
  {J.}~\bibnamefont {Yin}}, \bibinfo {author} {\bibfnamefont {Q.}~\bibnamefont
  {Shen}}, \bibinfo {author} {\bibfnamefont {Y.}~\bibnamefont {Cao}}, \bibinfo
  {author} {\bibfnamefont {Z.-P.}\ \bibnamefont {Li}}, \bibinfo {author}
  {\bibfnamefont {F.-Z.}\ \bibnamefont {Li}}, \bibinfo {author} {\bibfnamefont
  {X.-W.}\ \bibnamefont {Chen}}, \bibinfo {author} {\bibfnamefont {L.-H.}\
  \bibnamefont {Sun}}, \bibinfo {author} {\bibfnamefont {J.-J.}\ \bibnamefont
  {Jia}}, \bibinfo {author} {\bibfnamefont {J.-C.}\ \bibnamefont {Wu}},
  \bibinfo {author} {\bibfnamefont {X.-J.}\ \bibnamefont {Jiang}}, \bibinfo
  {author} {\bibfnamefont {J.-F.}\ \bibnamefont {Wang}}, \bibinfo {author}
  {\bibfnamefont {Y.-M.}\ \bibnamefont {Huang}}, \bibinfo {author}
  {\bibfnamefont {Q.}~\bibnamefont {Wang}}, \bibinfo {author} {\bibfnamefont
  {Y.-L.}\ \bibnamefont {Zhou}}, \bibinfo {author} {\bibfnamefont
  {L.}~\bibnamefont {Deng}}, \bibinfo {author} {\bibfnamefont {T.}~\bibnamefont
  {Xi}}, \bibinfo {author} {\bibfnamefont {L.}~\bibnamefont {Ma}}, \bibinfo
  {author} {\bibfnamefont {T.}~\bibnamefont {Hu}}, \bibinfo {author}
  {\bibfnamefont {Q.}~\bibnamefont {Zhang}}, \bibinfo {author} {\bibfnamefont
  {Y.-A.}\ \bibnamefont {Chen}}, \bibinfo {author} {\bibfnamefont {N.-L.}\
  \bibnamefont {Liu}}, \bibinfo {author} {\bibfnamefont {X.-B.}\ \bibnamefont
  {Wang}}, \bibinfo {author} {\bibfnamefont {Z.-C.}\ \bibnamefont {Zhu}},
  \bibinfo {author} {\bibfnamefont {C.-Y.}\ \bibnamefont {Lu}}, \bibinfo
  {author} {\bibfnamefont {R.}~\bibnamefont {Shu}}, \bibinfo {author}
  {\bibfnamefont {C.-Z.}\ \bibnamefont {Peng}}, \bibinfo {author}
  {\bibfnamefont {J.-Y.}\ \bibnamefont {Wang}}, \ and\ \bibinfo {author}
  {\bibfnamefont {J.-W.}\ \bibnamefont {Pan}},\ }\href@noop {} {\bibfield
  {journal} {\bibinfo  {journal} {Nature}\ }\textbf {\bibinfo {volume} {549}},\
  \bibinfo {pages} {43} (\bibinfo {year} {2017})}\BibitemShut {NoStop}%
\bibitem [{\citenamefont {Chen}\ and\ \citenamefont {Lo}(2005)}]{Chen_ISIT05}%
  \BibitemOpen
  \bibfield  {author} {\bibinfo {author} {\bibfnamefont {K.}~\bibnamefont
  {Chen}}\ and\ \bibinfo {author} {\bibfnamefont {H.-K.}\ \bibnamefont {Lo}},\
  }in\ \href {\doibase 10.1109/ISIT.2005.1523616} {\emph {\bibinfo {booktitle}
  {Proceedings. International Symposium on Information Theory, 2005. ISIT
  2005.}}}\ (\bibinfo {year} {2005})\ pp.\ \bibinfo {pages}
  {1607--1611}\BibitemShut {NoStop}%
\bibitem [{\citenamefont {Epping}\ \emph {et~al.}(2017)\citenamefont {Epping},
  \citenamefont {Kampermann}, \citenamefont {Macchiavello},\ and\ \citenamefont
  {Bru{\ss}}}]{Epping_NJP17}%
  \BibitemOpen
  \bibfield  {author} {\bibinfo {author} {\bibfnamefont {M.}~\bibnamefont
  {Epping}}, \bibinfo {author} {\bibfnamefont {H.}~\bibnamefont {Kampermann}},
  \bibinfo {author} {\bibfnamefont {C.}~\bibnamefont {Macchiavello}}, \ and\
  \bibinfo {author} {\bibfnamefont {D.}~\bibnamefont {Bru{\ss}}},\ }\href
  {http://stacks.iop.org/1367-2630/19/i=9/a=093012} {\bibfield  {journal}
  {\bibinfo  {journal} {New Journal of Physics}\ }\textbf {\bibinfo {volume}
  {19}},\ \bibinfo {pages} {093012} (\bibinfo {year} {2017})}\BibitemShut
  {NoStop}%
\bibitem [{\citenamefont {Beals}\ \emph {et~al.}(2013)\citenamefont {Beals},
  \citenamefont {Brierley}, \citenamefont {Gray}, \citenamefont {Harrow},
  \citenamefont {Kutin}, \citenamefont {Linden}, \citenamefont {Shepherd},\
  and\ \citenamefont {Stather}}]{Beals_PRSA13}%
  \BibitemOpen
  \bibfield  {author} {\bibinfo {author} {\bibfnamefont {R.}~\bibnamefont
  {Beals}}, \bibinfo {author} {\bibfnamefont {S.}~\bibnamefont {Brierley}},
  \bibinfo {author} {\bibfnamefont {O.}~\bibnamefont {Gray}}, \bibinfo {author}
  {\bibfnamefont {A.~W.}\ \bibnamefont {Harrow}}, \bibinfo {author}
  {\bibfnamefont {S.}~\bibnamefont {Kutin}}, \bibinfo {author} {\bibfnamefont
  {N.}~\bibnamefont {Linden}}, \bibinfo {author} {\bibfnamefont
  {D.}~\bibnamefont {Shepherd}}, \ and\ \bibinfo {author} {\bibfnamefont
  {M.}~\bibnamefont {Stather}},\ }\href {\doibase 10.1098/rspa.2012.0686}
  {\bibfield  {journal} {\bibinfo  {journal} {Proc. R. Soc. A}\ }\textbf
  {\bibinfo {volume} {469}},\ \bibinfo {pages} {20120686} (\bibinfo {year}
  {2013})}\BibitemShut {NoStop}%
\bibitem [{\citenamefont {Meter}\ and\ \citenamefont
  {Devitt}(2016)}]{VanMeter_Computer16}%
  \BibitemOpen
  \bibfield  {author} {\bibinfo {author} {\bibfnamefont {R.~V.}\ \bibnamefont
  {Meter}}\ and\ \bibinfo {author} {\bibfnamefont {S.~J.}\ \bibnamefont
  {Devitt}},\ }\href@noop {} {\bibfield  {journal} {\bibinfo  {journal}
  {Computer}\ }\textbf {\bibinfo {volume} {49}},\ \bibinfo {pages} {31}
  (\bibinfo {year} {2016})}\BibitemShut {NoStop}%
\bibitem [{\citenamefont {Sheikholeslami}\ \emph {et~al.}(2016)\citenamefont
  {Sheikholeslami}, \citenamefont {Bash}, \citenamefont {Towsley},
  \citenamefont {Goeckel},\ and\ \citenamefont {Guha}}]{Sheikholeslami_IEEE16}%
  \BibitemOpen
  \bibfield  {author} {\bibinfo {author} {\bibfnamefont {A.}~\bibnamefont
  {Sheikholeslami}}, \bibinfo {author} {\bibfnamefont {B.~A.}\ \bibnamefont
  {Bash}}, \bibinfo {author} {\bibfnamefont {D.}~\bibnamefont {Towsley}},
  \bibinfo {author} {\bibfnamefont {D.}~\bibnamefont {Goeckel}}, \ and\
  \bibinfo {author} {\bibfnamefont {S.}~\bibnamefont {Guha}},\ }in\ \href@noop
  {} {\emph {\bibinfo {booktitle} {2016 IEEE International Symposium on
  Information Theory (ISIT)}}}\ (\bibinfo {year} {2016})\ pp.\ \bibinfo {pages}
  {2064--2068}\BibitemShut {NoStop}%
\bibitem [{\citenamefont {Fitzsimons}(2017)}]{Fitzsimons_npjQI17}%
  \BibitemOpen
  \bibfield  {author} {\bibinfo {author} {\bibfnamefont {J.~F.}\ \bibnamefont
  {Fitzsimons}},\ }\href@noop {} {\bibfield  {journal} {\bibinfo  {journal}
  {npj Quant. Inf.}\ }\textbf {\bibinfo {volume} {3}},\ \bibinfo {pages} {23}
  (\bibinfo {year} {2017})}\BibitemShut {NoStop}%
\bibitem [{\citenamefont {Kimble}(2008)}]{Kimble_Nature08}%
  \BibitemOpen
  \bibfield  {author} {\bibinfo {author} {\bibfnamefont {H.~J.}\ \bibnamefont
  {Kimble}},\ }\href@noop {} {\bibfield  {journal} {\bibinfo  {journal}
  {Nature}\ }\textbf {\bibinfo {volume} {453}},\ \bibinfo {pages} {1023}
  (\bibinfo {year} {2008})}\BibitemShut {NoStop}%
\bibitem [{\citenamefont {{Pant}}\ \emph {et~al.}(2017)\citenamefont {{Pant}},
  \citenamefont {{Krovi}}, \citenamefont {{Towsley}}, \citenamefont
  {{Tassiulas}}, \citenamefont {{Jiang}}, \citenamefont {{Basu}}, \citenamefont
  {{Englund}},\ and\ \citenamefont {{Guha}}}]{Pant_arxiv17}%
  \BibitemOpen
  \bibfield  {author} {\bibinfo {author} {\bibfnamefont {M.}~\bibnamefont
  {{Pant}}}, \bibinfo {author} {\bibfnamefont {H.}~\bibnamefont {{Krovi}}},
  \bibinfo {author} {\bibfnamefont {D.}~\bibnamefont {{Towsley}}}, \bibinfo
  {author} {\bibfnamefont {L.}~\bibnamefont {{Tassiulas}}}, \bibinfo {author}
  {\bibfnamefont {L.}~\bibnamefont {{Jiang}}}, \bibinfo {author} {\bibfnamefont
  {P.}~\bibnamefont {{Basu}}}, \bibinfo {author} {\bibfnamefont
  {D.}~\bibnamefont {{Englund}}}, \ and\ \bibinfo {author} {\bibfnamefont
  {S.}~\bibnamefont {{Guha}}},\ }\href@noop {} {\bibfield  {journal} {\bibinfo
  {journal} {ArXiv e-prints}\ } (\bibinfo {year} {2017})},\ \Eprint
  {http://arxiv.org/abs/1708.07142} {arXiv:1708.07142 [quant-ph]} \BibitemShut
  {NoStop}%
\bibitem [{\citenamefont {Pirandola}(2017)}]{1601.00966v4}%
  \BibitemOpen
  \bibfield  {author} {\bibinfo {author} {\bibfnamefont {S.}~\bibnamefont
  {Pirandola}},\ }\href {http://arxiv.org/abs/1601.00966v4;
  http://arxiv.org/pdf/1601.00966v4} {\enquote {\bibinfo {title} {Capacities of
  repeater-assisted quantum communications},}\ } (\bibinfo {year} {2017}),\
  \Eprint {http://arxiv.org/abs/1601.00966v4} {arXiv:1601.00966v4 [quant-ph]}
  \BibitemShut {NoStop}%
\bibitem [{\citenamefont {Azuma}\ \emph {et~al.}(2016)\citenamefont {Azuma},
  \citenamefont {Mizutani},\ and\ \citenamefont {Lo}}]{Azuma_2016}%
  \BibitemOpen
  \bibfield  {author} {\bibinfo {author} {\bibfnamefont {K.}~\bibnamefont
  {Azuma}}, \bibinfo {author} {\bibfnamefont {A.}~\bibnamefont {Mizutani}}, \
  and\ \bibinfo {author} {\bibfnamefont {H.-K.}\ \bibnamefont {Lo}},\ }\href
  {\doibase 10.1038/ncomms13523} {\bibfield  {journal} {\bibinfo  {journal}
  {Nature Communications}\ }\textbf {\bibinfo {volume} {7}},\ \bibinfo {pages}
  {13523} (\bibinfo {year} {2016})}\BibitemShut {NoStop}%
\bibitem [{\citenamefont {Azuma}\ and\ \citenamefont
  {Kato}(2017)}]{PhysRevA.96.032332}%
  \BibitemOpen
  \bibfield  {author} {\bibinfo {author} {\bibfnamefont {K.}~\bibnamefont
  {Azuma}}\ and\ \bibinfo {author} {\bibfnamefont {G.}~\bibnamefont {Kato}},\
  }\href {\doibase 10.1103/PhysRevA.96.032332} {\bibfield  {journal} {\bibinfo
  {journal} {Phys. Rev. A}\ }\textbf {\bibinfo {volume} {96}},\ \bibinfo
  {pages} {032332} (\bibinfo {year} {2017})}\BibitemShut {NoStop}%
\bibitem [{\citenamefont {Zwerger}\ \emph {et~al.}(2012)\citenamefont
  {Zwerger}, \citenamefont {D{\"u}r},\ and\ \citenamefont
  {Briegel}}]{PhysRevA.85.062326}%
  \BibitemOpen
  \bibfield  {author} {\bibinfo {author} {\bibfnamefont {M.}~\bibnamefont
  {Zwerger}}, \bibinfo {author} {\bibfnamefont {W.}~\bibnamefont {D{\"u}r}}, \
  and\ \bibinfo {author} {\bibfnamefont {H.~J.}\ \bibnamefont {Briegel}},\
  }\href {\doibase 10.1103/PhysRevA.85.062326} {\bibfield  {journal} {\bibinfo
  {journal} {Phys. Rev. A}\ }\textbf {\bibinfo {volume} {85}},\ \bibinfo
  {pages} {062326} (\bibinfo {year} {2012})}\BibitemShut {NoStop}%
\bibitem [{\citenamefont {Azuma}\ \emph {et~al.}(2015)\citenamefont {Azuma},
  \citenamefont {Tamaki},\ and\ \citenamefont {Lo}}]{Azuma_2015}%
  \BibitemOpen
  \bibfield  {author} {\bibinfo {author} {\bibfnamefont {K.}~\bibnamefont
  {Azuma}}, \bibinfo {author} {\bibfnamefont {K.}~\bibnamefont {Tamaki}}, \
  and\ \bibinfo {author} {\bibfnamefont {H.-K.}\ \bibnamefont {Lo}},\ }\href
  {\doibase 10.1038/ncomms7787} {\bibfield  {journal} {\bibinfo  {journal}
  {Nat. Commun.}\ }\textbf {\bibinfo {volume} {6}},\ \bibinfo {pages} {6787}
  (\bibinfo {year} {2015})}\BibitemShut {NoStop}%
\bibitem [{\citenamefont {Muralidharan}\ \emph {et~al.}(2016)\citenamefont
  {Muralidharan}, \citenamefont {Li}, \citenamefont {Kim}, \citenamefont
  {Lutkenhaus}, \citenamefont {Lukin},\ and\ \citenamefont
  {Jiang}}]{Muralidharan_SciRep16}%
  \BibitemOpen
  \bibfield  {author} {\bibinfo {author} {\bibfnamefont {S.}~\bibnamefont
  {Muralidharan}}, \bibinfo {author} {\bibfnamefont {L.}~\bibnamefont {Li}},
  \bibinfo {author} {\bibfnamefont {J.}~\bibnamefont {Kim}}, \bibinfo {author}
  {\bibfnamefont {N.}~\bibnamefont {Lutkenhaus}}, \bibinfo {author}
  {\bibfnamefont {M.~D.}\ \bibnamefont {Lukin}}, \ and\ \bibinfo {author}
  {\bibfnamefont {L.}~\bibnamefont {Jiang}},\ }\href@noop {} {\bibfield
  {journal} {\bibinfo  {journal} {Sci. Rep.}\ }\textbf {\bibinfo {volume}
  {6}},\ \bibinfo {pages} {20463} (\bibinfo {year} {2016})}\BibitemShut
  {NoStop}%
\bibitem [{\citenamefont {Raussendorf}\ and\ \citenamefont
  {Briegel}(2001)}]{PhysRevLett.86.5188}%
  \BibitemOpen
  \bibfield  {author} {\bibinfo {author} {\bibfnamefont {R.}~\bibnamefont
  {Raussendorf}}\ and\ \bibinfo {author} {\bibfnamefont {H.~J.}\ \bibnamefont
  {Briegel}},\ }\href {\doibase 10.1103/PhysRevLett.86.5188} {\bibfield
  {journal} {\bibinfo  {journal} {Phys. Rev. Lett.}\ }\textbf {\bibinfo
  {volume} {86}},\ \bibinfo {pages} {5188} (\bibinfo {year}
  {2001})}\BibitemShut {NoStop}%
\bibitem [{\citenamefont {Briegel}\ and\ \citenamefont
  {Raussendorf}(2001)}]{PhysRevLett.86.910}%
  \BibitemOpen
  \bibfield  {author} {\bibinfo {author} {\bibfnamefont {H.~J.}\ \bibnamefont
  {Briegel}}\ and\ \bibinfo {author} {\bibfnamefont {R.}~\bibnamefont
  {Raussendorf}},\ }\href {\doibase 10.1103/PhysRevLett.86.910} {\bibfield
  {journal} {\bibinfo  {journal} {Phys. Rev. Lett.}\ }\textbf {\bibinfo
  {volume} {86}},\ \bibinfo {pages} {910} (\bibinfo {year} {2001})}\BibitemShut
  {NoStop}%
\bibitem [{\citenamefont {Browne}\ and\ \citenamefont
  {Rudolph}(2005)}]{PhysRevLett.95.010501}%
  \BibitemOpen
  \bibfield  {author} {\bibinfo {author} {\bibfnamefont {D.~E.}\ \bibnamefont
  {Browne}}\ and\ \bibinfo {author} {\bibfnamefont {T.}~\bibnamefont
  {Rudolph}},\ }\href {\doibase 10.1103/PhysRevLett.95.010501} {\bibfield
  {journal} {\bibinfo  {journal} {Phys. Rev. Lett.}\ }\textbf {\bibinfo
  {volume} {95}},\ \bibinfo {pages} {010501} (\bibinfo {year}
  {2005})}\BibitemShut {NoStop}%
\bibitem [{\citenamefont {Varnava}\ \emph {et~al.}(2006)\citenamefont
  {Varnava}, \citenamefont {Browne},\ and\ \citenamefont
  {Rudolph}}]{PhysRevLett.97.120501}%
  \BibitemOpen
  \bibfield  {author} {\bibinfo {author} {\bibfnamefont {M.}~\bibnamefont
  {Varnava}}, \bibinfo {author} {\bibfnamefont {D.~E.}\ \bibnamefont {Browne}},
  \ and\ \bibinfo {author} {\bibfnamefont {T.}~\bibnamefont {Rudolph}},\ }\href
  {\doibase 10.1103/PhysRevLett.97.120501} {\bibfield  {journal} {\bibinfo
  {journal} {Phys. Rev. Lett.}\ }\textbf {\bibinfo {volume} {97}},\ \bibinfo
  {pages} {120501} (\bibinfo {year} {2006})}\BibitemShut {NoStop}%
\bibitem [{\citenamefont {Munro}\ \emph {et~al.}(2005)\citenamefont {Munro},
  \citenamefont {Nemoto},\ and\ \citenamefont {Spiller}}]{Munro_NJP2005}%
  \BibitemOpen
  \bibfield  {author} {\bibinfo {author} {\bibfnamefont {W.~J.}\ \bibnamefont
  {Munro}}, \bibinfo {author} {\bibfnamefont {K.}~\bibnamefont {Nemoto}}, \
  and\ \bibinfo {author} {\bibfnamefont {T.~P.}\ \bibnamefont {Spiller}},\
  }\href {http://stacks.iop.org/1367-2630/7/i=1/a=137} {\bibfield  {journal}
  {\bibinfo  {journal} {New J. Phys.}\ }\textbf {\bibinfo {volume} {7}},\
  \bibinfo {pages} {137} (\bibinfo {year} {2005})}\BibitemShut {NoStop}%
\bibitem [{\citenamefont {Shapiro}(2006)}]{PhysRevA.73.062305}%
  \BibitemOpen
  \bibfield  {author} {\bibinfo {author} {\bibfnamefont {J.~H.}\ \bibnamefont
  {Shapiro}},\ }\href {\doibase 10.1103/PhysRevA.73.062305} {\bibfield
  {journal} {\bibinfo  {journal} {Phys. Rev. A}\ }\textbf {\bibinfo {volume}
  {73}},\ \bibinfo {pages} {062305} (\bibinfo {year} {2006})}\BibitemShut
  {NoStop}%
\bibitem [{\citenamefont {Hacker}\ \emph {et~al.}(2016)\citenamefont {Hacker},
  \citenamefont {Welte}, \citenamefont {Rempe},\ and\ \citenamefont
  {Ritter}}]{Hacker_Nature2016}%
  \BibitemOpen
  \bibfield  {author} {\bibinfo {author} {\bibfnamefont {B.}~\bibnamefont
  {Hacker}}, \bibinfo {author} {\bibfnamefont {S.}~\bibnamefont {Welte}},
  \bibinfo {author} {\bibfnamefont {G.}~\bibnamefont {Rempe}}, \ and\ \bibinfo
  {author} {\bibfnamefont {S.}~\bibnamefont {Ritter}},\ }\href {\doibase
  10.1038/nature18592} {\bibfield  {journal} {\bibinfo  {journal} {Nature}\
  }\textbf {\bibinfo {volume} {536}},\ \bibinfo {pages} {193} (\bibinfo {year}
  {2016})}\BibitemShut {NoStop}%
\bibitem [{\citenamefont {Zhao}\ \emph {et~al.}(2004)\citenamefont {Zhao},
  \citenamefont {Chen}, \citenamefont {Zhang}, \citenamefont {Yang},
  \citenamefont {Briegel},\ and\ \citenamefont {Pan}}]{Zhao_2004}%
  \BibitemOpen
  \bibfield  {author} {\bibinfo {author} {\bibfnamefont {Z.}~\bibnamefont
  {Zhao}}, \bibinfo {author} {\bibfnamefont {Y.-A.}\ \bibnamefont {Chen}},
  \bibinfo {author} {\bibfnamefont {A.-N.}\ \bibnamefont {Zhang}}, \bibinfo
  {author} {\bibfnamefont {T.}~\bibnamefont {Yang}}, \bibinfo {author}
  {\bibfnamefont {H.~J.}\ \bibnamefont {Briegel}}, \ and\ \bibinfo {author}
  {\bibfnamefont {J.-W.}\ \bibnamefont {Pan}},\ }\href {\doibase
  10.1038/nature02643} {\bibfield  {journal} {\bibinfo  {journal} {Nature}\
  }\textbf {\bibinfo {volume} {430}},\ \bibinfo {pages} {54} (\bibinfo {year}
  {2004})}\BibitemShut {NoStop}%
\bibitem [{\citenamefont {Gao}\ \emph {et~al.}(2010{\natexlab{a}})\citenamefont
  {Gao}, \citenamefont {Xu}, \citenamefont {Yao}, \citenamefont {G{\"u}hne},
  \citenamefont {Cabello}, \citenamefont {Lu}, \citenamefont {Peng},
  \citenamefont {Chen},\ and\ \citenamefont {Pan}}]{PhysRevLett.104.020501}%
  \BibitemOpen
  \bibfield  {author} {\bibinfo {author} {\bibfnamefont {W.-B.}\ \bibnamefont
  {Gao}}, \bibinfo {author} {\bibfnamefont {P.}~\bibnamefont {Xu}}, \bibinfo
  {author} {\bibfnamefont {X.-C.}\ \bibnamefont {Yao}}, \bibinfo {author}
  {\bibfnamefont {O.}~\bibnamefont {G{\"u}hne}}, \bibinfo {author}
  {\bibfnamefont {A.}~\bibnamefont {Cabello}}, \bibinfo {author} {\bibfnamefont
  {C.-Y.}\ \bibnamefont {Lu}}, \bibinfo {author} {\bibfnamefont {C.-Z.}\
  \bibnamefont {Peng}}, \bibinfo {author} {\bibfnamefont {Z.-B.}\ \bibnamefont
  {Chen}}, \ and\ \bibinfo {author} {\bibfnamefont {J.-W.}\ \bibnamefont
  {Pan}},\ }\href {\doibase 10.1103/PhysRevLett.104.020501} {\bibfield
  {journal} {\bibinfo  {journal} {Phys. Rev. Lett.}\ }\textbf {\bibinfo
  {volume} {104}},\ \bibinfo {pages} {020501} (\bibinfo {year}
  {2010}{\natexlab{a}})}\BibitemShut {NoStop}%
\bibitem [{\citenamefont {Gao}\ \emph {et~al.}(2010{\natexlab{b}})\citenamefont
  {Gao}, \citenamefont {Lu}, \citenamefont {Yao}, \citenamefont {Xu},
  \citenamefont {G{\"u}hne}, \citenamefont {Goebel}, \citenamefont {Chen},
  \citenamefont {Peng}, \citenamefont {Chen},\ and\ \citenamefont
  {Pan}}]{Gao_2010}%
  \BibitemOpen
  \bibfield  {author} {\bibinfo {author} {\bibfnamefont {W.-B.}\ \bibnamefont
  {Gao}}, \bibinfo {author} {\bibfnamefont {C.-Y.}\ \bibnamefont {Lu}},
  \bibinfo {author} {\bibfnamefont {X.-C.}\ \bibnamefont {Yao}}, \bibinfo
  {author} {\bibfnamefont {P.}~\bibnamefont {Xu}}, \bibinfo {author}
  {\bibfnamefont {O.}~\bibnamefont {G{\"u}hne}}, \bibinfo {author}
  {\bibfnamefont {A.}~\bibnamefont {Goebel}}, \bibinfo {author} {\bibfnamefont
  {Y.-A.}\ \bibnamefont {Chen}}, \bibinfo {author} {\bibfnamefont {C.-Z.}\
  \bibnamefont {Peng}}, \bibinfo {author} {\bibfnamefont {Z.-B.}\ \bibnamefont
  {Chen}}, \ and\ \bibinfo {author} {\bibfnamefont {J.-W.}\ \bibnamefont
  {Pan}},\ }\href {\doibase 10.1038/nphys1603} {\bibfield  {journal} {\bibinfo
  {journal} {Nature Physics}\ }\textbf {\bibinfo {volume} {6}},\ \bibinfo
  {pages} {331} (\bibinfo {year} {2010}{\natexlab{b}})}\BibitemShut {NoStop}%
\bibitem [{\citenamefont {Lindner}\ and\ \citenamefont
  {Rudolph}(2009)}]{PhysRevLett.103.113602}%
  \BibitemOpen
  \bibfield  {author} {\bibinfo {author} {\bibfnamefont {N.~H.}\ \bibnamefont
  {Lindner}}\ and\ \bibinfo {author} {\bibfnamefont {T.}~\bibnamefont
  {Rudolph}},\ }\href {\doibase 10.1103/PhysRevLett.103.113602} {\bibfield
  {journal} {\bibinfo  {journal} {Phys. Rev. Lett.}\ }\textbf {\bibinfo
  {volume} {103}},\ \bibinfo {pages} {113602} (\bibinfo {year}
  {2009})}\BibitemShut {NoStop}%
\bibitem [{\citenamefont {Greenberger}\ \emph {et~al.}(1989)\citenamefont
  {Greenberger}, \citenamefont {Horne},\ and\ \citenamefont
  {Zeilinger}}]{Greenberger_1989}%
  \BibitemOpen
  \bibfield  {author} {\bibinfo {author} {\bibfnamefont {D.~M.}\ \bibnamefont
  {Greenberger}}, \bibinfo {author} {\bibfnamefont {M.~A.}\ \bibnamefont
  {Horne}}, \ and\ \bibinfo {author} {\bibfnamefont {A.}~\bibnamefont
  {Zeilinger}},\ }\href {\doibase 10.1007/978-94-017-0849-4_10} {\bibfield
  {journal} {\bibinfo  {journal} {Bell{\rq}s Theorem, Quantum Theory and
  Conceptions of the Universe}\ ,\ \bibinfo {pages} {69}} (\bibinfo {year}
  {1989})}\BibitemShut {NoStop}%
\bibitem [{\citenamefont {Schwartz}\ \emph {et~al.}(2016)\citenamefont
  {Schwartz}, \citenamefont {Cogan}, \citenamefont {Schmidgall}, \citenamefont
  {Don}, \citenamefont {Gantz}, \citenamefont {Kenneth}, \citenamefont
  {Lindner},\ and\ \citenamefont {Gershoni}}]{Schwartz_2016}%
  \BibitemOpen
  \bibfield  {author} {\bibinfo {author} {\bibfnamefont {I.}~\bibnamefont
  {Schwartz}}, \bibinfo {author} {\bibfnamefont {D.}~\bibnamefont {Cogan}},
  \bibinfo {author} {\bibfnamefont {E.~R.}\ \bibnamefont {Schmidgall}},
  \bibinfo {author} {\bibfnamefont {Y.}~\bibnamefont {Don}}, \bibinfo {author}
  {\bibfnamefont {L.}~\bibnamefont {Gantz}}, \bibinfo {author} {\bibfnamefont
  {O.}~\bibnamefont {Kenneth}}, \bibinfo {author} {\bibfnamefont {N.~H.}\
  \bibnamefont {Lindner}}, \ and\ \bibinfo {author} {\bibfnamefont
  {D.}~\bibnamefont {Gershoni}},\ }\href {\doibase 10.1126/science.aah4758}
  {\bibfield  {journal} {\bibinfo  {journal} {Science}\ }\textbf {\bibinfo
  {volume} {354}},\ \bibinfo {pages} {434} (\bibinfo {year}
  {2016})}\BibitemShut {NoStop}%
\bibitem [{\citenamefont {Economou}\ \emph {et~al.}(2010)\citenamefont
  {Economou}, \citenamefont {Lindner},\ and\ \citenamefont
  {Rudolph}}]{PhysRevLett.105.093601}%
  \BibitemOpen
  \bibfield  {author} {\bibinfo {author} {\bibfnamefont {S.~E.}\ \bibnamefont
  {Economou}}, \bibinfo {author} {\bibfnamefont {N.}~\bibnamefont {Lindner}}, \
  and\ \bibinfo {author} {\bibfnamefont {T.}~\bibnamefont {Rudolph}},\ }\href
  {\doibase 10.1103/PhysRevLett.105.093601} {\bibfield  {journal} {\bibinfo
  {journal} {Phys. Rev. Lett.}\ }\textbf {\bibinfo {volume} {105}},\ \bibinfo
  {pages} {093601} (\bibinfo {year} {2010})}\BibitemShut {NoStop}%
\bibitem [{\citenamefont {Buterakos}\ \emph {et~al.}(2017)\citenamefont
  {Buterakos}, \citenamefont {Barnes},\ and\ \citenamefont
  {Economou}}]{PhysRevX.7.041023}%
  \BibitemOpen
  \bibfield  {author} {\bibinfo {author} {\bibfnamefont {D.}~\bibnamefont
  {Buterakos}}, \bibinfo {author} {\bibfnamefont {E.}~\bibnamefont {Barnes}}, \
  and\ \bibinfo {author} {\bibfnamefont {S.~E.}\ \bibnamefont {Economou}},\
  }\href {\doibase 10.1103/PhysRevX.7.041023} {\bibfield  {journal} {\bibinfo
  {journal} {Phys. Rev. X}\ }\textbf {\bibinfo {volume} {7}},\ \bibinfo {pages}
  {041023} (\bibinfo {year} {2017})}\BibitemShut {NoStop}%
\bibitem [{\citenamefont {Toishi}\ \emph {et~al.}(2009)\citenamefont {Toishi},
  \citenamefont {Englund}, \citenamefont {Faraon},\ and\ \citenamefont {Vu{\v
  c}kovi{\'c}}}]{Toishi_OE09}%
  \BibitemOpen
  \bibfield  {author} {\bibinfo {author} {\bibfnamefont {M.}~\bibnamefont
  {Toishi}}, \bibinfo {author} {\bibfnamefont {D.}~\bibnamefont {Englund}},
  \bibinfo {author} {\bibfnamefont {A.}~\bibnamefont {Faraon}}, \ and\ \bibinfo
  {author} {\bibfnamefont {J.}~\bibnamefont {Vu{\v c}kovi{\'c}}},\ }\href
  {\doibase 10.1364/OE.17.014618} {\bibfield  {journal} {\bibinfo  {journal}
  {Opt. Express}\ }\textbf {\bibinfo {volume} {17}},\ \bibinfo {pages} {14618}
  (\bibinfo {year} {2009})}\BibitemShut {NoStop}%
\bibitem [{\citenamefont {Englund}\ \emph {et~al.}(2010)\citenamefont
  {Englund}, \citenamefont {Shields}, \citenamefont {Rivoire}, \citenamefont
  {Hatami}, \citenamefont {Vuckovic}, \citenamefont {Park},\ and\ \citenamefont
  {Lukin}}]{Englund_NanoLett10}%
  \BibitemOpen
  \bibfield  {author} {\bibinfo {author} {\bibfnamefont {D.}~\bibnamefont
  {Englund}}, \bibinfo {author} {\bibfnamefont {B.}~\bibnamefont {Shields}},
  \bibinfo {author} {\bibfnamefont {K.}~\bibnamefont {Rivoire}}, \bibinfo
  {author} {\bibfnamefont {F.}~\bibnamefont {Hatami}}, \bibinfo {author}
  {\bibfnamefont {J.}~\bibnamefont {Vuckovic}}, \bibinfo {author}
  {\bibfnamefont {H.}~\bibnamefont {Park}}, \ and\ \bibinfo {author}
  {\bibfnamefont {M.~D.}\ \bibnamefont {Lukin}},\ }\href {\doibase
  10.1021/nl101662v} {\bibfield  {journal} {\bibinfo  {journal} {Nano Letters}\
  }\textbf {\bibinfo {volume} {10}},\ \bibinfo {pages} {3922} (\bibinfo {year}
  {2010})},\ \bibinfo {note} {pMID: 20825160},\ \Eprint
  {http://arxiv.org/abs/https://doi.org/10.1021/nl101662v}
  {https://doi.org/10.1021/nl101662v} \BibitemShut {NoStop}%
\bibitem [{\citenamefont {Calusine}\ \emph {et~al.}(2014)\citenamefont
  {Calusine}, \citenamefont {Politi},\ and\ \citenamefont
  {Awschalom}}]{Calusine_APL14}%
  \BibitemOpen
  \bibfield  {author} {\bibinfo {author} {\bibfnamefont {G.}~\bibnamefont
  {Calusine}}, \bibinfo {author} {\bibfnamefont {A.}~\bibnamefont {Politi}}, \
  and\ \bibinfo {author} {\bibfnamefont {D.~D.}\ \bibnamefont {Awschalom}},\
  }\href@noop {} {\bibfield  {journal} {\bibinfo  {journal} {Appl. Phys.
  Lett.}\ }\textbf {\bibinfo {volume} {105}},\ \bibinfo {pages} {011123}
  (\bibinfo {year} {2014})}\BibitemShut {NoStop}%
\bibitem [{\citenamefont {Grange}\ \emph {et~al.}(2015)\citenamefont {Grange},
  \citenamefont {Hornecker}, \citenamefont {Hunger}, \citenamefont {Poizat},
  \citenamefont {G{\'e}rard}, \citenamefont {Senellart},\ and\ \citenamefont
  {Auff{\`e}ves}}]{Grange_PRL15}%
  \BibitemOpen
  \bibfield  {author} {\bibinfo {author} {\bibfnamefont {T.}~\bibnamefont
  {Grange}}, \bibinfo {author} {\bibfnamefont {G.}~\bibnamefont {Hornecker}},
  \bibinfo {author} {\bibfnamefont {D.}~\bibnamefont {Hunger}}, \bibinfo
  {author} {\bibfnamefont {J.-P.}\ \bibnamefont {Poizat}}, \bibinfo {author}
  {\bibfnamefont {J.-M.}\ \bibnamefont {G{\'e}rard}}, \bibinfo {author}
  {\bibfnamefont {P.}~\bibnamefont {Senellart}}, \ and\ \bibinfo {author}
  {\bibfnamefont {A.}~\bibnamefont {Auff{\`e}ves}},\ }\href {\doibase
  10.1103/PhysRevLett.114.193601} {\bibfield  {journal} {\bibinfo  {journal}
  {Phys. Rev. Lett.}\ }\textbf {\bibinfo {volume} {114}},\ \bibinfo {pages}
  {193601} (\bibinfo {year} {2015})}\BibitemShut {NoStop}%
\bibitem [{\citenamefont {Vora}\ \emph {et~al.}(2015)\citenamefont {Vora},
  \citenamefont {Bracker}, \citenamefont {Carter}, \citenamefont {Sweeney},
  \citenamefont {Kim}, \citenamefont {Kim}, \citenamefont {Yang}, \citenamefont
  {Brereton}, \citenamefont {Economou},\ and\ \citenamefont
  {Gammon}}]{Vora_NC15}%
  \BibitemOpen
  \bibfield  {author} {\bibinfo {author} {\bibfnamefont {P.~M.}\ \bibnamefont
  {Vora}}, \bibinfo {author} {\bibfnamefont {A.~S.}\ \bibnamefont {Bracker}},
  \bibinfo {author} {\bibfnamefont {S.~G.}\ \bibnamefont {Carter}}, \bibinfo
  {author} {\bibfnamefont {T.~M.}\ \bibnamefont {Sweeney}}, \bibinfo {author}
  {\bibfnamefont {M.}~\bibnamefont {Kim}}, \bibinfo {author} {\bibfnamefont
  {C.~S.}\ \bibnamefont {Kim}}, \bibinfo {author} {\bibfnamefont
  {L.}~\bibnamefont {Yang}}, \bibinfo {author} {\bibfnamefont {P.~G.}\
  \bibnamefont {Brereton}}, \bibinfo {author} {\bibfnamefont {S.~E.}\
  \bibnamefont {Economou}}, \ and\ \bibinfo {author} {\bibfnamefont
  {D.}~\bibnamefont {Gammon}},\ }\href@noop {} {\bibfield  {journal} {\bibinfo
  {journal} {Nat. Commun.}\ }\textbf {\bibinfo {volume} {6}},\ \bibinfo {pages}
  {7665} (\bibinfo {year} {2015})}\BibitemShut {NoStop}%
\bibitem [{\citenamefont {Somaschi}\ \emph {et~al.}(2016)\citenamefont
  {Somaschi}, \citenamefont {Giesz}, \citenamefont {Santis}, \citenamefont
  {Loredo}, \citenamefont {Almeida}, \citenamefont {Hornecker}, \citenamefont
  {Portalupi}, \citenamefont {Grange}, \citenamefont {Anton}, \citenamefont
  {Demory}, \citenamefont {Gomez}, \citenamefont {Sagnes}, \citenamefont
  {Lanzillotti-Kimura}, \citenamefont {Lemaitre}, \citenamefont {Auffeves},
  \citenamefont {White}, \citenamefont {Lanco},\ and\ \citenamefont
  {Senellart}}]{Somaschi_NatPhoton16}%
  \BibitemOpen
  \bibfield  {author} {\bibinfo {author} {\bibfnamefont {N.}~\bibnamefont
  {Somaschi}}, \bibinfo {author} {\bibfnamefont {V.}~\bibnamefont {Giesz}},
  \bibinfo {author} {\bibfnamefont {L.~D.}\ \bibnamefont {Santis}}, \bibinfo
  {author} {\bibfnamefont {J.~C.}\ \bibnamefont {Loredo}}, \bibinfo {author}
  {\bibfnamefont {M.~P.}\ \bibnamefont {Almeida}}, \bibinfo {author}
  {\bibfnamefont {G.}~\bibnamefont {Hornecker}}, \bibinfo {author}
  {\bibfnamefont {S.~L.}\ \bibnamefont {Portalupi}}, \bibinfo {author}
  {\bibfnamefont {T.}~\bibnamefont {Grange}}, \bibinfo {author} {\bibfnamefont
  {C.}~\bibnamefont {Anton}}, \bibinfo {author} {\bibfnamefont
  {J.}~\bibnamefont {Demory}}, \bibinfo {author} {\bibfnamefont
  {C.}~\bibnamefont {Gomez}}, \bibinfo {author} {\bibfnamefont
  {I.}~\bibnamefont {Sagnes}}, \bibinfo {author} {\bibfnamefont {N.~D.}\
  \bibnamefont {Lanzillotti-Kimura}}, \bibinfo {author} {\bibfnamefont
  {A.}~\bibnamefont {Lemaitre}}, \bibinfo {author} {\bibfnamefont
  {A.}~\bibnamefont {Auffeves}}, \bibinfo {author} {\bibfnamefont {A.~G.}\
  \bibnamefont {White}}, \bibinfo {author} {\bibfnamefont {L.}~\bibnamefont
  {Lanco}}, \ and\ \bibinfo {author} {\bibfnamefont {P.}~\bibnamefont
  {Senellart}},\ }\href@noop {} {\bibfield  {journal} {\bibinfo  {journal}
  {Nat. Photon.}\ }\textbf {\bibinfo {volume} {10}},\ \bibinfo {pages} {340}
  (\bibinfo {year} {2016})}\BibitemShut {NoStop}%
\bibitem [{\citenamefont {Schr{\"o}der}\ \emph {et~al.}(2017)\citenamefont
  {Schr{\"o}der}, \citenamefont {Walsh}, \citenamefont {Zheng}, \citenamefont
  {Mouradian}, \citenamefont {Li}, \citenamefont {Malladi}, \citenamefont
  {Bakhru}, \citenamefont {Lu}, \citenamefont {Stein}, \citenamefont {Heuck},\
  and\ \citenamefont {Englund}}]{Schroder_OME17}%
  \BibitemOpen
  \bibfield  {author} {\bibinfo {author} {\bibfnamefont {T.}~\bibnamefont
  {Schr{\"o}der}}, \bibinfo {author} {\bibfnamefont {M.}~\bibnamefont {Walsh}},
  \bibinfo {author} {\bibfnamefont {J.}~\bibnamefont {Zheng}}, \bibinfo
  {author} {\bibfnamefont {S.}~\bibnamefont {Mouradian}}, \bibinfo {author}
  {\bibfnamefont {L.}~\bibnamefont {Li}}, \bibinfo {author} {\bibfnamefont
  {G.}~\bibnamefont {Malladi}}, \bibinfo {author} {\bibfnamefont
  {H.}~\bibnamefont {Bakhru}}, \bibinfo {author} {\bibfnamefont
  {M.}~\bibnamefont {Lu}}, \bibinfo {author} {\bibfnamefont {A.}~\bibnamefont
  {Stein}}, \bibinfo {author} {\bibfnamefont {M.}~\bibnamefont {Heuck}}, \ and\
  \bibinfo {author} {\bibfnamefont {D.}~\bibnamefont {Englund}},\ }\href
  {\doibase 10.1364/OME.7.001514} {\bibfield  {journal} {\bibinfo  {journal}
  {Opt. Mater. Express}\ }\textbf {\bibinfo {volume} {7}},\ \bibinfo {pages}
  {1514} (\bibinfo {year} {2017})}\BibitemShut {NoStop}%
\bibitem [{\citenamefont {Bracher}\ \emph {et~al.}(2017)\citenamefont
  {Bracher}, \citenamefont {Zhang},\ and\ \citenamefont {Hu}}]{Bracher_PNAS17}%
  \BibitemOpen
  \bibfield  {author} {\bibinfo {author} {\bibfnamefont {D.~O.}\ \bibnamefont
  {Bracher}}, \bibinfo {author} {\bibfnamefont {X.}~\bibnamefont {Zhang}}, \
  and\ \bibinfo {author} {\bibfnamefont {E.~L.}\ \bibnamefont {Hu}},\ }\href
  {\doibase 10.1073/pnas.1704219114} {\bibfield  {journal} {\bibinfo  {journal}
  {Proceedings of the National Academy of Sciences}\ }\textbf {\bibinfo
  {volume} {114}},\ \bibinfo {pages} {4060} (\bibinfo {year} {2017})},\ \Eprint
  {http://arxiv.org/abs/http://www.pnas.org/content/114/16/4060.full.pdf}
  {http://www.pnas.org/content/114/16/4060.full.pdf} \BibitemShut {NoStop}%
\bibitem [{\citenamefont {Luxmoore}\ \emph {et~al.}(2013)\citenamefont
  {Luxmoore}, \citenamefont {Wasley}, \citenamefont {Ramsay}, \citenamefont
  {Thijssen}, \citenamefont {Oulton}, \citenamefont {Hugues}, \citenamefont
  {Kasture}, \citenamefont {Achanta}, \citenamefont {Fox},\ and\ \citenamefont
  {Skolnick}}]{Luxmoore_PRL13}%
  \BibitemOpen
  \bibfield  {author} {\bibinfo {author} {\bibfnamefont {I.~J.}\ \bibnamefont
  {Luxmoore}}, \bibinfo {author} {\bibfnamefont {N.~A.}\ \bibnamefont
  {Wasley}}, \bibinfo {author} {\bibfnamefont {A.~J.}\ \bibnamefont {Ramsay}},
  \bibinfo {author} {\bibfnamefont {A.~C.~T.}\ \bibnamefont {Thijssen}},
  \bibinfo {author} {\bibfnamefont {R.}~\bibnamefont {Oulton}}, \bibinfo
  {author} {\bibfnamefont {M.}~\bibnamefont {Hugues}}, \bibinfo {author}
  {\bibfnamefont {S.}~\bibnamefont {Kasture}}, \bibinfo {author} {\bibfnamefont
  {V.~G.}\ \bibnamefont {Achanta}}, \bibinfo {author} {\bibfnamefont {A.~M.}\
  \bibnamefont {Fox}}, \ and\ \bibinfo {author} {\bibfnamefont {M.~S.}\
  \bibnamefont {Skolnick}},\ }\href {\doibase 10.1103/PhysRevLett.110.037402}
  {\bibfield  {journal} {\bibinfo  {journal} {Phys. Rev. Lett.}\ }\textbf
  {\bibinfo {volume} {110}},\ \bibinfo {pages} {037402} (\bibinfo {year}
  {2013})}\BibitemShut {NoStop}%
\bibitem [{\citenamefont {Arcari}\ \emph {et~al.}(2014)\citenamefont {Arcari},
  \citenamefont {S{\"o}llner}, \citenamefont {Javadi}, \citenamefont {{Lindskov
  Hansen}}, \citenamefont {Mahmoodian}, \citenamefont {Liu}, \citenamefont
  {Thyrrestrup}, \citenamefont {Lee}, \citenamefont {Song}, \citenamefont
  {Stobbe},\ and\ \citenamefont {Lodahl}}]{PhysRevLett.113.093603}%
  \BibitemOpen
  \bibfield  {author} {\bibinfo {author} {\bibfnamefont {M.}~\bibnamefont
  {Arcari}}, \bibinfo {author} {\bibfnamefont {I.}~\bibnamefont {S{\"o}llner}},
  \bibinfo {author} {\bibfnamefont {A.}~\bibnamefont {Javadi}}, \bibinfo
  {author} {\bibfnamefont {S.}~\bibnamefont {{Lindskov Hansen}}}, \bibinfo
  {author} {\bibfnamefont {S.}~\bibnamefont {Mahmoodian}}, \bibinfo {author}
  {\bibfnamefont {J.}~\bibnamefont {Liu}}, \bibinfo {author} {\bibfnamefont
  {H.}~\bibnamefont {Thyrrestrup}}, \bibinfo {author} {\bibfnamefont {E.~H.}\
  \bibnamefont {Lee}}, \bibinfo {author} {\bibfnamefont {J.~D.}\ \bibnamefont
  {Song}}, \bibinfo {author} {\bibfnamefont {S.}~\bibnamefont {Stobbe}}, \ and\
  \bibinfo {author} {\bibfnamefont {P.}~\bibnamefont {Lodahl}},\ }\href
  {\doibase 10.1103/PhysRevLett.113.093603} {\bibfield  {journal} {\bibinfo
  {journal} {Phys. Rev. Lett.}\ }\textbf {\bibinfo {volume} {113}},\ \bibinfo
  {pages} {093603} (\bibinfo {year} {2014})}\BibitemShut {NoStop}%
\bibitem [{\citenamefont {Lodahl}\ \emph {et~al.}(2015)\citenamefont {Lodahl},
  \citenamefont {Mahmoodian},\ and\ \citenamefont {Stobbe}}]{Lodahl_RMP15}%
  \BibitemOpen
  \bibfield  {author} {\bibinfo {author} {\bibfnamefont {P.}~\bibnamefont
  {Lodahl}}, \bibinfo {author} {\bibfnamefont {S.}~\bibnamefont {Mahmoodian}},
  \ and\ \bibinfo {author} {\bibfnamefont {S.}~\bibnamefont {Stobbe}},\ }\href
  {\doibase 10.1103/RevModPhys.87.347} {\bibfield  {journal} {\bibinfo
  {journal} {Rev. Mod. Phys.}\ }\textbf {\bibinfo {volume} {87}},\ \bibinfo
  {pages} {347} (\bibinfo {year} {2015})}\BibitemShut {NoStop}%
\bibitem [{\citenamefont {Lodahl}\ \emph {et~al.}(2017)\citenamefont {Lodahl},
  \citenamefont {Mahmoodian}, \citenamefont {Stobbe}, \citenamefont
  {Rauschenbeutel}, \citenamefont {Schneeweiss}, \citenamefont {Volz},
  \citenamefont {Pichler},\ and\ \citenamefont {Zoller}}]{Lodahl_Nature17}%
  \BibitemOpen
  \bibfield  {author} {\bibinfo {author} {\bibfnamefont {P.}~\bibnamefont
  {Lodahl}}, \bibinfo {author} {\bibfnamefont {S.}~\bibnamefont {Mahmoodian}},
  \bibinfo {author} {\bibfnamefont {S.}~\bibnamefont {Stobbe}}, \bibinfo
  {author} {\bibfnamefont {A.}~\bibnamefont {Rauschenbeutel}}, \bibinfo
  {author} {\bibfnamefont {P.}~\bibnamefont {Schneeweiss}}, \bibinfo {author}
  {\bibfnamefont {J.}~\bibnamefont {Volz}}, \bibinfo {author} {\bibfnamefont
  {H.}~\bibnamefont {Pichler}}, \ and\ \bibinfo {author} {\bibfnamefont
  {P.}~\bibnamefont {Zoller}},\ }\href {\doibase 10.1038/nature21037}
  {\bibfield  {journal} {\bibinfo  {journal} {Nature}\ }\textbf {\bibinfo
  {volume} {541}},\ \bibinfo {pages} {473} (\bibinfo {year}
  {2017})}\BibitemShut {NoStop}%
\bibitem [{\citenamefont {Lang}\ \emph {et~al.}(2017)\citenamefont {Lang},
  \citenamefont {Oulton},\ and\ \citenamefont {Beggs}}]{Lang_JOpt17}%
  \BibitemOpen
  \bibfield  {author} {\bibinfo {author} {\bibfnamefont {B.}~\bibnamefont
  {Lang}}, \bibinfo {author} {\bibfnamefont {R.}~\bibnamefont {Oulton}}, \ and\
  \bibinfo {author} {\bibfnamefont {D.~M.}\ \bibnamefont {Beggs}},\ }\href@noop
  {} {\bibfield  {journal} {\bibinfo  {journal} {J. Opt.}\ }\textbf {\bibinfo
  {volume} {19}},\ \bibinfo {pages} {045001} (\bibinfo {year}
  {2017})}\BibitemShut {NoStop}%
\bibitem [{\citenamefont {Togan}\ \emph {et~al.}(2010)\citenamefont {Togan},
  \citenamefont {Chu}, \citenamefont {Trifonov}, \citenamefont {Jiang},
  \citenamefont {Maze}, \citenamefont {Childress}, \citenamefont {Dutt},
  \citenamefont {S{\o}rensen}, \citenamefont {Hemmer}, \citenamefont {Zibrov},\
  and\ \citenamefont {Lukin}}]{Togan_2010}%
  \BibitemOpen
  \bibfield  {author} {\bibinfo {author} {\bibfnamefont {E.}~\bibnamefont
  {Togan}}, \bibinfo {author} {\bibfnamefont {Y.}~\bibnamefont {Chu}}, \bibinfo
  {author} {\bibfnamefont {A.~S.}\ \bibnamefont {Trifonov}}, \bibinfo {author}
  {\bibfnamefont {L.}~\bibnamefont {Jiang}}, \bibinfo {author} {\bibfnamefont
  {J.}~\bibnamefont {Maze}}, \bibinfo {author} {\bibfnamefont {L.}~\bibnamefont
  {Childress}}, \bibinfo {author} {\bibfnamefont {M.~V.~G.}\ \bibnamefont
  {Dutt}}, \bibinfo {author} {\bibfnamefont {A.~S.}\ \bibnamefont
  {S{\o}rensen}}, \bibinfo {author} {\bibfnamefont {P.~R.}\ \bibnamefont
  {Hemmer}}, \bibinfo {author} {\bibfnamefont {A.~S.}\ \bibnamefont {Zibrov}},
  \ and\ \bibinfo {author} {\bibfnamefont {M.~D.}\ \bibnamefont {Lukin}},\
  }\href {\doibase 10.1038/nature09256} {\bibfield  {journal} {\bibinfo
  {journal} {Nature}\ }\textbf {\bibinfo {volume} {466}},\ \bibinfo {pages}
  {730} (\bibinfo {year} {2010})}\BibitemShut {NoStop}%
\bibitem [{\citenamefont {Greve}\ \emph {et~al.}(2012)\citenamefont {Greve},
  \citenamefont {Yu}, \citenamefont {McMahon}, \citenamefont {Pelc},
  \citenamefont {Natarajan}, \citenamefont {Kim}, \citenamefont {Abe},
  \citenamefont {Maier}, \citenamefont {Schneider}, \citenamefont {Kamp},
  \citenamefont {Hofling}, \citenamefont {Hadfield}, \citenamefont {Forchel},
  \citenamefont {Fejer},\ and\ \citenamefont {Yamamoto}}]{DeGreve_Nature12}%
  \BibitemOpen
  \bibfield  {author} {\bibinfo {author} {\bibfnamefont {K.~D.}\ \bibnamefont
  {Greve}}, \bibinfo {author} {\bibfnamefont {L.}~\bibnamefont {Yu}}, \bibinfo
  {author} {\bibfnamefont {P.~L.}\ \bibnamefont {McMahon}}, \bibinfo {author}
  {\bibfnamefont {J.~S.}\ \bibnamefont {Pelc}}, \bibinfo {author}
  {\bibfnamefont {C.~M.}\ \bibnamefont {Natarajan}}, \bibinfo {author}
  {\bibfnamefont {N.~Y.}\ \bibnamefont {Kim}}, \bibinfo {author} {\bibfnamefont
  {E.}~\bibnamefont {Abe}}, \bibinfo {author} {\bibfnamefont {S.}~\bibnamefont
  {Maier}}, \bibinfo {author} {\bibfnamefont {C.}~\bibnamefont {Schneider}},
  \bibinfo {author} {\bibfnamefont {M.}~\bibnamefont {Kamp}}, \bibinfo {author}
  {\bibfnamefont {S.}~\bibnamefont {Hofling}}, \bibinfo {author} {\bibfnamefont
  {R.~H.}\ \bibnamefont {Hadfield}}, \bibinfo {author} {\bibfnamefont
  {A.}~\bibnamefont {Forchel}}, \bibinfo {author} {\bibfnamefont {M.~M.}\
  \bibnamefont {Fejer}}, \ and\ \bibinfo {author} {\bibfnamefont
  {Y.}~\bibnamefont {Yamamoto}},\ }\href@noop {} {\bibfield  {journal}
  {\bibinfo  {journal} {Nature}\ }\textbf {\bibinfo {volume} {491}},\ \bibinfo
  {pages} {421} (\bibinfo {year} {2012})}\BibitemShut {NoStop}%
\bibitem [{\citenamefont {Gao}\ \emph {et~al.}(2012)\citenamefont {Gao},
  \citenamefont {Fallahi}, \citenamefont {Togan}, \citenamefont
  {Miguel-Sanchez},\ and\ \citenamefont {Imamoglu}}]{Gao_Nature12}%
  \BibitemOpen
  \bibfield  {author} {\bibinfo {author} {\bibfnamefont {W.~B.}\ \bibnamefont
  {Gao}}, \bibinfo {author} {\bibfnamefont {P.}~\bibnamefont {Fallahi}},
  \bibinfo {author} {\bibfnamefont {E.}~\bibnamefont {Togan}}, \bibinfo
  {author} {\bibfnamefont {J.}~\bibnamefont {Miguel-Sanchez}}, \ and\ \bibinfo
  {author} {\bibfnamefont {A.}~\bibnamefont {Imamoglu}},\ }\href@noop {}
  {\bibfield  {journal} {\bibinfo  {journal} {Nature}\ }\textbf {\bibinfo
  {volume} {491}},\ \bibinfo {pages} {426} (\bibinfo {year}
  {2012})}\BibitemShut {NoStop}%
\bibitem [{\citenamefont {Hensen}\ \emph {et~al.}(2015)\citenamefont {Hensen},
  \citenamefont {Bernien}, \citenamefont {Dr{\'e}au}, \citenamefont {Reiserer},
  \citenamefont {Kalb}, \citenamefont {Blok}, \citenamefont {Ruitenberg},
  \citenamefont {Vermeulen}, \citenamefont {Schouten}, \citenamefont
  {Abell{\'a}n},\ and\ \citenamefont {et~al.}}]{Hensen_2015}%
  \BibitemOpen
  \bibfield  {author} {\bibinfo {author} {\bibfnamefont {B.}~\bibnamefont
  {Hensen}}, \bibinfo {author} {\bibfnamefont {H.}~\bibnamefont {Bernien}},
  \bibinfo {author} {\bibfnamefont {A.~E.}\ \bibnamefont {Dr{\'e}au}}, \bibinfo
  {author} {\bibfnamefont {A.}~\bibnamefont {Reiserer}}, \bibinfo {author}
  {\bibfnamefont {N.}~\bibnamefont {Kalb}}, \bibinfo {author} {\bibfnamefont
  {M.~S.}\ \bibnamefont {Blok}}, \bibinfo {author} {\bibfnamefont
  {J.}~\bibnamefont {Ruitenberg}}, \bibinfo {author} {\bibfnamefont {R.~F.~L.}\
  \bibnamefont {Vermeulen}}, \bibinfo {author} {\bibfnamefont {R.~N.}\
  \bibnamefont {Schouten}}, \bibinfo {author} {\bibfnamefont {C.}~\bibnamefont
  {Abell{\'a}n}}, \ and\ \bibinfo {author} {\bibnamefont {et~al.}},\ }\href
  {\doibase 10.1038/nature15759} {\bibfield  {journal} {\bibinfo  {journal}
  {Nature}\ }\textbf {\bibinfo {volume} {526}},\ \bibinfo {pages} {682}
  (\bibinfo {year} {2015})}\BibitemShut {NoStop}%
\bibitem [{\citenamefont {Stockill}\ \emph {et~al.}(2017)\citenamefont
  {Stockill}, \citenamefont {Stanley}, \citenamefont {Huthmacher},
  \citenamefont {Clarke}, \citenamefont {Hugues}, \citenamefont {Miller},
  \citenamefont {Matthiesen}, \citenamefont {{Le Gall}},\ and\ \citenamefont
  {Atat{\"u}re}}]{Stockill_PRL17}%
  \BibitemOpen
  \bibfield  {author} {\bibinfo {author} {\bibfnamefont {R.}~\bibnamefont
  {Stockill}}, \bibinfo {author} {\bibfnamefont {M.~J.}\ \bibnamefont
  {Stanley}}, \bibinfo {author} {\bibfnamefont {L.}~\bibnamefont {Huthmacher}},
  \bibinfo {author} {\bibfnamefont {E.}~\bibnamefont {Clarke}}, \bibinfo
  {author} {\bibfnamefont {M.}~\bibnamefont {Hugues}}, \bibinfo {author}
  {\bibfnamefont {A.~J.}\ \bibnamefont {Miller}}, \bibinfo {author}
  {\bibfnamefont {C.}~\bibnamefont {Matthiesen}}, \bibinfo {author}
  {\bibfnamefont {C.}~\bibnamefont {{Le Gall}}}, \ and\ \bibinfo {author}
  {\bibfnamefont {M.}~\bibnamefont {Atat{\"u}re}},\ }\href {\doibase
  10.1103/PhysRevLett.119.010503} {\bibfield  {journal} {\bibinfo  {journal}
  {Phys. Rev. Lett.}\ }\textbf {\bibinfo {volume} {119}},\ \bibinfo {pages}
  {010503} (\bibinfo {year} {2017})}\BibitemShut {NoStop}%
\bibitem [{\citenamefont {Hein}\ \emph {et~al.}(2004)\citenamefont {Hein},
  \citenamefont {Eisert},\ and\ \citenamefont {Briegel}}]{Hein_2004}%
  \BibitemOpen
  \bibfield  {author} {\bibinfo {author} {\bibfnamefont {M.}~\bibnamefont
  {Hein}}, \bibinfo {author} {\bibfnamefont {J.}~\bibnamefont {Eisert}}, \ and\
  \bibinfo {author} {\bibfnamefont {H.~J.}\ \bibnamefont {Briegel}},\ }\href
  {\doibase 10.1103/PhysRevA.69.062311} {\bibfield  {journal} {\bibinfo
  {journal} {Phys. Rev. A}\ }\textbf {\bibinfo {volume} {69}},\ \bibinfo
  {pages} {062311} (\bibinfo {year} {2004})}\BibitemShut {NoStop}%
\bibitem [{\citenamefont {Raussendorf}\ \emph {et~al.}(2003)\citenamefont
  {Raussendorf}, \citenamefont {Browne},\ and\ \citenamefont
  {Briegel}}]{PhysRevA.68.022312}%
  \BibitemOpen
  \bibfield  {author} {\bibinfo {author} {\bibfnamefont {R.}~\bibnamefont
  {Raussendorf}}, \bibinfo {author} {\bibfnamefont {D.~E.}\ \bibnamefont
  {Browne}}, \ and\ \bibinfo {author} {\bibfnamefont {H.~J.}\ \bibnamefont
  {Briegel}},\ }\href {\doibase 10.1103/PhysRevA.68.022312} {\bibfield
  {journal} {\bibinfo  {journal} {Phys. Rev. A}\ }\textbf {\bibinfo {volume}
  {68}},\ \bibinfo {pages} {022312} (\bibinfo {year} {2003})}\BibitemShut
  {NoStop}%
\bibitem [{\citenamefont {Nielsen}\ and\ \citenamefont
  {Dawson}(2005)}]{PhysRevA.71.042323}%
  \BibitemOpen
  \bibfield  {author} {\bibinfo {author} {\bibfnamefont {M.~A.}\ \bibnamefont
  {Nielsen}}\ and\ \bibinfo {author} {\bibfnamefont {C.~M.}\ \bibnamefont
  {Dawson}},\ }\href {\doibase 10.1103/PhysRevA.71.042323} {\bibfield
  {journal} {\bibinfo  {journal} {Phys. Rev. A}\ }\textbf {\bibinfo {volume}
  {71}},\ \bibinfo {pages} {042323} (\bibinfo {year} {2005})}\BibitemShut
  {NoStop}%
\bibitem [{\citenamefont {D{\"u}r}\ \emph {et~al.}(2003)\citenamefont
  {D{\"u}r}, \citenamefont {Aschauer},\ and\ \citenamefont
  {Briegel}}]{PhysRevLett.91.107903}%
  \BibitemOpen
  \bibfield  {author} {\bibinfo {author} {\bibfnamefont {W.}~\bibnamefont
  {D{\"u}r}}, \bibinfo {author} {\bibfnamefont {H.}~\bibnamefont {Aschauer}}, \
  and\ \bibinfo {author} {\bibfnamefont {H.-J.}\ \bibnamefont {Briegel}},\
  }\href {\doibase 10.1103/PhysRevLett.91.107903} {\bibfield  {journal}
  {\bibinfo  {journal} {Phys. Rev. Lett.}\ }\textbf {\bibinfo {volume} {91}},\
  \bibinfo {pages} {107903} (\bibinfo {year} {2003})}\BibitemShut {NoStop}%
\bibitem [{\citenamefont {Schlingemann}\ and\ \citenamefont
  {Werner}(2001)}]{PhysRevA.65.012308}%
  \BibitemOpen
  \bibfield  {author} {\bibinfo {author} {\bibfnamefont {D.}~\bibnamefont
  {Schlingemann}}\ and\ \bibinfo {author} {\bibfnamefont {R.~F.}\ \bibnamefont
  {Werner}},\ }\href {\doibase 10.1103/PhysRevA.65.012308} {\bibfield
  {journal} {\bibinfo  {journal} {Phys. Rev. A}\ }\textbf {\bibinfo {volume}
  {65}},\ \bibinfo {pages} {012308} (\bibinfo {year} {2001})}\BibitemShut
  {NoStop}%
\bibitem [{\citenamefont {Knill}\ \emph {et~al.}(2001)\citenamefont {Knill},
  \citenamefont {Laflamme},\ and\ \citenamefont {Milburn}}]{Knill_2001}%
  \BibitemOpen
  \bibfield  {author} {\bibinfo {author} {\bibfnamefont {E.}~\bibnamefont
  {Knill}}, \bibinfo {author} {\bibfnamefont {R.}~\bibnamefont {Laflamme}}, \
  and\ \bibinfo {author} {\bibfnamefont {G.~J.}\ \bibnamefont {Milburn}},\
  }\href {\doibase 10.1038/35051009} {\bibfield  {journal} {\bibinfo  {journal}
  {Nature}\ }\textbf {\bibinfo {volume} {409}},\ \bibinfo {pages} {46}
  (\bibinfo {year} {2001})}\BibitemShut {NoStop}%
\bibitem [{\citenamefont {Pan}\ \emph {et~al.}(2012)\citenamefont {Pan},
  \citenamefont {Chen}, \citenamefont {Lu}, \citenamefont {Weinfurter},
  \citenamefont {Zeilinger},\ and\ \citenamefont {{\ifmmode \dot{Z}\else
  {\.Z}\fi{}ukowski}}}]{Pan_RMP12}%
  \BibitemOpen
  \bibfield  {author} {\bibinfo {author} {\bibfnamefont {J.-W.}\ \bibnamefont
  {Pan}}, \bibinfo {author} {\bibfnamefont {Z.-B.}\ \bibnamefont {Chen}},
  \bibinfo {author} {\bibfnamefont {C.-Y.}\ \bibnamefont {Lu}}, \bibinfo
  {author} {\bibfnamefont {H.}~\bibnamefont {Weinfurter}}, \bibinfo {author}
  {\bibfnamefont {A.}~\bibnamefont {Zeilinger}}, \ and\ \bibinfo {author}
  {\bibfnamefont {M.}~\bibnamefont {{\ifmmode \dot{Z}\else
  {\.Z}\fi{}ukowski}}},\ }\href {\doibase 10.1103/RevModPhys.84.777} {\bibfield
   {journal} {\bibinfo  {journal} {Rev. Mod. Phys.}\ }\textbf {\bibinfo
  {volume} {84}},\ \bibinfo {pages} {777} (\bibinfo {year} {2012})}\BibitemShut
  {NoStop}%
\bibitem [{\citenamefont {Wang}\ \emph {et~al.}(2016)\citenamefont {Wang},
  \citenamefont {Chen}, \citenamefont {Li}, \citenamefont {Huang},
  \citenamefont {Liu}, \citenamefont {Chen}, \citenamefont {Luo}, \citenamefont
  {Su}, \citenamefont {Wu}, \citenamefont {Li}, \citenamefont {Lu},
  \citenamefont {Hu}, \citenamefont {Jiang}, \citenamefont {Peng},
  \citenamefont {Li}, \citenamefont {Liu}, \citenamefont {Chen}, \citenamefont
  {Lu},\ and\ \citenamefont {Pan}}]{PhysRevLett.117.210502}%
  \BibitemOpen
  \bibfield  {author} {\bibinfo {author} {\bibfnamefont {X.-L.}\ \bibnamefont
  {Wang}}, \bibinfo {author} {\bibfnamefont {L.-K.}\ \bibnamefont {Chen}},
  \bibinfo {author} {\bibfnamefont {W.}~\bibnamefont {Li}}, \bibinfo {author}
  {\bibfnamefont {H.-L.}\ \bibnamefont {Huang}}, \bibinfo {author}
  {\bibfnamefont {C.}~\bibnamefont {Liu}}, \bibinfo {author} {\bibfnamefont
  {C.}~\bibnamefont {Chen}}, \bibinfo {author} {\bibfnamefont {Y.-H.}\
  \bibnamefont {Luo}}, \bibinfo {author} {\bibfnamefont {Z.-E.}\ \bibnamefont
  {Su}}, \bibinfo {author} {\bibfnamefont {D.}~\bibnamefont {Wu}}, \bibinfo
  {author} {\bibfnamefont {Z.-D.}\ \bibnamefont {Li}}, \bibinfo {author}
  {\bibfnamefont {H.}~\bibnamefont {Lu}}, \bibinfo {author} {\bibfnamefont
  {Y.}~\bibnamefont {Hu}}, \bibinfo {author} {\bibfnamefont {X.}~\bibnamefont
  {Jiang}}, \bibinfo {author} {\bibfnamefont {C.-Z.}\ \bibnamefont {Peng}},
  \bibinfo {author} {\bibfnamefont {L.}~\bibnamefont {Li}}, \bibinfo {author}
  {\bibfnamefont {N.-L.}\ \bibnamefont {Liu}}, \bibinfo {author} {\bibfnamefont
  {Y.-A.}\ \bibnamefont {Chen}}, \bibinfo {author} {\bibfnamefont {C.-Y.}\
  \bibnamefont {Lu}}, \ and\ \bibinfo {author} {\bibfnamefont {J.-W.}\
  \bibnamefont {Pan}},\ }\href {\doibase 10.1103/PhysRevLett.117.210502}
  {\bibfield  {journal} {\bibinfo  {journal} {Phys. Rev. Lett.}\ }\textbf
  {\bibinfo {volume} {117}},\ \bibinfo {pages} {210502} (\bibinfo {year}
  {2016})}\BibitemShut {NoStop}%
\bibitem [{\citenamefont {Senellart}\ \emph {et~al.}(2017)\citenamefont
  {Senellart}, \citenamefont {Solomon},\ and\ \citenamefont
  {White}}]{Senellart_NatNano17}%
  \BibitemOpen
  \bibfield  {author} {\bibinfo {author} {\bibfnamefont {P.}~\bibnamefont
  {Senellart}}, \bibinfo {author} {\bibfnamefont {G.}~\bibnamefont {Solomon}},
  \ and\ \bibinfo {author} {\bibfnamefont {A.}~\bibnamefont {White}},\
  }\href@noop {} {\bibfield  {journal} {\bibinfo  {journal} {Nat.
  Nanotechnol.}\ }\textbf {\bibinfo {volume} {12}},\ \bibinfo {pages} {1026}
  (\bibinfo {year} {2017})}\BibitemShut {NoStop}%
\bibitem [{\citenamefont {Krenner}\ \emph {et~al.}(2005)\citenamefont
  {Krenner}, \citenamefont {Sabathil}, \citenamefont {Clark}, \citenamefont
  {Kress}, \citenamefont {Schuh}, \citenamefont {Bichler}, \citenamefont
  {Abstreiter},\ and\ \citenamefont {Finley}}]{PhysRevLett.94.057402}%
  \BibitemOpen
  \bibfield  {author} {\bibinfo {author} {\bibfnamefont {H.~J.}\ \bibnamefont
  {Krenner}}, \bibinfo {author} {\bibfnamefont {M.}~\bibnamefont {Sabathil}},
  \bibinfo {author} {\bibfnamefont {E.~C.}\ \bibnamefont {Clark}}, \bibinfo
  {author} {\bibfnamefont {A.}~\bibnamefont {Kress}}, \bibinfo {author}
  {\bibfnamefont {D.}~\bibnamefont {Schuh}}, \bibinfo {author} {\bibfnamefont
  {M.}~\bibnamefont {Bichler}}, \bibinfo {author} {\bibfnamefont
  {G.}~\bibnamefont {Abstreiter}}, \ and\ \bibinfo {author} {\bibfnamefont
  {J.~J.}\ \bibnamefont {Finley}},\ }\href {\doibase
  10.1103/PhysRevLett.94.057402} {\bibfield  {journal} {\bibinfo  {journal}
  {Phys. Rev. Lett.}\ }\textbf {\bibinfo {volume} {94}},\ \bibinfo {pages}
  {057402} (\bibinfo {year} {2005})}\BibitemShut {NoStop}%
\bibitem [{\citenamefont {Stinaff}\ \emph {et~al.}(2006)\citenamefont
  {Stinaff}, \citenamefont {Scheibner}, \citenamefont {Bracker}, \citenamefont
  {Ponomarev}, \citenamefont {Korenev}, \citenamefont {Ware}, \citenamefont
  {Doty}, \citenamefont {Reinecke},\ and\ \citenamefont {Gammon}}]{Stinaff636}%
  \BibitemOpen
  \bibfield  {author} {\bibinfo {author} {\bibfnamefont {E.~A.}\ \bibnamefont
  {Stinaff}}, \bibinfo {author} {\bibfnamefont {M.}~\bibnamefont {Scheibner}},
  \bibinfo {author} {\bibfnamefont {A.~S.}\ \bibnamefont {Bracker}}, \bibinfo
  {author} {\bibfnamefont {I.~V.}\ \bibnamefont {Ponomarev}}, \bibinfo {author}
  {\bibfnamefont {V.~L.}\ \bibnamefont {Korenev}}, \bibinfo {author}
  {\bibfnamefont {M.~E.}\ \bibnamefont {Ware}}, \bibinfo {author}
  {\bibfnamefont {M.~F.}\ \bibnamefont {Doty}}, \bibinfo {author}
  {\bibfnamefont {T.~L.}\ \bibnamefont {Reinecke}}, \ and\ \bibinfo {author}
  {\bibfnamefont {D.}~\bibnamefont {Gammon}},\ }\href {\doibase
  10.1126/science.1121189} {\bibfield  {journal} {\bibinfo  {journal}
  {Science}\ }\textbf {\bibinfo {volume} {311}},\ \bibinfo {pages} {636}
  (\bibinfo {year} {2006})}\BibitemShut {NoStop}%
\bibitem [{\citenamefont {Bayer}\ \emph {et~al.}(2001)\citenamefont {Bayer},
  \citenamefont {Hawrylak}, \citenamefont {Hinzer}, \citenamefont {Fafard},
  \citenamefont {Korkusinski}, \citenamefont {Wasilewski}, \citenamefont
  {Stern},\ and\ \citenamefont {Forchel}}]{Bayer451}%
  \BibitemOpen
  \bibfield  {author} {\bibinfo {author} {\bibfnamefont {M.}~\bibnamefont
  {Bayer}}, \bibinfo {author} {\bibfnamefont {P.}~\bibnamefont {Hawrylak}},
  \bibinfo {author} {\bibfnamefont {K.}~\bibnamefont {Hinzer}}, \bibinfo
  {author} {\bibfnamefont {S.}~\bibnamefont {Fafard}}, \bibinfo {author}
  {\bibfnamefont {M.}~\bibnamefont {Korkusinski}}, \bibinfo {author}
  {\bibfnamefont {Z.}~\bibnamefont {Wasilewski}}, \bibinfo {author}
  {\bibfnamefont {O.}~\bibnamefont {Stern}}, \ and\ \bibinfo {author}
  {\bibfnamefont {A.}~\bibnamefont {Forchel}},\ }\href {\doibase
  10.1126/science.291.5503.451} {\bibfield  {journal} {\bibinfo  {journal}
  {Science}\ }\textbf {\bibinfo {volume} {291}},\ \bibinfo {pages} {451}
  (\bibinfo {year} {2001})}\BibitemShut {NoStop}%
\bibitem [{\citenamefont {Gimeno-Segovia}\ \emph {et~al.}(2018)\citenamefont
  {Gimeno-Segovia}, \citenamefont {Rudolph},\ and\ \citenamefont
  {Economou}}]{Gimeno_Segovia_2017}%
  \BibitemOpen
  \bibfield  {author} {\bibinfo {author} {\bibfnamefont {M.}~\bibnamefont
  {Gimeno-Segovia}}, \bibinfo {author} {\bibfnamefont {T.}~\bibnamefont
  {Rudolph}}, \ and\ \bibinfo {author} {\bibfnamefont {S.~E.}\ \bibnamefont
  {Economou}},\ }\href@noop {} {\enquote {\bibinfo {title} {Deterministic
  generation of large-scale entangled photonic cluster state from interacting
  solid state emitters},}\ } (\bibinfo {year} {2018}),\ \Eprint
  {http://arxiv.org/abs/arXiv:1801.02599} {arXiv:1801.02599} \BibitemShut
  {NoStop}%
\bibitem [{\citenamefont {Economou}\ \emph {et~al.}(2006)\citenamefont
  {Economou}, \citenamefont {Sham}, \citenamefont {Wu},\ and\ \citenamefont
  {Steel}}]{PhysRevB.74.205415}%
  \BibitemOpen
  \bibfield  {author} {\bibinfo {author} {\bibfnamefont {S.~E.}\ \bibnamefont
  {Economou}}, \bibinfo {author} {\bibfnamefont {L.~J.}\ \bibnamefont {Sham}},
  \bibinfo {author} {\bibfnamefont {Y.}~\bibnamefont {Wu}}, \ and\ \bibinfo
  {author} {\bibfnamefont {D.~G.}\ \bibnamefont {Steel}},\ }\href {\doibase
  10.1103/PhysRevB.74.205415} {\bibfield  {journal} {\bibinfo  {journal} {Phys.
  Rev. B}\ }\textbf {\bibinfo {volume} {74}},\ \bibinfo {pages} {205415}
  (\bibinfo {year} {2006})}\BibitemShut {NoStop}%
\bibitem [{\citenamefont {Press}\ \emph {et~al.}(2008)\citenamefont {Press},
  \citenamefont {Ladd}, \citenamefont {Zhang},\ and\ \citenamefont
  {Yamamoto}}]{Press_Nature08}%
  \BibitemOpen
  \bibfield  {author} {\bibinfo {author} {\bibfnamefont {D.}~\bibnamefont
  {Press}}, \bibinfo {author} {\bibfnamefont {T.~D.}\ \bibnamefont {Ladd}},
  \bibinfo {author} {\bibfnamefont {B.}~\bibnamefont {Zhang}}, \ and\ \bibinfo
  {author} {\bibfnamefont {Y.}~\bibnamefont {Yamamoto}},\ }\href@noop {}
  {\bibfield  {journal} {\bibinfo  {journal} {Nature}\ }\textbf {\bibinfo
  {volume} {456}},\ \bibinfo {pages} {218} (\bibinfo {year}
  {2008})}\BibitemShut {NoStop}%
\bibitem [{\citenamefont {Greilich}\ \emph {et~al.}(2009)\citenamefont
  {Greilich}, \citenamefont {Economou}, \citenamefont {Spatzek}, \citenamefont
  {Yakovlev}, \citenamefont {Reuter}, \citenamefont {Wieck}, \citenamefont
  {Reinecke},\ and\ \citenamefont {Bayer}}]{Greilich_NP09}%
  \BibitemOpen
  \bibfield  {author} {\bibinfo {author} {\bibfnamefont {A.}~\bibnamefont
  {Greilich}}, \bibinfo {author} {\bibfnamefont {S.~E.}\ \bibnamefont
  {Economou}}, \bibinfo {author} {\bibfnamefont {S.}~\bibnamefont {Spatzek}},
  \bibinfo {author} {\bibfnamefont {D.~R.}\ \bibnamefont {Yakovlev}}, \bibinfo
  {author} {\bibfnamefont {D.}~\bibnamefont {Reuter}}, \bibinfo {author}
  {\bibfnamefont {A.~D.}\ \bibnamefont {Wieck}}, \bibinfo {author}
  {\bibfnamefont {T.~L.}\ \bibnamefont {Reinecke}}, \ and\ \bibinfo {author}
  {\bibfnamefont {M.}~\bibnamefont {Bayer}},\ }\href@noop {} {\bibfield
  {journal} {\bibinfo  {journal} {Nat. Phys.}\ }\textbf {\bibinfo {volume}
  {5}},\ \bibinfo {pages} {262} (\bibinfo {year} {2009})}\BibitemShut {NoStop}%
\bibitem [{\citenamefont {Madsen}\ \emph {et~al.}(2014)\citenamefont {Madsen},
  \citenamefont {Ates}, \citenamefont {Liu}, \citenamefont {Javadi},
  \citenamefont {Albrecht}, \citenamefont {Yeo}, \citenamefont {Stobbe},\ and\
  \citenamefont {Lodahl}}]{PhysRevB.90.155303}%
  \BibitemOpen
  \bibfield  {author} {\bibinfo {author} {\bibfnamefont {K.~H.}\ \bibnamefont
  {Madsen}}, \bibinfo {author} {\bibfnamefont {S.}~\bibnamefont {Ates}},
  \bibinfo {author} {\bibfnamefont {J.}~\bibnamefont {Liu}}, \bibinfo {author}
  {\bibfnamefont {A.}~\bibnamefont {Javadi}}, \bibinfo {author} {\bibfnamefont
  {S.~M.}\ \bibnamefont {Albrecht}}, \bibinfo {author} {\bibfnamefont
  {I.}~\bibnamefont {Yeo}}, \bibinfo {author} {\bibfnamefont {S.}~\bibnamefont
  {Stobbe}}, \ and\ \bibinfo {author} {\bibfnamefont {P.}~\bibnamefont
  {Lodahl}},\ }\href {\doibase 10.1103/PhysRevB.90.155303} {\bibfield
  {journal} {\bibinfo  {journal} {Phys. Rev. B}\ }\textbf {\bibinfo {volume}
  {90}},\ \bibinfo {pages} {155303} (\bibinfo {year} {2014})}\BibitemShut
  {NoStop}%
\bibitem [{\citenamefont {Birowosuto}\ \emph {et~al.}(2012)\citenamefont
  {Birowosuto}, \citenamefont {Sumikura}, \citenamefont {Matsuo}, \citenamefont
  {Taniyama}, \citenamefont {van Veldhoven}, \citenamefont {N{\"o}tzel},\ and\
  \citenamefont {Notomi}}]{Birowosuto_2012}%
  \BibitemOpen
  \bibfield  {author} {\bibinfo {author} {\bibfnamefont {M.~D.}\ \bibnamefont
  {Birowosuto}}, \bibinfo {author} {\bibfnamefont {H.}~\bibnamefont
  {Sumikura}}, \bibinfo {author} {\bibfnamefont {S.}~\bibnamefont {Matsuo}},
  \bibinfo {author} {\bibfnamefont {H.}~\bibnamefont {Taniyama}}, \bibinfo
  {author} {\bibfnamefont {P.~J.}\ \bibnamefont {van Veldhoven}}, \bibinfo
  {author} {\bibfnamefont {R.}~\bibnamefont {N{\"o}tzel}}, \ and\ \bibinfo
  {author} {\bibfnamefont {M.}~\bibnamefont {Notomi}},\ }\href {\doibase
  10.1038/srep00321} {\bibfield  {journal} {\bibinfo  {journal} {Scientific
  Reports}\ }\textbf {\bibinfo {volume} {2}} (\bibinfo {year} {2012}),\
  10.1038/srep00321}\BibitemShut {NoStop}%
\bibitem [{\citenamefont {Kelaita}\ \emph {et~al.}(2016)\citenamefont
  {Kelaita}, \citenamefont {Fischer}, \citenamefont {Babinec}, \citenamefont
  {Lagoudakis}, \citenamefont {Sarmiento}, \citenamefont {Rundquist},
  \citenamefont {Majumdar},\ and\ \citenamefont {Vu{\v
  c}kovi{\'c}}}]{Kelaita_2016}%
  \BibitemOpen
  \bibfield  {author} {\bibinfo {author} {\bibfnamefont {Y.~A.}\ \bibnamefont
  {Kelaita}}, \bibinfo {author} {\bibfnamefont {K.~A.}\ \bibnamefont
  {Fischer}}, \bibinfo {author} {\bibfnamefont {T.~M.}\ \bibnamefont
  {Babinec}}, \bibinfo {author} {\bibfnamefont {K.~G.}\ \bibnamefont
  {Lagoudakis}}, \bibinfo {author} {\bibfnamefont {T.}~\bibnamefont
  {Sarmiento}}, \bibinfo {author} {\bibfnamefont {A.}~\bibnamefont
  {Rundquist}}, \bibinfo {author} {\bibfnamefont {A.}~\bibnamefont {Majumdar}},
  \ and\ \bibinfo {author} {\bibfnamefont {J.}~\bibnamefont {Vu{\v
  c}kovi{\'c}}},\ }\href {\doibase 10.1364/ome.7.000231} {\bibfield  {journal}
  {\bibinfo  {journal} {Optical Materials Express}\ }\textbf {\bibinfo {volume}
  {7}},\ \bibinfo {pages} {231} (\bibinfo {year} {2016})}\BibitemShut {NoStop}%
\bibitem [{\citenamefont {Carter}\ \emph {et~al.}(2013)\citenamefont {Carter},
  \citenamefont {Sweeney}, \citenamefont {Kim}, \citenamefont {Kim},
  \citenamefont {Solenov}, \citenamefont {Economou}, \citenamefont {Reinecke},
  \citenamefont {Yang}, \citenamefont {Bracker},\ and\ \citenamefont
  {Gammon}}]{Carter_2013}%
  \BibitemOpen
  \bibfield  {author} {\bibinfo {author} {\bibfnamefont {S.~G.}\ \bibnamefont
  {Carter}}, \bibinfo {author} {\bibfnamefont {T.~M.}\ \bibnamefont {Sweeney}},
  \bibinfo {author} {\bibfnamefont {M.}~\bibnamefont {Kim}}, \bibinfo {author}
  {\bibfnamefont {C.~S.}\ \bibnamefont {Kim}}, \bibinfo {author} {\bibfnamefont
  {D.}~\bibnamefont {Solenov}}, \bibinfo {author} {\bibfnamefont {S.~E.}\
  \bibnamefont {Economou}}, \bibinfo {author} {\bibfnamefont {T.~L.}\
  \bibnamefont {Reinecke}}, \bibinfo {author} {\bibfnamefont {L.}~\bibnamefont
  {Yang}}, \bibinfo {author} {\bibfnamefont {A.~S.}\ \bibnamefont {Bracker}}, \
  and\ \bibinfo {author} {\bibfnamefont {D.}~\bibnamefont {Gammon}},\ }\href
  {\doibase 10.1038/nphoton.2013.41} {\bibfield  {journal} {\bibinfo  {journal}
  {Nature Photonics}\ }\textbf {\bibinfo {volume} {7}},\ \bibinfo {pages} {329}
  (\bibinfo {year} {2013})}\BibitemShut {NoStop}%
\bibitem [{\citenamefont {Fotso}\ \emph {et~al.}(2016)\citenamefont {Fotso},
  \citenamefont {Feiguin}, \citenamefont {Awschalom},\ and\ \citenamefont
  {Dobrovitski}}]{PhysRevLett.116.033603}%
  \BibitemOpen
  \bibfield  {author} {\bibinfo {author} {\bibfnamefont {H.~F.}\ \bibnamefont
  {Fotso}}, \bibinfo {author} {\bibfnamefont {A.~E.}\ \bibnamefont {Feiguin}},
  \bibinfo {author} {\bibfnamefont {D.~D.}\ \bibnamefont {Awschalom}}, \ and\
  \bibinfo {author} {\bibfnamefont {V.~V.}\ \bibnamefont {Dobrovitski}},\
  }\href {\doibase 10.1103/PhysRevLett.116.033603} {\bibfield  {journal}
  {\bibinfo  {journal} {Phys. Rev. Lett.}\ }\textbf {\bibinfo {volume} {116}},\
  \bibinfo {pages} {033603} (\bibinfo {year} {2016})}\BibitemShut {NoStop}%
\bibitem [{\citenamefont {{Androvitsaneas}}\ \emph {et~al.}(2016)\citenamefont
  {{Androvitsaneas}}, \citenamefont {{Young}}, \citenamefont {{Lennon}},
  \citenamefont {{Schneider}}, \citenamefont {{Maier}}, \citenamefont
  {{Hinchliff}}, \citenamefont {{Atkinson}}, \citenamefont {{Kamp}},
  \citenamefont {{H{\"o}fling}}, \citenamefont {{Rarity}},\ and\ \citenamefont
  {{Oulton}}}]{2016arXiv160902851A}%
  \BibitemOpen
  \bibfield  {author} {\bibinfo {author} {\bibfnamefont {P.}~\bibnamefont
  {{Androvitsaneas}}}, \bibinfo {author} {\bibfnamefont {A.~B.}\ \bibnamefont
  {{Young}}}, \bibinfo {author} {\bibfnamefont {J.~M.}\ \bibnamefont
  {{Lennon}}}, \bibinfo {author} {\bibfnamefont {C.}~\bibnamefont
  {{Schneider}}}, \bibinfo {author} {\bibfnamefont {S.}~\bibnamefont
  {{Maier}}}, \bibinfo {author} {\bibfnamefont {J.~J.}\ \bibnamefont
  {{Hinchliff}}}, \bibinfo {author} {\bibfnamefont {G.}~\bibnamefont
  {{Atkinson}}}, \bibinfo {author} {\bibfnamefont {M.}~\bibnamefont {{Kamp}}},
  \bibinfo {author} {\bibfnamefont {S.}~\bibnamefont {{H{\"o}fling}}}, \bibinfo
  {author} {\bibfnamefont {J.~G.}\ \bibnamefont {{Rarity}}}, \ and\ \bibinfo
  {author} {\bibfnamefont {R.}~\bibnamefont {{Oulton}}},\ }\href@noop {}
  {\bibfield  {journal} {\bibinfo  {journal} {ArXiv e-prints}\ } (\bibinfo
  {year} {2016})},\ \Eprint {http://arxiv.org/abs/1609.02851} {arXiv:1609.02851
  [quant-ph]} \BibitemShut {NoStop}%
\bibitem [{\citenamefont {Gazzano}\ \emph {et~al.}(2013)\citenamefont
  {Gazzano}, \citenamefont {de~Vasconcellos}, \citenamefont {Arnold},
  \citenamefont {Nowak}, \citenamefont {Galopin}, \citenamefont {Sagnes},
  \citenamefont {Lanco}, \citenamefont {Lemaitre},\ and\ \citenamefont
  {Senellart}}]{Gazzano_NC13}%
  \BibitemOpen
  \bibfield  {author} {\bibinfo {author} {\bibfnamefont {O.}~\bibnamefont
  {Gazzano}}, \bibinfo {author} {\bibfnamefont {S.~M.}\ \bibnamefont
  {de~Vasconcellos}}, \bibinfo {author} {\bibfnamefont {C.}~\bibnamefont
  {Arnold}}, \bibinfo {author} {\bibfnamefont {A.}~\bibnamefont {Nowak}},
  \bibinfo {author} {\bibfnamefont {E.}~\bibnamefont {Galopin}}, \bibinfo
  {author} {\bibfnamefont {I.}~\bibnamefont {Sagnes}}, \bibinfo {author}
  {\bibfnamefont {L.}~\bibnamefont {Lanco}}, \bibinfo {author} {\bibfnamefont
  {A.}~\bibnamefont {Lemaitre}}, \ and\ \bibinfo {author} {\bibfnamefont
  {P.}~\bibnamefont {Senellart}},\ }\href@noop {} {\bibfield  {journal}
  {\bibinfo  {journal} {Nat. Commun.}\ }\textbf {\bibinfo {volume} {4}},\
  \bibinfo {pages} {1425} (\bibinfo {year} {2013})}\BibitemShut {NoStop}%
\bibitem [{\citenamefont {Arnold}\ \emph {et~al.}(2015)\citenamefont {Arnold},
  \citenamefont {Demory}, \citenamefont {Loo}, \citenamefont {Lemaitre},
  \citenamefont {Sagnes}, \citenamefont {Glazov}, \citenamefont {Krebs},
  \citenamefont {Voisin}, \citenamefont {Senellart},\ and\ \citenamefont
  {Lanco}}]{Arnold_NC15}%
  \BibitemOpen
  \bibfield  {author} {\bibinfo {author} {\bibfnamefont {C.}~\bibnamefont
  {Arnold}}, \bibinfo {author} {\bibfnamefont {J.}~\bibnamefont {Demory}},
  \bibinfo {author} {\bibfnamefont {V.}~\bibnamefont {Loo}}, \bibinfo {author}
  {\bibfnamefont {A.}~\bibnamefont {Lemaitre}}, \bibinfo {author}
  {\bibfnamefont {I.}~\bibnamefont {Sagnes}}, \bibinfo {author} {\bibfnamefont
  {M.}~\bibnamefont {Glazov}}, \bibinfo {author} {\bibfnamefont
  {O.}~\bibnamefont {Krebs}}, \bibinfo {author} {\bibfnamefont
  {P.}~\bibnamefont {Voisin}}, \bibinfo {author} {\bibfnamefont
  {P.}~\bibnamefont {Senellart}}, \ and\ \bibinfo {author} {\bibfnamefont
  {L.}~\bibnamefont {Lanco}},\ }\href@noop {} {\bibfield  {journal} {\bibinfo
  {journal} {Nat. Commun.}\ }\textbf {\bibinfo {volume} {6}},\ \bibinfo {pages}
  {6236} (\bibinfo {year} {2015})}\BibitemShut {NoStop}%
\bibitem [{\citenamefont {Englund}\ \emph {et~al.}(2005)\citenamefont
  {Englund}, \citenamefont {Fattal}, \citenamefont {Waks}, \citenamefont
  {Solomon}, \citenamefont {Zhang}, \citenamefont {Nakaoka}, \citenamefont
  {Arakawa}, \citenamefont {Yamamoto},\ and\ \citenamefont {{Vu\ifmmode
  \check{c}\else {\v c}\fi{}kovi\ifmmode \acute{c}\else
  {\'c}\fi{}}}}]{PhysRevLett.95.013904}%
  \BibitemOpen
  \bibfield  {author} {\bibinfo {author} {\bibfnamefont {D.}~\bibnamefont
  {Englund}}, \bibinfo {author} {\bibfnamefont {D.}~\bibnamefont {Fattal}},
  \bibinfo {author} {\bibfnamefont {E.}~\bibnamefont {Waks}}, \bibinfo {author}
  {\bibfnamefont {G.}~\bibnamefont {Solomon}}, \bibinfo {author} {\bibfnamefont
  {B.}~\bibnamefont {Zhang}}, \bibinfo {author} {\bibfnamefont
  {T.}~\bibnamefont {Nakaoka}}, \bibinfo {author} {\bibfnamefont
  {Y.}~\bibnamefont {Arakawa}}, \bibinfo {author} {\bibfnamefont
  {Y.}~\bibnamefont {Yamamoto}}, \ and\ \bibinfo {author} {\bibfnamefont
  {J.}~\bibnamefont {{Vu\ifmmode \check{c}\else {\v c}\fi{}kovi\ifmmode
  \acute{c}\else {\'c}\fi{}}}},\ }\href {\doibase
  10.1103/PhysRevLett.95.013904} {\bibfield  {journal} {\bibinfo  {journal}
  {Phys. Rev. Lett.}\ }\textbf {\bibinfo {volume} {95}},\ \bibinfo {pages}
  {013904} (\bibinfo {year} {2005})}\BibitemShut {NoStop}%
\bibitem [{\citenamefont {Lodahl}\ \emph {et~al.}(2004)\citenamefont {Lodahl},
  \citenamefont {{Floris van Driel}}, \citenamefont {Nikolaev}, \citenamefont
  {Irman}, \citenamefont {Overgaag}, \citenamefont {Vanmaekelbergh},\ and\
  \citenamefont {Vos}}]{Lodahl_Nature2004}%
  \BibitemOpen
  \bibfield  {author} {\bibinfo {author} {\bibfnamefont {P.}~\bibnamefont
  {Lodahl}}, \bibinfo {author} {\bibfnamefont {A.}~\bibnamefont {{Floris van
  Driel}}}, \bibinfo {author} {\bibfnamefont {I.~S.}\ \bibnamefont {Nikolaev}},
  \bibinfo {author} {\bibfnamefont {A.}~\bibnamefont {Irman}}, \bibinfo
  {author} {\bibfnamefont {K.}~\bibnamefont {Overgaag}}, \bibinfo {author}
  {\bibfnamefont {D.}~\bibnamefont {Vanmaekelbergh}}, \ and\ \bibinfo {author}
  {\bibfnamefont {W.~L.}\ \bibnamefont {Vos}},\ }\href {\doibase
  10.1038/nature02772} {\bibfield  {journal} {\bibinfo  {journal} {Nature}\
  }\textbf {\bibinfo {volume} {430}},\ \bibinfo {pages} {654} (\bibinfo {year}
  {2004})}\BibitemShut {NoStop}%
\bibitem [{\citenamefont {Faraon}\ \emph {et~al.}(2007)\citenamefont {Faraon},
  \citenamefont {Englund}, \citenamefont {Fushman}, \citenamefont {Vu{\v
  c}kovi{\'c}}, \citenamefont {Stoltz},\ and\ \citenamefont
  {Petroff}}]{Faraon_APL2007}%
  \BibitemOpen
  \bibfield  {author} {\bibinfo {author} {\bibfnamefont {A.}~\bibnamefont
  {Faraon}}, \bibinfo {author} {\bibfnamefont {D.}~\bibnamefont {Englund}},
  \bibinfo {author} {\bibfnamefont {I.}~\bibnamefont {Fushman}}, \bibinfo
  {author} {\bibfnamefont {J.}~\bibnamefont {Vu{\v c}kovi{\'c}}}, \bibinfo
  {author} {\bibfnamefont {N.}~\bibnamefont {Stoltz}}, \ and\ \bibinfo {author}
  {\bibfnamefont {P.}~\bibnamefont {Petroff}},\ }\href {\doibase
  10.1063/1.2742789} {\bibfield  {journal} {\bibinfo  {journal} {Applied
  Physics Letters}\ }\textbf {\bibinfo {volume} {90}},\ \bibinfo {pages}
  {213110} (\bibinfo {year} {2007})}\BibitemShut {NoStop}%
\bibitem [{\citenamefont {Mahmoodian}\ \emph {et~al.}(2016)\citenamefont
  {Mahmoodian}, \citenamefont {Lodahl},\ and\ \citenamefont
  {S{\o}rensen}}]{PhysRevLett.117.240501}%
  \BibitemOpen
  \bibfield  {author} {\bibinfo {author} {\bibfnamefont {S.}~\bibnamefont
  {Mahmoodian}}, \bibinfo {author} {\bibfnamefont {P.}~\bibnamefont {Lodahl}},
  \ and\ \bibinfo {author} {\bibfnamefont {A.~S.}\ \bibnamefont
  {S{\o}rensen}},\ }\href {\doibase 10.1103/PhysRevLett.117.240501} {\bibfield
  {journal} {\bibinfo  {journal} {Phys. Rev. Lett.}\ }\textbf {\bibinfo
  {volume} {117}},\ \bibinfo {pages} {240501} (\bibinfo {year}
  {2016})}\BibitemShut {NoStop}%
\bibitem [{\citenamefont {Bernien}\ \emph {et~al.}(2013)\citenamefont
  {Bernien}, \citenamefont {Hensen}, \citenamefont {Pfaff}, \citenamefont
  {Koolstra}, \citenamefont {Blok}, \citenamefont {Robledo}, \citenamefont
  {Taminiau}, \citenamefont {Markham}, \citenamefont {Twitchen}, \citenamefont
  {Childress},\ and\ \citenamefont {et~al.}}]{Bernien_2013}%
  \BibitemOpen
  \bibfield  {author} {\bibinfo {author} {\bibfnamefont {H.}~\bibnamefont
  {Bernien}}, \bibinfo {author} {\bibfnamefont {B.}~\bibnamefont {Hensen}},
  \bibinfo {author} {\bibfnamefont {W.}~\bibnamefont {Pfaff}}, \bibinfo
  {author} {\bibfnamefont {G.}~\bibnamefont {Koolstra}}, \bibinfo {author}
  {\bibfnamefont {M.~S.}\ \bibnamefont {Blok}}, \bibinfo {author}
  {\bibfnamefont {L.}~\bibnamefont {Robledo}}, \bibinfo {author} {\bibfnamefont
  {T.~H.}\ \bibnamefont {Taminiau}}, \bibinfo {author} {\bibfnamefont
  {M.}~\bibnamefont {Markham}}, \bibinfo {author} {\bibfnamefont {D.~J.}\
  \bibnamefont {Twitchen}}, \bibinfo {author} {\bibfnamefont {L.}~\bibnamefont
  {Childress}}, \ and\ \bibinfo {author} {\bibnamefont {et~al.}},\ }\href
  {\doibase 10.1038/nature12016} {\bibfield  {journal} {\bibinfo  {journal}
  {Nature}\ }\textbf {\bibinfo {volume} {497}},\ \bibinfo {pages} {86}
  (\bibinfo {year} {2013})}\BibitemShut {NoStop}%
\bibitem [{\citenamefont {Taminiau}\ \emph {et~al.}(2014)\citenamefont
  {Taminiau}, \citenamefont {Cramer}, \citenamefont {van~der Sar},
  \citenamefont {Dobrovitski},\ and\ \citenamefont
  {Hanson}}]{Taminiau_NatNano2014}%
  \BibitemOpen
  \bibfield  {author} {\bibinfo {author} {\bibfnamefont {T.~H.}\ \bibnamefont
  {Taminiau}}, \bibinfo {author} {\bibfnamefont {J.}~\bibnamefont {Cramer}},
  \bibinfo {author} {\bibfnamefont {T.}~\bibnamefont {van~der Sar}}, \bibinfo
  {author} {\bibfnamefont {V.~V.}\ \bibnamefont {Dobrovitski}}, \ and\ \bibinfo
  {author} {\bibfnamefont {R.}~\bibnamefont {Hanson}},\ }\href {\doibase
  10.1038/nnano.2014.2} {\bibfield  {journal} {\bibinfo  {journal} {Nature
  Nanotechnology}\ }\textbf {\bibinfo {volume} {9}},\ \bibinfo {pages} {171}
  (\bibinfo {year} {2014})}\BibitemShut {NoStop}%
\bibitem [{\citenamefont {Dutt}\ \emph {et~al.}(2007)\citenamefont {Dutt},
  \citenamefont {Childress}, \citenamefont {Jiang}, \citenamefont {Togan},
  \citenamefont {Maze}, \citenamefont {Jelezko}, \citenamefont {Zibrov},
  \citenamefont {Hemmer},\ and\ \citenamefont {Lukin}}]{Dutt1312}%
  \BibitemOpen
  \bibfield  {author} {\bibinfo {author} {\bibfnamefont {M.~V.~G.}\
  \bibnamefont {Dutt}}, \bibinfo {author} {\bibfnamefont {L.}~\bibnamefont
  {Childress}}, \bibinfo {author} {\bibfnamefont {L.}~\bibnamefont {Jiang}},
  \bibinfo {author} {\bibfnamefont {E.}~\bibnamefont {Togan}}, \bibinfo
  {author} {\bibfnamefont {J.}~\bibnamefont {Maze}}, \bibinfo {author}
  {\bibfnamefont {F.}~\bibnamefont {Jelezko}}, \bibinfo {author} {\bibfnamefont
  {A.~S.}\ \bibnamefont {Zibrov}}, \bibinfo {author} {\bibfnamefont {P.~R.}\
  \bibnamefont {Hemmer}}, \ and\ \bibinfo {author} {\bibfnamefont {M.~D.}\
  \bibnamefont {Lukin}},\ }\href {\doibase 10.1126/science.1139831} {\bibfield
  {journal} {\bibinfo  {journal} {Science}\ }\textbf {\bibinfo {volume}
  {316}},\ \bibinfo {pages} {1312} (\bibinfo {year} {2007})}\BibitemShut
  {NoStop}%
\bibitem [{\citenamefont {Neumann}\ \emph {et~al.}(2008)\citenamefont
  {Neumann}, \citenamefont {Mizuochi}, \citenamefont {Rempp}, \citenamefont
  {Hemmer}, \citenamefont {Watanabe}, \citenamefont {Yamasaki}, \citenamefont
  {Jacques}, \citenamefont {Gaebel}, \citenamefont {Jelezko},\ and\
  \citenamefont {Wrachtrup}}]{Neumann1326}%
  \BibitemOpen
  \bibfield  {author} {\bibinfo {author} {\bibfnamefont {P.}~\bibnamefont
  {Neumann}}, \bibinfo {author} {\bibfnamefont {N.}~\bibnamefont {Mizuochi}},
  \bibinfo {author} {\bibfnamefont {F.}~\bibnamefont {Rempp}}, \bibinfo
  {author} {\bibfnamefont {P.}~\bibnamefont {Hemmer}}, \bibinfo {author}
  {\bibfnamefont {H.}~\bibnamefont {Watanabe}}, \bibinfo {author}
  {\bibfnamefont {S.}~\bibnamefont {Yamasaki}}, \bibinfo {author}
  {\bibfnamefont {V.}~\bibnamefont {Jacques}}, \bibinfo {author} {\bibfnamefont
  {T.}~\bibnamefont {Gaebel}}, \bibinfo {author} {\bibfnamefont
  {F.}~\bibnamefont {Jelezko}}, \ and\ \bibinfo {author} {\bibfnamefont
  {J.}~\bibnamefont {Wrachtrup}},\ }\href {\doibase 10.1126/science.1157233}
  {\bibfield  {journal} {\bibinfo  {journal} {Science}\ }\textbf {\bibinfo
  {volume} {320}},\ \bibinfo {pages} {1326} (\bibinfo {year}
  {2008})}\BibitemShut {NoStop}%
\bibitem [{\citenamefont {Fuchs}\ \emph {et~al.}(2011)\citenamefont {Fuchs},
  \citenamefont {Burkard}, \citenamefont {Klimov},\ and\ \citenamefont
  {Awschalom}}]{Fuchs_NatPhys2011}%
  \BibitemOpen
  \bibfield  {author} {\bibinfo {author} {\bibfnamefont {G.~D.}\ \bibnamefont
  {Fuchs}}, \bibinfo {author} {\bibfnamefont {G.}~\bibnamefont {Burkard}},
  \bibinfo {author} {\bibfnamefont {P.~V.}\ \bibnamefont {Klimov}}, \ and\
  \bibinfo {author} {\bibfnamefont {D.~D.}\ \bibnamefont {Awschalom}},\ }\href
  {\doibase 10.1038/nphys2026} {\bibfield  {journal} {\bibinfo  {journal}
  {Nature Physics}\ }\textbf {\bibinfo {volume} {7}},\ \bibinfo {pages} {789}
  (\bibinfo {year} {2011})}\BibitemShut {NoStop}%
\bibitem [{\citenamefont {Cramer}\ \emph {et~al.}(2016)\citenamefont {Cramer},
  \citenamefont {Kalb}, \citenamefont {Rol}, \citenamefont {Hensen},
  \citenamefont {Blok}, \citenamefont {Markham}, \citenamefont {Twitchen},
  \citenamefont {Hanson},\ and\ \citenamefont {Taminiau}}]{Cramer_NatComm2016}%
  \BibitemOpen
  \bibfield  {author} {\bibinfo {author} {\bibfnamefont {J.}~\bibnamefont
  {Cramer}}, \bibinfo {author} {\bibfnamefont {N.}~\bibnamefont {Kalb}},
  \bibinfo {author} {\bibfnamefont {M.~A.}\ \bibnamefont {Rol}}, \bibinfo
  {author} {\bibfnamefont {B.}~\bibnamefont {Hensen}}, \bibinfo {author}
  {\bibfnamefont {M.~S.}\ \bibnamefont {Blok}}, \bibinfo {author}
  {\bibfnamefont {M.}~\bibnamefont {Markham}}, \bibinfo {author} {\bibfnamefont
  {D.~J.}\ \bibnamefont {Twitchen}}, \bibinfo {author} {\bibfnamefont
  {R.}~\bibnamefont {Hanson}}, \ and\ \bibinfo {author} {\bibfnamefont {T.~H.}\
  \bibnamefont {Taminiau}},\ }\href {\doibase 10.1038/ncomms11526} {\bibfield
  {journal} {\bibinfo  {journal} {Nature Communications}\ }\textbf {\bibinfo
  {volume} {7}},\ \bibinfo {pages} {11526} (\bibinfo {year}
  {2016})}\BibitemShut {NoStop}%
\bibitem [{\citenamefont {Balasubramanian}\ \emph {et~al.}(2009)\citenamefont
  {Balasubramanian}, \citenamefont {Neumann}, \citenamefont {Twitchen},
  \citenamefont {Markham}, \citenamefont {Kolesov}, \citenamefont {Mizuochi},
  \citenamefont {Isoya}, \citenamefont {Achard}, \citenamefont {Beck},
  \citenamefont {Tissler},\ and\ \citenamefont
  {et~al.}}]{Balasubramanian_2009}%
  \BibitemOpen
  \bibfield  {author} {\bibinfo {author} {\bibfnamefont {G.}~\bibnamefont
  {Balasubramanian}}, \bibinfo {author} {\bibfnamefont {P.}~\bibnamefont
  {Neumann}}, \bibinfo {author} {\bibfnamefont {D.}~\bibnamefont {Twitchen}},
  \bibinfo {author} {\bibfnamefont {M.}~\bibnamefont {Markham}}, \bibinfo
  {author} {\bibfnamefont {R.}~\bibnamefont {Kolesov}}, \bibinfo {author}
  {\bibfnamefont {N.}~\bibnamefont {Mizuochi}}, \bibinfo {author}
  {\bibfnamefont {J.}~\bibnamefont {Isoya}}, \bibinfo {author} {\bibfnamefont
  {J.}~\bibnamefont {Achard}}, \bibinfo {author} {\bibfnamefont
  {J.}~\bibnamefont {Beck}}, \bibinfo {author} {\bibfnamefont {J.}~\bibnamefont
  {Tissler}}, \ and\ \bibinfo {author} {\bibnamefont {et~al.}},\ }\href
  {\doibase 10.1038/nmat2420} {\bibfield  {journal} {\bibinfo  {journal} {Nat.
  Mater.}\ }\textbf {\bibinfo {volume} {8}},\ \bibinfo {pages} {383} (\bibinfo
  {year} {2009})}\BibitemShut {NoStop}%
\bibitem [{\citenamefont {Fuchs}\ \emph {et~al.}(2009)\citenamefont {Fuchs},
  \citenamefont {Dobrovitski}, \citenamefont {Toyli}, \citenamefont
  {Heremans},\ and\ \citenamefont {Awschalom}}]{Fuchs_2009}%
  \BibitemOpen
  \bibfield  {author} {\bibinfo {author} {\bibfnamefont {G.~D.}\ \bibnamefont
  {Fuchs}}, \bibinfo {author} {\bibfnamefont {V.~V.}\ \bibnamefont
  {Dobrovitski}}, \bibinfo {author} {\bibfnamefont {D.~M.}\ \bibnamefont
  {Toyli}}, \bibinfo {author} {\bibfnamefont {F.~J.}\ \bibnamefont {Heremans}},
  \ and\ \bibinfo {author} {\bibfnamefont {D.~D.}\ \bibnamefont {Awschalom}},\
  }\href {\doibase 10.1126/science.1181193} {\bibfield  {journal} {\bibinfo
  {journal} {Science}\ }\textbf {\bibinfo {volume} {326}},\ \bibinfo {pages}
  {1520} (\bibinfo {year} {2009})}\BibitemShut {NoStop}%
\bibitem [{\citenamefont {Economou}\ and\ \citenamefont
  {Dev}(2016)}]{Economou_2016}%
  \BibitemOpen
  \bibfield  {author} {\bibinfo {author} {\bibfnamefont {S.~E.}\ \bibnamefont
  {Economou}}\ and\ \bibinfo {author} {\bibfnamefont {P.}~\bibnamefont {Dev}},\
  }\href {\doibase 10.1088/0957-4484/27/50/504001} {\bibfield  {journal}
  {\bibinfo  {journal} {Nanotechnology}\ }\textbf {\bibinfo {volume} {27}},\
  \bibinfo {pages} {504001} (\bibinfo {year} {2016})}\BibitemShut {NoStop}%
\bibitem [{\citenamefont {Ku}\ \emph {et~al.}(2017)\citenamefont {Ku},
  \citenamefont {Long}, \citenamefont {Wu}, \citenamefont {Bal}, \citenamefont
  {Lake}, \citenamefont {Barnes}, \citenamefont {Economou},\ and\ \citenamefont
  {Pappas}}]{Ku_PRA17}%
  \BibitemOpen
  \bibfield  {author} {\bibinfo {author} {\bibfnamefont {H.~S.}\ \bibnamefont
  {Ku}}, \bibinfo {author} {\bibfnamefont {J.~L.}\ \bibnamefont {Long}},
  \bibinfo {author} {\bibfnamefont {X.}~\bibnamefont {Wu}}, \bibinfo {author}
  {\bibfnamefont {M.}~\bibnamefont {Bal}}, \bibinfo {author} {\bibfnamefont
  {R.~E.}\ \bibnamefont {Lake}}, \bibinfo {author} {\bibfnamefont
  {E.}~\bibnamefont {Barnes}}, \bibinfo {author} {\bibfnamefont {S.~E.}\
  \bibnamefont {Economou}}, \ and\ \bibinfo {author} {\bibfnamefont {D.~P.}\
  \bibnamefont {Pappas}},\ }\href {\doibase 10.1103/PhysRevA.96.042339}
  {\bibfield  {journal} {\bibinfo  {journal} {Phys. Rev. A}\ }\textbf {\bibinfo
  {volume} {96}},\ \bibinfo {pages} {042339} (\bibinfo {year}
  {2017})}\BibitemShut {NoStop}%
\bibitem [{\citenamefont {Alegre}\ \emph {et~al.}(2007)\citenamefont {Alegre},
  \citenamefont {Santori}, \citenamefont {Medeiros-Ribeiro},\ and\
  \citenamefont {Beausoleil}}]{PhysRevB.76.165205}%
  \BibitemOpen
  \bibfield  {author} {\bibinfo {author} {\bibfnamefont {T.~P.~M.}\
  \bibnamefont {Alegre}}, \bibinfo {author} {\bibfnamefont {C.}~\bibnamefont
  {Santori}}, \bibinfo {author} {\bibfnamefont {G.}~\bibnamefont
  {Medeiros-Ribeiro}}, \ and\ \bibinfo {author} {\bibfnamefont {R.~G.}\
  \bibnamefont {Beausoleil}},\ }\href {\doibase 10.1103/PhysRevB.76.165205}
  {\bibfield  {journal} {\bibinfo  {journal} {Phys. Rev. B}\ }\textbf {\bibinfo
  {volume} {76}},\ \bibinfo {pages} {165205} (\bibinfo {year}
  {2007})}\BibitemShut {NoStop}%
\bibitem [{\citenamefont {Pedersen}\ \emph {et~al.}(2007)\citenamefont
  {Pedersen}, \citenamefont {M{\"o}ller},\ and\ \citenamefont
  {M{\"o}lmer}}]{Pedersen_2007}%
  \BibitemOpen
  \bibfield  {author} {\bibinfo {author} {\bibfnamefont {L.~H.}\ \bibnamefont
  {Pedersen}}, \bibinfo {author} {\bibfnamefont {N.~M.}\ \bibnamefont
  {M{\"o}ller}}, \ and\ \bibinfo {author} {\bibfnamefont {K.}~\bibnamefont
  {M{\"o}lmer}},\ }\href {\doibase 10.1016/j.physleta.2007.02.069} {\bibfield
  {journal} {\bibinfo  {journal} {Physics Letters A}\ }\textbf {\bibinfo
  {volume} {367}},\ \bibinfo {pages} {47} (\bibinfo {year} {2007})}\BibitemShut
  {NoStop}%
\bibitem [{Note1()}]{Note1}%
  \BibitemOpen
  \bibinfo {note} {We allow for arbitrary local operations, chosen ahead of
  time, to maximize fidelity.}\BibitemShut {Stop}%
\bibitem [{\citenamefont {Economou}\ and\ \citenamefont
  {Reinecke}(2007)}]{Economou_PRL07}%
  \BibitemOpen
  \bibfield  {author} {\bibinfo {author} {\bibfnamefont {S.~E.}\ \bibnamefont
  {Economou}}\ and\ \bibinfo {author} {\bibfnamefont {T.~L.}\ \bibnamefont
  {Reinecke}},\ }\href@noop {} {\bibfield  {journal} {\bibinfo  {journal}
  {Phys. Rev. Lett.}\ }\textbf {\bibinfo {volume} {99}},\ \bibinfo {pages}
  {217401} (\bibinfo {year} {2007})}\BibitemShut {NoStop}%
\bibitem [{\citenamefont {Economou}(2012)}]{Economou_PRB12}%
  \BibitemOpen
  \bibfield  {author} {\bibinfo {author} {\bibfnamefont {S.~E.}\ \bibnamefont
  {Economou}},\ }\href@noop {} {\bibfield  {journal} {\bibinfo  {journal}
  {Phys. Rev. B}\ }\textbf {\bibinfo {volume} {85}},\ \bibinfo {pages}
  {241401(R)} (\bibinfo {year} {2012})}\BibitemShut {NoStop}%
\bibitem [{\citenamefont {Economou}\ and\ \citenamefont
  {Barnes}(2015)}]{PhysRevB.91.161405}%
  \BibitemOpen
  \bibfield  {author} {\bibinfo {author} {\bibfnamefont {S.~E.}\ \bibnamefont
  {Economou}}\ and\ \bibinfo {author} {\bibfnamefont {E.}~\bibnamefont
  {Barnes}},\ }\href {\doibase 10.1103/PhysRevB.91.161405} {\bibfield
  {journal} {\bibinfo  {journal} {Phys. Rev. B}\ }\textbf {\bibinfo {volume}
  {91}},\ \bibinfo {pages} {161405} (\bibinfo {year} {2015})}\BibitemShut
  {NoStop}%
\bibitem [{\citenamefont {Riedel}\ \emph {et~al.}(2017)\citenamefont {Riedel},
  \citenamefont {S{\"o}llner}, \citenamefont {Shields}, \citenamefont
  {Starosielec}, \citenamefont {Appel}, \citenamefont {Neu}, \citenamefont
  {Maletinsky},\ and\ \citenamefont {Warburton}}]{PhysRevX.7.031040}%
  \BibitemOpen
  \bibfield  {author} {\bibinfo {author} {\bibfnamefont {D.}~\bibnamefont
  {Riedel}}, \bibinfo {author} {\bibfnamefont {I.}~\bibnamefont {S{\"o}llner}},
  \bibinfo {author} {\bibfnamefont {B.~J.}\ \bibnamefont {Shields}}, \bibinfo
  {author} {\bibfnamefont {S.}~\bibnamefont {Starosielec}}, \bibinfo {author}
  {\bibfnamefont {P.}~\bibnamefont {Appel}}, \bibinfo {author} {\bibfnamefont
  {E.}~\bibnamefont {Neu}}, \bibinfo {author} {\bibfnamefont {P.}~\bibnamefont
  {Maletinsky}}, \ and\ \bibinfo {author} {\bibfnamefont {R.~J.}\ \bibnamefont
  {Warburton}},\ }\href {\doibase 10.1103/PhysRevX.7.031040} {\bibfield
  {journal} {\bibinfo  {journal} {Phys. Rev. X}\ }\textbf {\bibinfo {volume}
  {7}},\ \bibinfo {pages} {031040} (\bibinfo {year} {2017})}\BibitemShut
  {NoStop}%
\bibitem [{\citenamefont {Hausmann}\ \emph {et~al.}(2013)\citenamefont
  {Hausmann}, \citenamefont {Shields}, \citenamefont {Quan}, \citenamefont
  {Chu}, \citenamefont {de~Leon}, \citenamefont {Evans}, \citenamefont {Burek},
  \citenamefont {Zibrov}, \citenamefont {Markham}, \citenamefont {Twitchen},\
  and\ \citenamefont {et~al.}}]{Hausmann_2013}%
  \BibitemOpen
  \bibfield  {author} {\bibinfo {author} {\bibfnamefont {B.~J.~M.}\
  \bibnamefont {Hausmann}}, \bibinfo {author} {\bibfnamefont {B.~J.}\
  \bibnamefont {Shields}}, \bibinfo {author} {\bibfnamefont {Q.}~\bibnamefont
  {Quan}}, \bibinfo {author} {\bibfnamefont {Y.}~\bibnamefont {Chu}}, \bibinfo
  {author} {\bibfnamefont {N.~P.}\ \bibnamefont {de~Leon}}, \bibinfo {author}
  {\bibfnamefont {R.}~\bibnamefont {Evans}}, \bibinfo {author} {\bibfnamefont
  {M.~J.}\ \bibnamefont {Burek}}, \bibinfo {author} {\bibfnamefont {A.~S.}\
  \bibnamefont {Zibrov}}, \bibinfo {author} {\bibfnamefont {M.}~\bibnamefont
  {Markham}}, \bibinfo {author} {\bibfnamefont {D.~J.}\ \bibnamefont
  {Twitchen}}, \ and\ \bibinfo {author} {\bibnamefont {et~al.}},\ }\href
  {\doibase 10.1021/nl402174g} {\bibfield  {journal} {\bibinfo  {journal} {Nano
  Letters}\ }\textbf {\bibinfo {volume} {13}},\ \bibinfo {pages} {5791}
  (\bibinfo {year} {2013})}\BibitemShut {NoStop}%
\bibitem [{\citenamefont {Li}\ \emph {et~al.}(2015)\citenamefont {Li},
  \citenamefont {Schr{\"o}der}, \citenamefont {Chen}, \citenamefont {Walsh},
  \citenamefont {Bayn}, \citenamefont {Goldstein}, \citenamefont {Gaathon},
  \citenamefont {Trusheim}, \citenamefont {Lu}, \citenamefont {Mower},\ and\
  \citenamefont {et~al.}}]{Li_NatCommun15}%
  \BibitemOpen
  \bibfield  {author} {\bibinfo {author} {\bibfnamefont {L.}~\bibnamefont
  {Li}}, \bibinfo {author} {\bibfnamefont {T.}~\bibnamefont {Schr{\"o}der}},
  \bibinfo {author} {\bibfnamefont {E.~H.}\ \bibnamefont {Chen}}, \bibinfo
  {author} {\bibfnamefont {M.}~\bibnamefont {Walsh}}, \bibinfo {author}
  {\bibfnamefont {I.}~\bibnamefont {Bayn}}, \bibinfo {author} {\bibfnamefont
  {J.}~\bibnamefont {Goldstein}}, \bibinfo {author} {\bibfnamefont
  {O.}~\bibnamefont {Gaathon}}, \bibinfo {author} {\bibfnamefont {M.~E.}\
  \bibnamefont {Trusheim}}, \bibinfo {author} {\bibfnamefont {M.}~\bibnamefont
  {Lu}}, \bibinfo {author} {\bibfnamefont {J.}~\bibnamefont {Mower}}, \ and\
  \bibinfo {author} {\bibnamefont {et~al.}},\ }\href {\doibase
  10.1038/ncomms7173} {\bibfield  {journal} {\bibinfo  {journal} {Nature
  Communications}\ }\textbf {\bibinfo {volume} {6}},\ \bibinfo {pages} {6173}
  (\bibinfo {year} {2015})}\BibitemShut {NoStop}%
\bibitem [{\citenamefont {von Bardeleben}\ \emph {et~al.}(2015)\citenamefont
  {von Bardeleben}, \citenamefont {Cantin}, \citenamefont {Rauls},\ and\
  \citenamefont {Gerstmann}}]{PhysRevB.92.064104}%
  \BibitemOpen
  \bibfield  {author} {\bibinfo {author} {\bibfnamefont {H.~J.}\ \bibnamefont
  {von Bardeleben}}, \bibinfo {author} {\bibfnamefont {J.~L.}\ \bibnamefont
  {Cantin}}, \bibinfo {author} {\bibfnamefont {E.}~\bibnamefont {Rauls}}, \
  and\ \bibinfo {author} {\bibfnamefont {U.}~\bibnamefont {Gerstmann}},\ }\href
  {\doibase 10.1103/PhysRevB.92.064104} {\bibfield  {journal} {\bibinfo
  {journal} {Phys. Rev. B}\ }\textbf {\bibinfo {volume} {92}},\ \bibinfo
  {pages} {064104} (\bibinfo {year} {2015})}\BibitemShut {NoStop}%
\bibitem [{\citenamefont {Aharonovich}\ \emph {et~al.}(2016)\citenamefont
  {Aharonovich}, \citenamefont {Englund},\ and\ \citenamefont
  {Toth}}]{Aharonovich_2016}%
  \BibitemOpen
  \bibfield  {author} {\bibinfo {author} {\bibfnamefont {I.}~\bibnamefont
  {Aharonovich}}, \bibinfo {author} {\bibfnamefont {D.}~\bibnamefont
  {Englund}}, \ and\ \bibinfo {author} {\bibfnamefont {M.}~\bibnamefont
  {Toth}},\ }\href {\doibase 10.1038/nphoton.2016.186} {\bibfield  {journal}
  {\bibinfo  {journal} {Nat. Photonics}\ }\textbf {\bibinfo {volume} {10}},\
  \bibinfo {pages} {631} (\bibinfo {year} {2016})}\BibitemShut {NoStop}%
\end{thebibliography}

\end{document}